\documentclass[a4paper,fleqn,usenatbib,useAMS,times]{mnras}
\usepackage[T1]{fontenc}
\usepackage{ae,aecompl}
\usepackage{newtxtext}
\usepackage{newtxmath}
\usepackage{graphicx}	
\usepackage{rotfloat} 
\usepackage{rotating}
\usepackage{float} 
\usepackage{graphicx}
\usepackage{longtable}
\usepackage{url}
\usepackage{dcolumn}
\usepackage{placeins}
\usepackage{multirow} 
\usepackage{hyperref}
\usepackage{pdflscape}
\usepackage{bm}		
\usepackage{ragged2e}
\usepackage[T1]{fontenc}
\usepackage[none]{hyphenat}
\usepackage{color,soul}
\usepackage{adjustbox}
\usepackage{multicol}
\usepackage{caption}
\usepackage{subcaption}
\usepackage{supertabular}
\usepackage[english]{babel}
\usepackage{blindtext}
\usepackage{tabu}
\usepackage{booktabs,caption,fixltx2e}
\usepackage{threeparttable}
\usepackage{array}
\usepackage{afterpage}
\usepackage{capt-of}
\usepackage[stable]{footmisc} 
\usepackage{booktabs}
\usepackage{natbib} 
\usepackage{threeparttable}
\renewcommand{\arraystretch}{1} 
\usepackage{Custom_Functions}
\usepackage{placeins}

\newcommand{\lya}{Ly$\alpha$}

\newcommand{\HI}{\mbox{H\,{\sc i}}}

\newcommand{\HeII}{\mbox{He\,{\sc ii}}}

\newcommand{\OII}{\mbox{O\,{\sc ii}}}
\newcommand{\OIII}{\mbox{O\,{\sc iii}}}

\newcommand{\OVI}{\mbox{O\,{\sc vi}}}

\newcommand{\CII}{\mbox{C\,{\sc ii}}}
\newcommand{\CIII}{\mbox{C\,{\sc iii}}}
\newcommand{\CIV}{\mbox{C\,{\sc iv}}}

\newcommand{\SiII}{\mbox{Si\,{\sc ii}}}
\newcommand{\SiIII}{\mbox{Si\,{\sc iii}}}
\newcommand{\SiIV}{\mbox{Si\,{\sc iv}}}

\newcommand{\SIII}{\mbox{S\,{\sc iii}}}

\newcommand{\NeVIII}{\mbox{Ne\,{\sc viii}}}

\newcommand{\MgII}{\mbox{Mg\,{\sc ii}}}

\defcitealias{Schaye2007}{S07}
\defcitealias{Kim2016}{KIM16}
\defcitealias{Haardt2012}{HM12}
\defcitealias{Khaire2016}{KS16a}
\defcitealias{Khaire2017}{KS18}
\defcitealias{Khaire2018}{KS19}
\defcitealias{Mohapatra2019}{M19}
\newcommand{\zabs}{$z_{\rm abs}$}

\newcommand{\be}{\begin{equation}}
\newcommand{\en}{\end{equation}}

\def\kms{km~s$^{-1}$}

\def\cmsq{cm$^{-2}$}
\def\cmcb{cm$^{-3}$}

\def\nce{$N({\CIII})$}
\def\nc2{$N({\CII})$}
\def\nhi{$N({\HI})$}
\def\si{${\sim}$}

\def\HM12{\citetalias{Haardt2012}}
\def\ks18{\citetalias{Khaire2018}}
\def\AS19{\citetalias{Mohapatra2019}}
\def\KIM16{\citetalias{Kim2016}}
\def\KS18{\citetalias{Khaire2018}}
\def\L18{\citetalias{Lehner2018}}
\defcitealias{Lehner2019}{L19}
\def\L19{\citetalias{Lehner2019}}

\newcommand{\CLOUDY}{\mbox{\scriptsize{CLOUDY}}}

\newcommand{\pirate}{$\Gamma_{\rm HI}$}
\newcommand{\fesc}{$f_{\rm esc}$}
\newcommand{\nh}{$n_{\rm \tiny H}$}

\defcitealias{Lehner2018}{L18}
\def\L18{\citetalias{Lehner2018}}
\defcitealias{wotta19}{W19}

\newcommand{\mnnh}{n_{\rm H}}
\newcommand{\mnh}{N_{\rm H}}

\title[Low$-z$ C~{\sc iii} absorbers]{Physical conditions and redshift evolution of optically thin C~{\sc iii} absorbers: Low$-z$ sample
}

\author[Mohapatra, A. et al.]{Abhisek Mohapatra$^{1}$\thanks{Contact e-mail: \href{abhisekphy@gmail.com}{abhisekphy@gmail.com}}, R. Srianand$^{2}$, Ananta C. Pradhan$^{1}$\
\\
\\
$^{1}$Department of Physics and Astronomy, National Institute of Technology, Rourkela, Odisha 769\ 008, India\\
$^{2}$Inter-University Centre for Astronomy \& Astrophysics, Postbag 4, Ganeshkhind, Pune 411\ 007, India\\}
\date{Accepted\dotfill. Received\dotfill; in original form\dotfill}

\pubyear{2019}

\begin{document}
\label{firstpage}
\pagerange{\pageref{firstpage}--\pageref{lastpage1}}
\maketitle

\begin{abstract}
We present a detailed analysis of
99 optically thin \CIII\ absorption systems at redshift, $0.2 \le z \le 0.9$ associated with 
neutral hydrogen column densities in the range, $15 \le {\rm log}$ \nhi\ (\cmsq) $\le 16.2$.
Using photoionization models, we infer the number density (\nh), C-abundance ($[C/H]$)
and line-of-sight thickness ($L$) of these
systems in the ranges, $-3.4 \le$ log {\nh} (in \cmcb) $\le -1.6$, $-1.6 \le [C/H] \le 0.4$, 
and 1.3 pc $\le L \le$ 10 kpc, respectively with most of the systems having sub-kpc scale thickness.
We combine the low$-z$ and previously reported high$-z$ ($2.1\le z\le 3.3$) optically thin \CIII\ systems
to study the redshift evolution and various
correlation between the derived physical parameters. We see a significant redshift evolution in \nh,
$[C/H]$ and $L$. We compare the redshift evolution of metallicity in \CIII\ systems with
those of various types of
absorption systems. We find that the slope of $[C/H]$ vs. $z$ for \CIII\ absorbers is stepper
compared to the redshift evolution of cosmic metallicity of the damped \lya\ sample
(DLAs) but consistent with that of sub$-$DLAs. We find the existence of strong
anti-correlation between $L$ vs. $[C/H]$ for the combined sample with a
significance level of 8.39$\sigma$.
We see evidence of two distinct $[C/H]$ branch \CIII\ populations
(low$-[C/H]$ branch, $[C/H]$ $\le -1.2$ and high$-[C/H]$ branch, $[C/H]$ $> -1.2$) in the combined \CIII\ sample
when divided appropriately in the $L$ vs. \nce\ plane. Further studies of \CIII\ absorbers
in the redshift range, $1.0 \le z \le 2.0$ is important to map the
redshift evolution of these absorbers and gain insights
into the time evolution physical conditions of the circumgalactic medium.
\end{abstract}

\section{Introduction}
Galaxies accrete gas from its surrounding regions in order to sustain the star forming activity over
long timescales ($\sim$Gyr) \citep{Fox2017}. In return, various stellar activities
push back the gas into the halo of the galaxy and/or into the circumgalactic medium (CGM) through winds \citep{Heckman2000, Veilleux2005, Zhang2018, Lochhaas2018}. 
The gas flows around a galaxy provide crucial information on the
initial mass function and star formation rate (SFR) history
of the host galaxy.
So, the measurement of chemical abundances and physical parameters of the CGM gas are essential to explore processes driving the evolution of galaxies.
However, it is impossible to detect these inflows/outflows in emission via direct observations
with current telescopes. Therefore, these gas
flows are usually detected in absorption lines embedded in the spectra of luminous
background sources \citep[such as quasars and gamma-ray bursts (GRBs);][]{Ford2014, Hafen2020, Ho2020}. 
Currently, high-resolution
spectrographs have the ability to trace the inflow/outflow in the intervening galaxies
with significant sensitivity and resolve the individual components up to a few \kms.

Metals observed in quasar spectra originating from 
galactic haloes (or CGM) around the intervening galaxies
are found to evolve with cosmic time \citep{Bergeron1986,Adelberger2005,Steidel2010,Keeney2017, Tumlinson2017,Muzahid2018}. 
In the past few decades, absorption line studies show that the photoionized gas with a temperature of about $10^4K$ are ubiquitous throughout the gaseous halo
of the galaxies up to impact parameters of \si100 kpc \citep{Stocke2013, Werk2013}.
Using the low redshift COS-Halos sample, \citet{Stern2016} have shown that for \si\ $L_{\ast}$ galaxies,
the haloes are
reservoirs of cool gas with masses in the range \si\ $10^{10}-10^{11}$ \(M_\odot\). 
Despite having ample of CGM surveys 
using quasar absorption lines, the origin and distribution of halo gas in the CGM is still uncertain.
However, it has been shown from several hydrodynamic simulations that the enrichment of
heavy elements in the CGM is predominantly due to the galactic winds \citep{Veilleux2005, MCcourt2015, Muratov2015}. 
The massive superwinds of supernova explosions drive the metal enriched out-flowing gas into the haloes of galaxies. 
The wind material goes through
fragmentation due to thermal instabilities to form clumps of clouds \citep{Field1965, Thompson2016}.
The clump of gas then cools with time and reaches pressure equilibrium with the surroundings \citep{MCcourt2015, Heckman2017} or become
self-gravitating \citep{Dedikov2004}. 
Recently, several simulation results also suggest that the condensations under thermal instabilities and fragmentation of cool gas 
are responsible for forming small cloudlets (sizes \si\ pc) in the galactic halos
\citep{McCourt2018,Liang2020}.

Primarily, \MgII\ $\lambda\lambda$2976, 2803 doublet transitions have been 
used over the years to trace the relationship between metal ions and parent galaxies at low to intermediate redshifts (0.3 $\le z \le$ 2.5) 
\citep{Petitjean1990, Chen2010, Dutta2017,Joshi2017}.
However, the scenario has drastically changed with the use 
of other metal ions such as low ionized ions of \SiII, \CII, \SiIII,
\citep{Narayanan2005,Herenz2013,Muzahid2018, Bouma2019} 
intermediate to high ionized ions of \SiIV, \CIV, \OVI, \NeVIII\ \citep{Chen2001,Bergeron2005,Muzahid2014, Muzahid2015,  Hussain2017}, etc., with the help of high-quality and high-resolution
UV spectroscopy. 

Measurements of column densities of successive ionization stages of an element is advantageous to infer the ionization parameter (U) (alternatively \nh) of
photoionized cool absorbing gas with an assumed incident ultraviolet background (UVB) radiation. Additionally,
\nhi\ of
the absorbing gas is further used to constrain the physical parameters accurately. 
Hence, large surveys of successive ions of heavy elements
with additional coverage of \HI\ are potentially essential to identify the underlying mechanism of the stability and origin of the absorbing gas.
With large surveys, we wish to capture the broad statistics of absorption systems and study various correlations between the derived physical and chemical parameters. This can further enhance our understanding of the absorbers and provide essential ingredients to perform more realistic hydrodynamic simulations of galaxy formation and the evolution of IGM/CGM.

However, the luxury of finding such idealized absorbing gas sample is limited due to various constraints such as, 
wavelength coverage of all the ions, blending due to the other ionic transitions, detection limits 
due to signal to noise ratio (SNR), etc. Mostly, carbon is
used as a common tracer of metals associated with \HI\ absorbing gas originating from the IGM/CGM 
\citep{Cowie1995,Schaye2007, Boksenberg2015}.
The \CIV\ $\lambda\lambda$1548, 1550 doublet and \CII\ $\lambda$1334 lie
redward to the \lya\ emission peak of the quasar
making them easily detectable and relatively free from contamination.
On the other hand, \CIII\ $\lambda$977 falls inside the \lya\ forest usually blended by other higher-order \HI\ Lyman series transitions.
Due to difficulty in finding clean \CIII\ absorption lines and the dearth of \CII\ detection in the low$-$\nhi\ absorption systems, the physical and chemical properties of these absorbing gas often remain poorly constrained.
However, considering ionization state, oscillatory strength, and elemental abundance, the \CIII\ line is one of the best and strongest ultraviolet (UV) transition to study the optically thin gas absorbers.

\citet{Kim2016} have produced a systematic survey of optically thin \CIII\ absorber sample at high redshift, 
$2.1 \le z \le 3.3$. 
In \citet{Mohapatra2019} [hereafter, \citetalias{Mohapatra2019}], we have analysed 
this sample using the updated \citet{Khaire2018} UVBs to explore 
the redshift evolution of these 
\CIII\ absorbers. We also emphasized the use of accurate UVB in photoionization (PI) models and its 
impact on the derived parameters in \AS19.
We found various interesting correlations and anti-correlations of the inferred parameters. We suggested that
the anti-correlation between line-of-sight thickness ($L = \mnh/\mnnh$) and C-abundances ($[C/H]$) can be well explained using a
simple toy model where $L$ is driven by the cooling time ($t_{cool}$)
as seen in the fragmentation of clouds due to thermal instabilities \citep{McCourt2018}.
The \CIII\ absorbing gas studied in \AS19\ are expected to be predominantly originated from the galactic haloes or gas around the galactic discs via inflows or outflows. 
However, we do not find any confirmed detection of galaxy sources within one Mpc impact parameter in the archival spectroscopic or photometric data. 

Motivated by the aforementioned findings of \AS19, we have further extended our study of the 
\CIII\ absorbers to the low redshift range, 0.2 $\le z \le$ 0.9 
to explore whether the 
\CIII\ selected absorbers substantially trace the cool fragmented gas clouds in the CGM. 
Note that, the
\CIII\ absorption is expected to be relatively less contaminated by the higher-order \HI\ Lyman series absorption at 
low$-z$ than at high$-z$ and hence, are more reliable in deriving the metallicity of the gas.
To derive the physical and chemical parameters of the low$-z$ \CIII\ absorbers, we have used 
grids of PI models.
In the PI 
models, we have implemented the recently updated \citet{Khaire2018} [hereafter \citetalias{Khaire2018}]
UVBs using quasar spectral index ($\alpha$) values, $-1.6$, $-1.8$ and $-2.0$ to study the uncertainties in the derived parameters.
Combining the low$-z$ \CIII\ absorbers with the high$-z$ \CIII\ absorbers from \AS19, we 
present a comparative study of the \CIII\ absorbers in a wide redshift range of $0.0\le z\le 3.3$. 
Note that, we do not find any galaxy association for the high$-z$ \CIII\ absorbers in \AS19. In contrast, the low$-z$ \CIII\ absorbers are a part of the COS-CGM survey with at least one galaxy within one Mpc projected distance.

This paper is arranged as follows. In \S \ref{data},
we describe the details of the low$-z$ \CIII\ sample. We briefly explain the PI models
and present the physical and chemical properties of the low$-z$ optically thin \CIII\ absorbers in \S \ref{PImodel}.
In \S \ref{discuss}, we discuss the redshift
evolution of various
parameters and analyse various correlations between the absorber
properties. 
In \S \ref{result},
we discuss our results and emphasize the need for a large survey of homogeneous optically thin \CIII\ absorbers 
across the redshift range which can be used
as an essential tracer of the cool fragmented gas around the CGM.
Throughout this study we use $\Lambda CDM$ cosmology 
with $H_0 =70$ {\kms} $Mpc^{-1}$ , $\Omega_\Lambda$ = 0.7, and $\Omega_m$ = 0.3.
We denote $[A/B] = {\rm log}\,(A/B)- {\rm log}\,(A/B)_{\odot}$ as the abundances of metals and use solar relative abundances values
from \citet{Grevesse2010}.

\section{Data}
\label{data}
\citet{Lehner2018} [hereafter \citetalias{Lehner2018}] have produced a catalog of 224 \HI-selected absorption systems
referred to as COS CGM compendium (CCC) using 335 background quasar spectra. 
The quasar spectra of this CCC sample have been observed using the {\it{G130M and/or G160M}} gratings of the {\it{Cosmic Origins Spectrograph (COS)}} on the {\em Hubble Space Telescope (HST)}. 

There are 152 strong \lya\ forest systems (SLFs) detected in the CCC with \nhi\ (\cmsq) in range, $15.0 \le$ log \nhi\ $\le 16.2$ which is 
spanning over a redshift range,
$0.2 \le z \le 0.9$. 
The absorption systems are systematically analysed by \L18\ to
obtain the Doppler parameters ($b$), \nhi\ and the column densities of other observed heavy elements.
They have reported the integrated \nhi\ and the column densities of the other metal ions of each absorption system. The complete information on detection and column densities of the observed ions is provided by \L18. 
The detection of several metal ions along with \HI\ is helpful to constrain the physical parameters of the cool CGM gas absorbers.
\citetalias{Lehner2018} have reported that the lower column density SLFs trace
the diffuse CGM gas as well as their transition phase
from the CGM to \lya\ forest/IGM. The quasar spectra of the SLFs sample have coverage of the wavelength range of both \CII\ and \CIII\ ions which allows us
to perform PI modelling as explained in \citetalias{Mohapatra2019}.

We extract a total of 99 \CIII\ absorption systems from the SLFs in the CCC sample. Out of these 99 \CIII\ absorption systems,
we find 67 well detected \CIII\ systems and 32 systems with detections albeit of having saturated \CIII\ profile or
blended by
other metal line transitions which are considered as lower limit \CIII\ systems. 
Similarly, in case of \CII,
we have firm detections for 18 systems only whereas for the remaining 81 systems we obtain upper limits. The low$-z$ \CIII\ absorption systems have the following flags on the measurement of column density:
\begin{enumerate}
\item Nine clean systems where we could measure the column densities of both \CIII\ and \CII.
\item 58 systems with clean detection of \CIII\ and upper limits on \CII\ ions.
\item 23 systems where we could obtain lower limits on \CIII\ and upper limits on \CII\ ions.
\item Nine systems with clean detections of \CII\ ions and with lower limits on \CIII\ ions due to line saturation.
\end{enumerate}
Since the low$-z$\, \CIII\ sample has about 32\% absorbers with lower limits on \CIII, we use the column densities of other detected metal ions (for e.g., \SiIII, \SiII, \OIII\ and \OII) to further constrain the \nh\ in our PI models. We find 24 \SiIII\ $+$ \SiII\ and 24 \OIII\ $+$ \OII\ absorption associated with the low$-z$ \CIII\ sample. Hence, we have categorized total 99 low$-z$ \CIII\ absorption systems based on observed $N(\CIII)/N(\CII)$ as follows:\\
\noindent (i) ${\bf Sample\,A\, (SA)}${\bf :} It consists of 51 \CIII\ systems which is further divided into two sub-samples on the basis of the presence of other metal ions:\\
${\bf SA1}${\bf :} It consists of nine absorption systems with clean \CIII\ $+$ \CII\ detections (for such absorption systems, Col ID (8) on observed $N(\CIII)/N(\CII)$ is denoted by {\lq{0}\rq} in Table~\ref{pitable}). Out of which, seven systems have
only \CIII\ and \CII\ detection,
(IDs 18, 19, 30, 39, 56, 66 and 76 in Table~\ref{pitable}) and the other two systems also have \SiIII\ and \SiII\ detections (IDs 22 and 35 in Table~\ref{pitable} and \ref{tab:si}).\\
${\bf SA2}${\bf :} It consists of 42 absorption systems (shown in Table~\ref{tab:si} and \ref{tab:oxy}) with lower limits on observed $N(\CIII)/N(\CII)$ but with the presence of other metal ions (Col ID (8) denoted by {\lq{$-2*$}\rq} in Table~\ref{pitable}). 
Out of which,
22 absorption systems have \SiIII\ $+$ \SiII\ associations (Table~\ref{tab:si}) and 24 absorption systems have \OIII\ $+$ \OII\ associations (Table~\ref{tab:oxy}).
There are four absorption systems (marked with superscript {\lq{*}\rq} on the IDs in Table~\ref{tab:si} and ~\ref{tab:oxy}) which have both \SiIII\ $+$ \SiII\ and \OIII\ + \OII\ ions associations. For systems in this sample, we could get
stringent constraints on \nh\ and other parameters from our photoionization models.\\
(ii) ${\bf Sample\, B\, (SB)}${\bf:} It consists of 48 absorption systems which have lower limits on N(\CIII)/N(\CII) without presence of any other metal ion combinations. Out of these 48 absorption systems in SB, we have 39 absorbers with clean \CIII\ detections and upper limits on \CII. Other nine absorbers in this sample have lower limits on N(\CIII)/N(\CII) due to saturated or blended \CIII\ profiles and/or upper limits on \CII.
Since, the absorption systems in this sample have only lower limits on the observed \nce/$N(\CII)$ (Col ID (8) denoted by {\lq{$-2$}\rq} in Table~\ref{pitable}) without the presence of other detected metal ion combinations, 
we could only derive upper limits on \nh\ and other inferred parameters from our PI models for these absorbers.
\section{PI Models}
\label{PImodel} 
We have constructed PI
models for \CIII\ absorbing gas using the simulation code {\CLOUDY} version c17.02\footnote{{\CLOUDY} version c13.03 \citep{Ferland2013} was used in \AS19\ for PI models of the high$-z$ absorbers. However, we note that change in \CLOUDY\ version does not alter our inferred results obtained from the PI models.} \citep{Ferland2017}. 
The extragalactic background radiation spectrum of \KS18\ is inbuilt in this version of {\CLOUDY} and can be included in the model through
{``\textit {table input}''} command. The basic assumptions in our single phase
constant density photoionization models for the absorbing gas are: (i) it has a plane parallel geometry (ii) it is in ionization and thermal equilibrium with an assumed incident
UVB radiation and (iii) it has negligible dust depletion.
We consider solar relative metallicity measured by \citet{Grevesse2010} 
which is also the default metal abundances in \CLOUDY.

\begin{figure}
  \centering
  \includegraphics[totalheight=0.26\textheight, trim=.1cm 1.5cm 0cm 0.6cm, clip=true, angle=0]{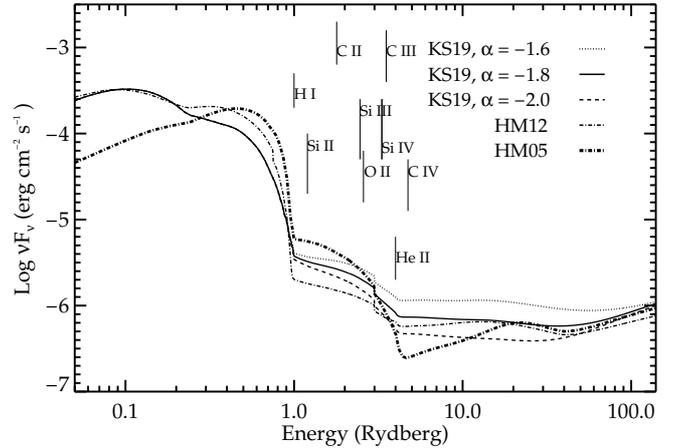}
\caption{
Background spectra predicted by different UVB models at $z =  0.5$ for comparison.
The \KS18\ UVBs with $\alpha$  = $-$2.0, $-$1.8 and $-$1.6 are shown
using dashed, solid and dotted lines, respectively. We also denote the 
\HM12\ and HM05 UVBs with dashed-dotted and thick dash-dotted lines, respectively. The
ionization energies of several ions are shown using vertical lines.}
\label{fig1}
\end{figure}
\subsection{The UVBs at low$-z$} 
\label{UVB}
The ionization states
of metals seen in the CGM/IGM are assumed to be maintained by PI equilibrium with the UVB radiation. 
In our PI models,
we have used the updated \KS18\ UVB contributed by quasars and galaxies (Q $+$ G) as input
ionizing source instead of the frequently used UVBs in the past (i.e., HM05 and
\citet{Haardt2012} (hereafter \citetalias{Haardt2012})).
In Fig.~\ref{fig1}, we show the \KS18\ UVB spectra for 
$\alpha$ = $-1.6$, $-1.8$ and $-2.0$ along-with the two HM UVBs (HM05 and \HM12 UVBs) at $z$ = 0.5. The ionization energies
of different ions (such as \HI,
\SiII, \CII, \SiIII, \CIII, \SiIV, \CIV, \HeII, etc.) are marked with vertical lines.
As seen in Fig.~\ref{fig1}, the \KS18\ UVBs differ significantly from the HM UVBs due to the different model parameters. The \KS18\ UVBs exhibit shallower slope in comparison to HM05 UVB
between 1.5 $-$ 4 Ryd.
Hence, use of HM05 UVB in the PI models as compared to the \KS18\ UVBs will under produce model predicted column densities of ions with
ionization potential (IP) in the range, $1.5-4$ Ryd (e.g., \SiIV, \CIII\ and \CIV)
relative to the ions with lower IP (e.g., \SiII\ and \CII).
\KS18\ have incorporated the updated quasar emissivity from \citet{Khaire2015} and the SFR density of the low$-z$ 
galaxies from \citet{KhaireApJ2015}. 
As a result at low$-z$, the \KS18\ UVB models have higher \HI\ ionizing photons than the \HM12\ UVB.
Also, the \HI\ PI rates (\pirate) 
estimated at low$-z$ using the \KS18\ UVBs reproduce
the measurements of \citet[]{Shull2015,Kollmeier2014,Gaikwad2017} and \citet[]{UVGaikwad2017} consistently at different redshifts. \KS18\ have also used the updated \nhi\
distribution reported in \citet{Inoue2014} to calculate the opacity of the IGM.

The shape of the UVB mostly relies on (i) the spectral energy distribution (SED) of quasars (basically depends on the value of the spectral index $\alpha$) and (ii) the escape fraction({\fesc}) of \HI\ ionizing photons escaping from the galaxies.
However, at redshift $z<1$, the integrated intensity of the UVB (at $E>13.6$ eV) is mainly dominated by quasars.
Hence, the extragalactic UVB radiation at low$-z$ is mostly contributed by quasar SED.
The quasar SED is usually calculated by a power law, 
$ f_{\nu} \propto \nu^\alpha $, with $\alpha$ values in Far-UV (FUV) are 
in the range, $-0.72$ to $-1.96$ \citep{Vanden2001,Telfer2002, Scott2004, Shull2012, Stevans2014, Lusso2015, Tilton2016}.
Recently, \citet{Khaire2017} has shown that the UVB model with $\alpha = -1.8$
reproduces the
{\HeII} {\lya} optical depth measurements in different redshift ranges \citep[also see][]{Prakash2019}.
Hence, we use \KS18\ UVB with $\alpha = -1.8$ as our fiducial UVB throughout this work. 
We also calculate the ranges of different cloud parameters 
[density ({\nh}), metallicity($[C/H]$), over-density ($\Delta$), and line-of-sight length, (L = $N$(H)/\nh)] considering \KS18\ UVBs with $\alpha = -1.6$ and $-2.0 $.

\subsection{PI modeling of the low$-z$ {\CIII} absorbers}
\label{s3}
The average gas temperature corresponding to the \CIII\ sample obtained from \citetalias{Lehner2018} is $T \le 5 \times 10^4$ K which is estimated from the
$b$ parameter of \HI\ absorption. At this temperature, photoionization is the dominant ionization mechanism for these gas absorbers.
Hence, the required ionization correction is obtained using the ratio of the column density of metal ions having well-aligned successive ionization stages.
The observed column density ratio of
two consecutive
ionized states of an element 
(e.g., $\CII/\CIII$, $\SiII/\SiIII$, \OII/\OIII\, etc.) mildly depends on the metallicity but it varies significantly with the shape and intensity of the assumed UVB radiation field, temperature and the density of the absorbing gas \citep{Lehner2013, Werk2014,Hussain2017,Muzahid2018}.
The ionization parameter, $U$ (or \nh) of an absorbing gas is constrained by matching the observed column density ratios of successive ionized states of metal ions with the corresponding PI model predicted ratios for an assumed ionizing UVB.
Moreover, detection of \HI\ absorption originating from these 
systems help us to fix the metallicity by matching individual observed column densities heavy elements and \nhi\ with the PI model predicted column densities.
The necessary initial conditions of PI models for each cloud are: (1) the ionizing UVB, (2) assumed hydrogen density (\nh), (3) the stopping criteria and (4) the chemical composition of the gas.

In our PI models, we use the fiducial \KS18\ UVB (with $\alpha=-1.8$) corresponding to \zabs\ of each absorber as the input ionizing radiation.
We use the observed \nhi\ for each \CIII\ absorption systems as the stopping criteria to terminate the \CLOUDY\ computations.
We generate grids of PI models
for \nh, $-5 \le$ log \nh\ (\cmcb) $\le -1$, with a grid size of 0.05 dex.
Then, we equate the observed column density ratio of the consecutive ionization stages of all the available metal ions
(\CII/\CIII, \SiII/\SiIII\ and \OII/\OIII)
with the model predicted column density ratio to uniquely constrain the U (or \nh) of the absorber. 
We use this constrained \nh\ 
to generate another grid of PI models varying the metallicity $[X/H]$ (X = C, Si and O)
in the range, $[X/H]$ = [$-$2, 0.5] (grid size of 0.05 dex).
Then we match the individual column densities of
respective metal ions (\CII, \CIII, \SiII, \SiIII, \OII\ and \OIII) with the model predicted column densities to
uniquely fix the respective ion abundances (i.e.,
C$-$abundance $[C/H]$, Si$-$abundance $[Si/H]$ and O$-$abundance $[O/H]$) for an absorption system.
We allow \CLOUDY\ to
determine the kinetic gas temperature self-consistently
under PI equilibrium.

The PI models for the sub-sample SA1 are ideal as these systems have clean detections of \CIII\ and \CII. For SA2, we use other metal ion combinations to put stringent constraint on the derived \nh\ and hence, the \nh\ obtained from these
systems are consistent with other observed ion ratios (for e.g., $N(\SiIII/N(\SiII)$ and/or $N(\OIII/N(\OII)$) as discussed in the following section. As mentioned in \S \ref{data}, for sub-sample SA (SA1 + SA2), we stress that the physical and chemical parameters obtained using our PI models are well constrained.
However, for the sub-sample SB which has 48 lower
limit systems, we could only put limits on the derived parameters from our PI models. 

\begin{figure*}
\centering
\begin{subfigure}{.5\textwidth}
  \centering
  \includegraphics[totalheight=0.325\textheight, trim=.1cm 0.5cm 0cm 0.6cm, clip=true, angle=0]{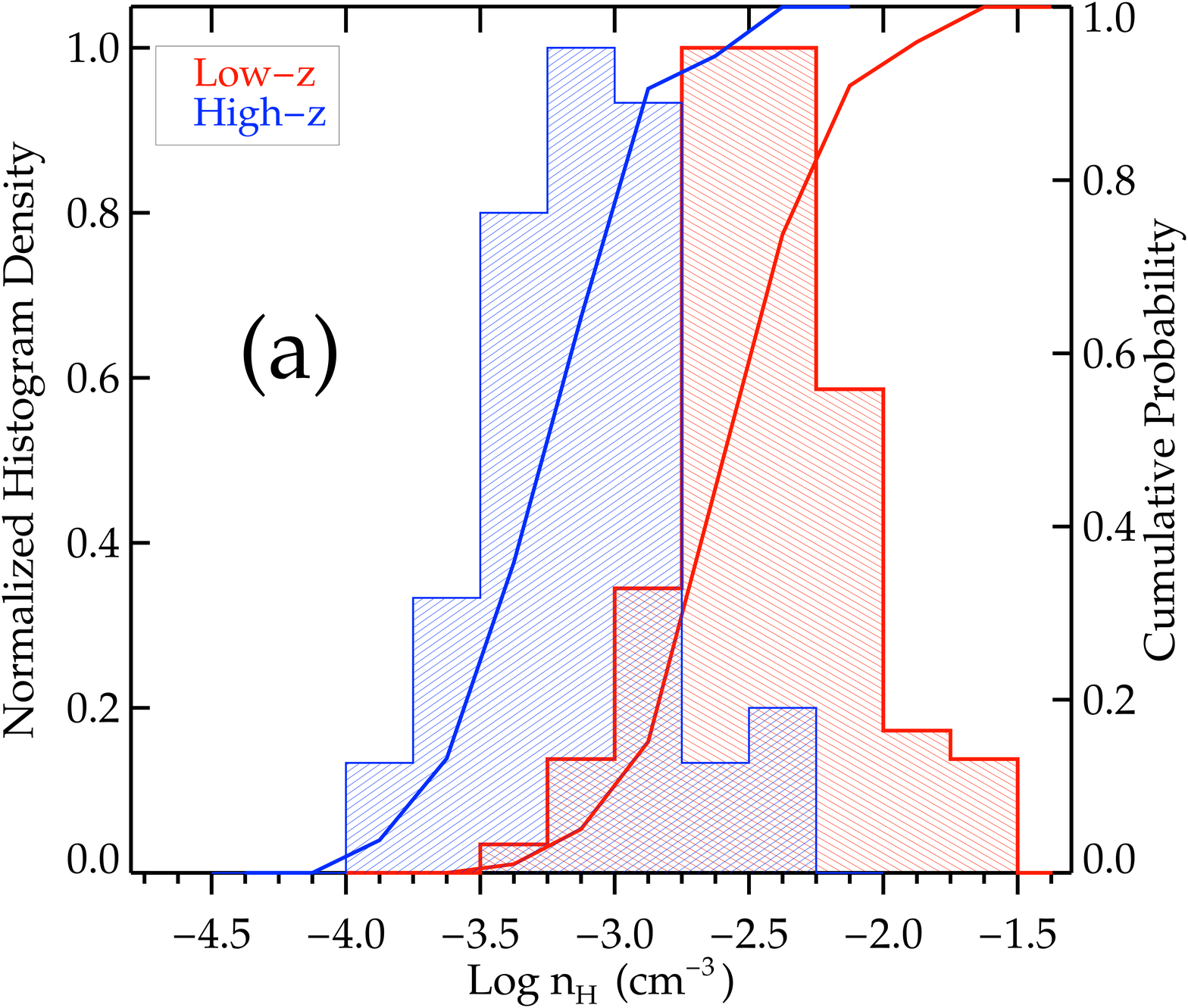}
\end{subfigure}%
\begin{subfigure}{.5\textwidth}
  \centering
 \includegraphics[totalheight=0.325\textheight, trim=.1cm 0.5cm 0cm 0.6cm, clip=true, angle=0]{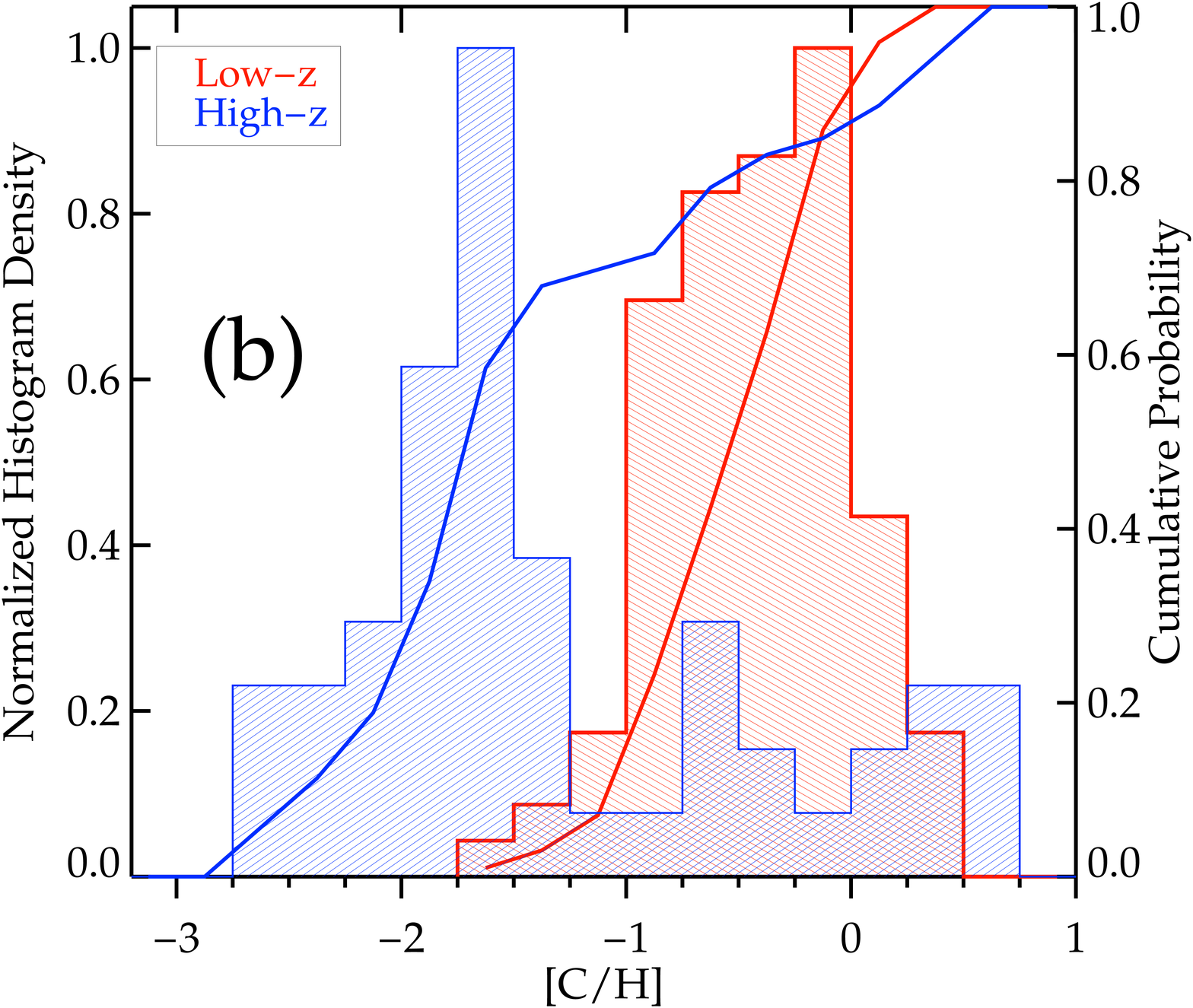}
\end{subfigure}
\begin{subfigure}{.5\textwidth}
  \centering
  \includegraphics[totalheight=0.325\textheight, trim=.1cm 0.5cm 0cm 0.6cm, clip=true, angle=0]{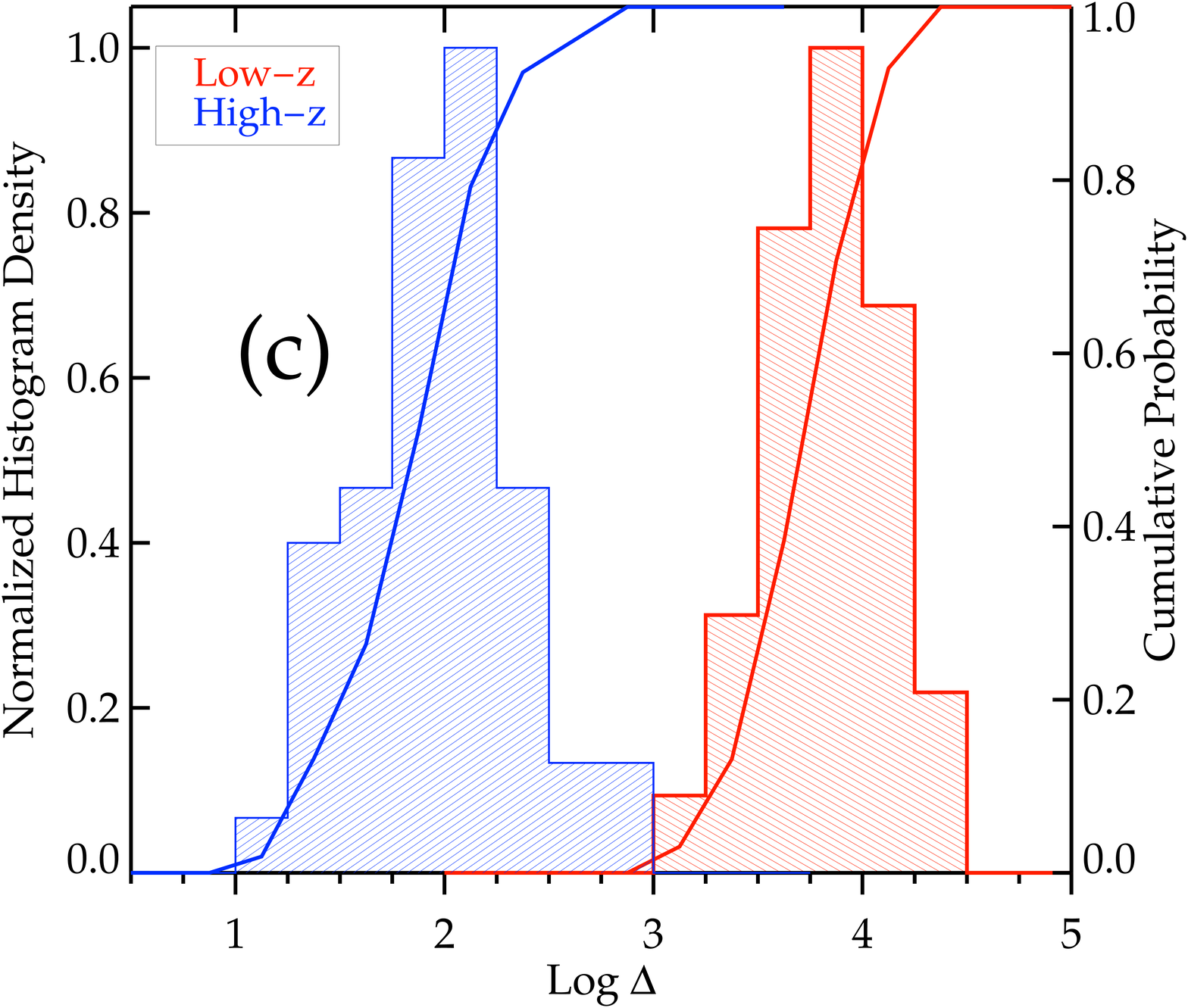}
\end{subfigure}
\begin{subfigure}{.5\textwidth}
  \centering
 \includegraphics[totalheight=0.325\textheight, trim=.1cm 0.5cm 0cm 0.6cm, clip=true, angle=0]{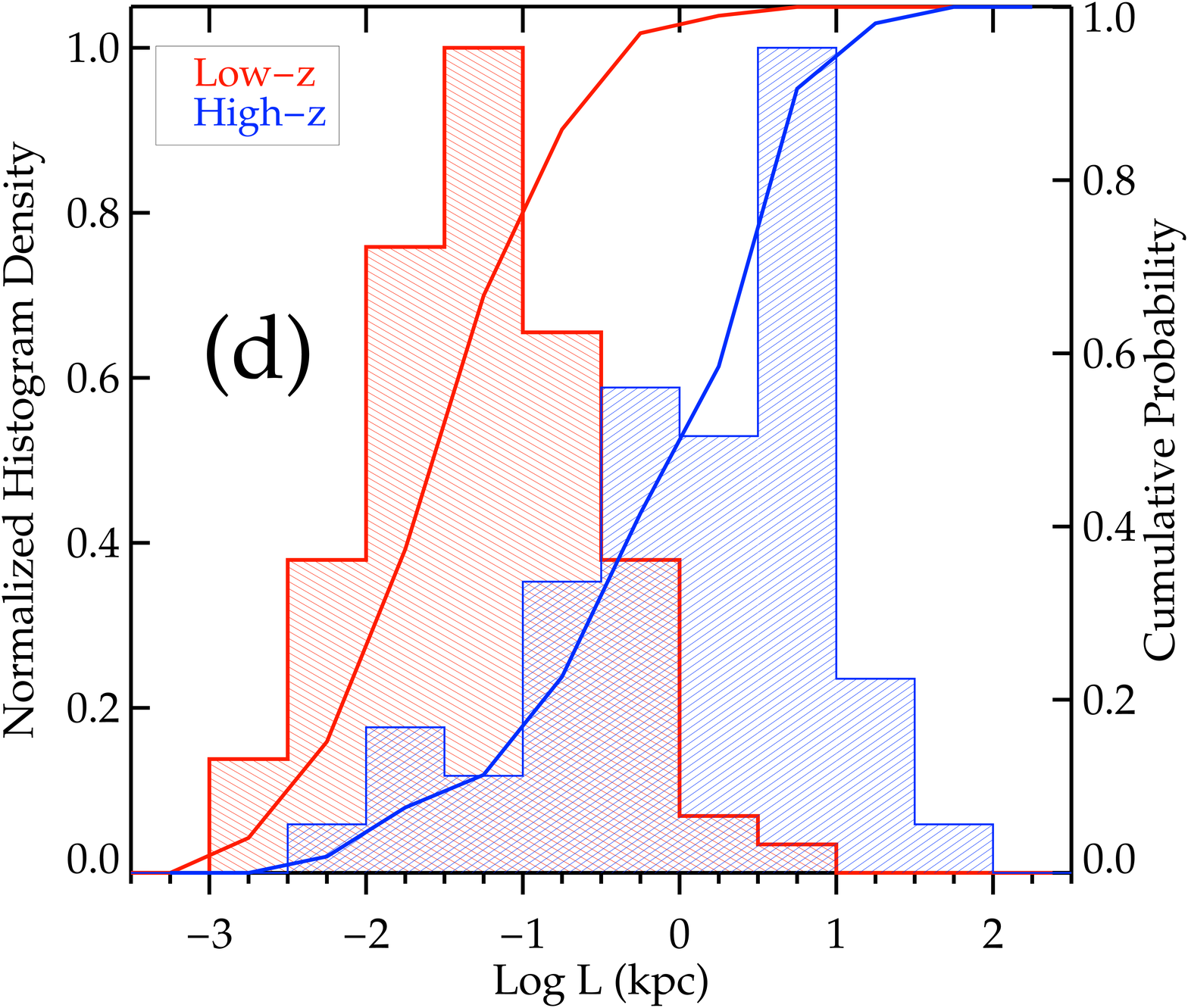}
\end{subfigure}
\caption{Normalized histogram density (NHD) and cumulative probability distribution (CPD) of
derived parameters ({\nh},\, $[C/H]$,\, $\Delta$,\, and L) of the low$-z$ \CIII\ absorption systems obtained
using our PI models with the fiducial \KS18\ UVB. We also 
show the NHD and CPD of derived parameters for
high$-z$ \CIII\ system-wide analysed data, S3 (for details of the samples, see \AS19). We have used
red colour for the data analysed in this paper and blue for high$-z$ sample S3 from \AS19, respectively.
}
\label{fig45}
\end{figure*}

\begin{figure*}
  \centering
   \includegraphics[totalheight=0.28\textheight, trim=0cm 0.0cm 0cm .0cm, clip=true, angle=0]{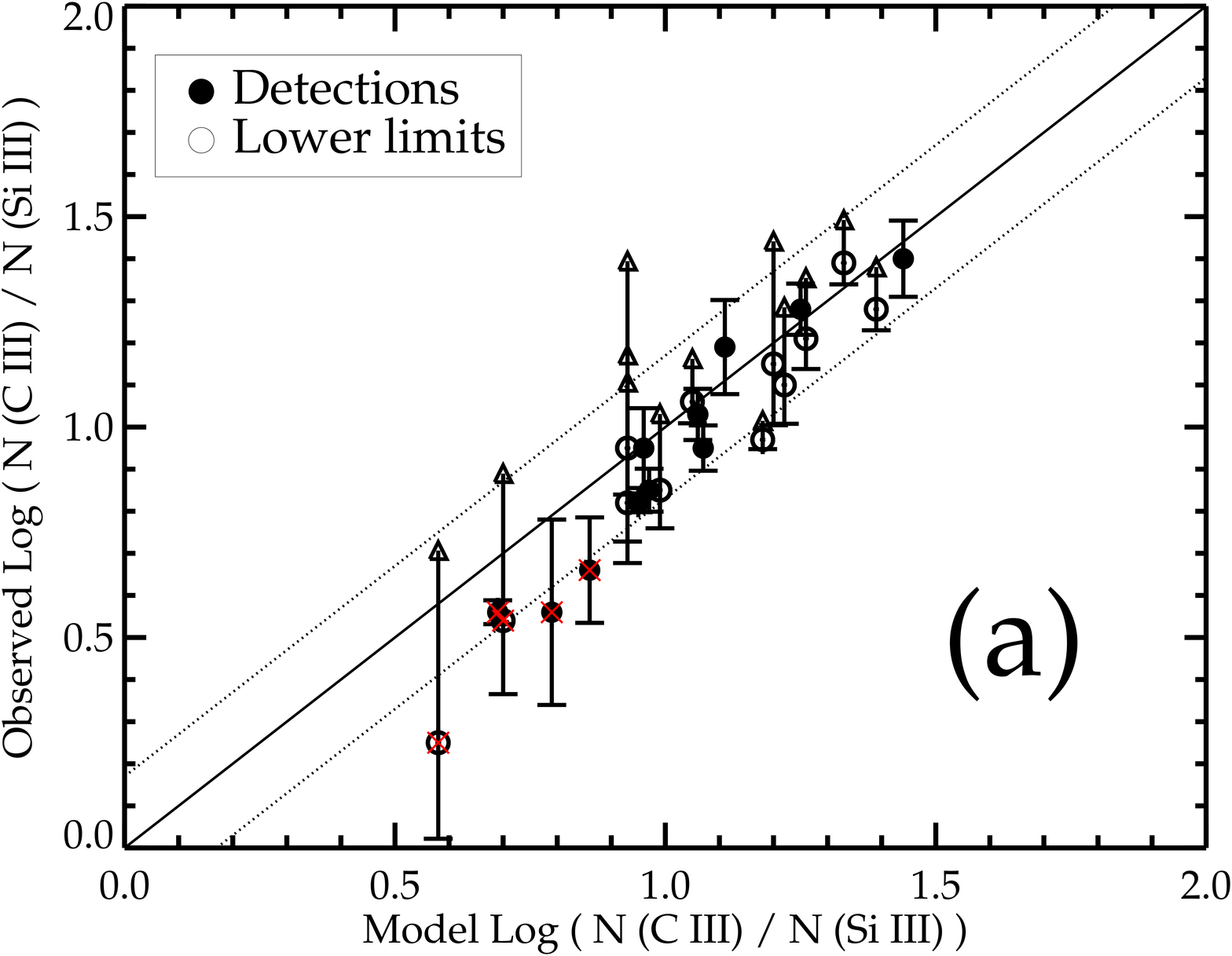}
   \includegraphics[totalheight=0.28\textheight, trim=0cm 0.0cm 0cm .0cm, clip=true, angle=0]{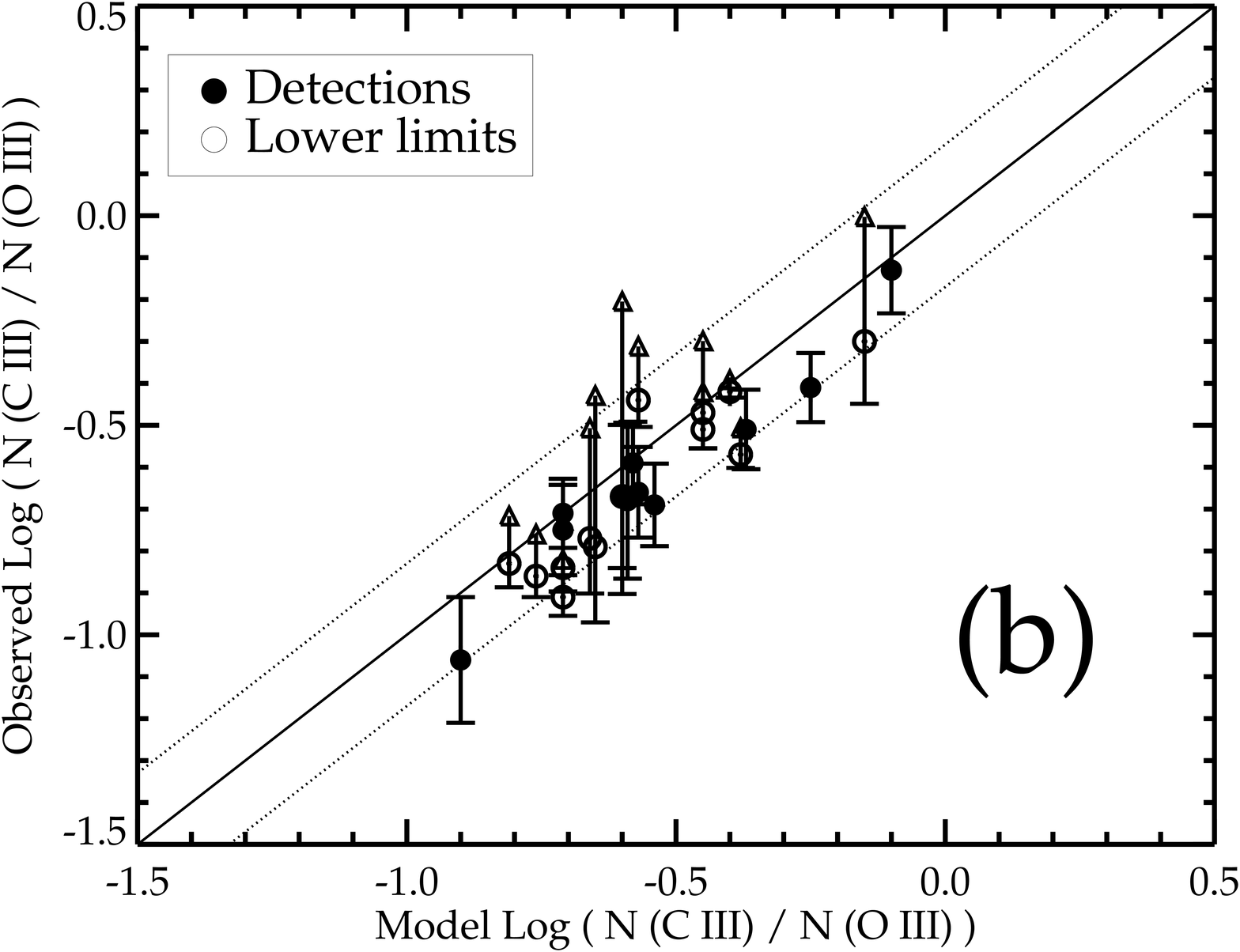}
   \caption{ Left panel: Observed N(\CIII)/N(\SiIII) is plotted against the model predictions in logarithmic scale.
   Filled and open circles represent the measurements and lower limits on column density ratios, respectively. The errors in the observed column densities are shown by vertical bars. The solid line represents the `y $=$ x' equality line, whereas the dotted line shows the uncertainty range of $\pm$ 0.17 dex within this equality line owing to the model uncertainties. Absorbers with super solar $[Si/C]$ values ($[Si/C] \ge 0.3$) are denoted with red cross marks. 
   Right panel: same as the left panel except for the observed N(\CIII)/N(\OIII) is plotted against the model values in logarithmic scale.}
\label{fig_osi}
\end{figure*}

\subsubsection{Results from the PI models}
\label{piresult}
Here, we discuss about various parameters constrained using our PI models with the fiducial \KS18\ UVB.
Also, we
systematically analyse the consistency
of the derived parameters of the two sub-samples, $SA$ and $SB$. 
We have provided the details of the PI model predicted parameters for all the 99 low$-z$ \CIII\ absorption systems in Table~\ref{pitable}.

\vskip 0.1in
\noindent{\bf Hydrogen density (\nh):} In panel (a) of Fig.~\ref{fig45}, we have shown the normalized histogram
density (NHD) and the cumulative probability density
(CPD) distributions of \nh\ for the low$-z$ \CIII\ sample in red colour. 
The \nh\ values for the low$-z$ \CIII\ absorption systems derived from the PI models are in the range, 
$-3.4 <$ log {\nh} (in \cmcb) $< -1.6$, with the mean and median values being $-2.45$ and $-2.48$ dex, respectively.
Similarly, the \nh\ ranges for the two sub-samples SA and SB are $-3.4 \le$ log {\nh} (in \cmcb) $\le -1.8$
and $-3.1 \le$ log {\nh} (in \cmcb) $\le -1.6$, respectively.
As the sub-sample SB ( $\simeq$ 48\% of the absorption systems of the low$-z$ \CIII\ sample)
has only limits on the column density ratio of \nce/\nc2,
we could only derive the upper limits on \nh\ for these absorption systems from our PI models.
The median \nh\ values of the two sub-samples differ only by 0.14 dex which indicates that the derived \nh\ for both the samples $SA$ and $SB$ are similar.
We also notice that 39 systems 
from SB show similar \nh\ distribution (i.e considering \nh\ lower limits as detections)  like SA with a median density difference of 0.19 dex.
We find a median difference of 0.3 dex in \nh\ between our prediction and previously reported
values for these absorbers in \citet{Lehner2019} [hereafter \citetalias{Lehner2019}]. This discrepancy is due
to the use of updated UVB in our PI models \citep[see,][for similar discussions for Ne~{\sc viii} absorbers]{Hussain2017}.
We see that the use of \KS18\ UVBs with $\alpha = -1.6$ and $-2.0$ for any absorber
causes a maximum uncertainty of 0.34 dex in derived log \nh\ values.
Hence, we associate an error of $\pm 0.17$ dex in the derived log \nh\ for our fiducial model, mostly due to the uncertainty of UVBs irrespective of the observational errors.

For comparison, we have also shown the \nh\ distribution of the high$-z$ \CIII\ sample (S3) from \AS19\ in panel (a) of Fig.~\ref{fig45} in blue colour.
Since we are using total integrated column densities for low$-z$ absorbers, the appropriate comparison will
be with the system-wide analysed sample S3 rather than component-wise analysed sample S1 $+$ S2 of \AS19.
As seen in the figure, the peaks of the \nh\ distribution show a significant difference between low$-z$ and high$-z$ \CIII\ samples.
We perform z-test on the \nh\ distribution of the low$-$ and high$-z$ samples and find that the null hypothesis of the two samples being same is rejected with a negligibly small p-value.
This clearly indicates either a redshift evolution of \nh\ between the two samples or different biases in the way the two samples are constructed. 
The median \nh\ of the low$-z$ \CIII\ sample is 0.6 dex higher than the high$-z$ \CIII\ systems (S3) in \AS19.
The CPD of \nh\ distributions for the low$-z$ \CIII\ sample is also significantly different when compared to the high$-z$ sample S3 of \AS19.

\vskip 0.1in
\par\noindent{\bf C-abundance [{\it C/H}]:}
The C-abundance [{\it C/H}] of each absorption system is obtained by matching the individual column densities of
C ions (\CII\ and \CIII) with the model predicted column densities at a fixed density. However,
we have only nine
systems $(SA1)$, which have clean detections of
\CIII\ and \CII\ absorption.
So, for most of the absorption systems (for sub-sample SA2 and SB), we have obtained $[C/H]$ by
matching
the model predicted column density with that of
at least one of the observed
carbon ion (\CIII\ or \CII) while making sure the column density of the other ion is consistent with the limiting value obtained from observations.
As discussed in the PI models, the different $\alpha$ values ($-1.6$ and $-2.0$ in \KS18\ UVBs) also
causes uncertainty of $\pm$0.12 dex in the measurement of $[C/H]$ with respect to our fiducial \KS18\ UVB model.
The median $[C/H]$ of SA2 (SB) is 0.10 (0.04) dex lower (higher) than the median $[C/H]$ of sub-sample SA1, where we have only nine clean \CIII\ $+$ \CII\ systems. For the two sub-samples SA and SB, the $[C/H]$ ranges are, $-1.3\, \le [C/H]\, \le 0.4$ and $-1.6\, \le [C/H]\, \le 0.3$ with the median values $-0.51$ and $-0.41$, respectively.
We also find that 39 systems with \CIII\ detections and limits on \CII\ from SB show identical $[C/H]$ distribution as SA with a negligible median difference of 0.06 dex. Hence, we conclude that  different sub-samples of low$-z$ \CIII\ absorbers show similar $[C/H]$ distributions irrespective of sample choices.

As we have \nh\ measurements using other ion ratios for \CIII\ systems in SA2 without \CII\ detections,
we use the ionization corrections for \CIII\ ($f_{\CIII}$) and \HI\ ($f_{\HI}$) predicted by the PI models 
to check the consistency of our derived $[C/H]$.
In this case, we calculate C-abundance as, 
\begin{equation} \label{emt}
[C/H] = {\rm log}~\nce/\nhi - {\rm log}~~f_{\CIII}/f_{\HI} - {\rm log}~(C/H)_{\odot}
\end{equation}
for each absorption system using
the observed \nce\ and \nhi.
We find that the $[C/H]$ obtained using this method is well within error range ($\pm0.12$ dex
due to uncertainty in the quasar SED while modeling the UVBs) of the derived $[C/H]$ from our PI models
with a median difference 
of 0.05 dex. Thus, even if we have only upper limits for \nc2\ for the sub-sample SA2, our PI model inferred
$[C/H]$ are robust and reliable.

In panel (b) of Fig.~\ref{fig45}, we have shown the NHDs and CPDs of [{\it C/H}] distribution for both the low$-z$ (red) and high$-z$ (blue) samples. The derived [{\it C/H}] for low$-z$ \CIII\ systems is in the range,
$-1.6 \le [C/H] \le 0.4$ with a median value of $-0.46$.
We find that the overall $[C/H]$ distribution for low$-z$ is unimodal.
Moreover, the \CIII\ selected sample seems to be
associated with high metallicity end of its parent SLF sample of \citetalias{Lehner2019}.
The median $[C/H]$ of the low$-z$ \CIII\ systems is found to be 1.1 dex higher than the high$-z$ \CIII\ systems in \AS19. 
There are total 14 ($\simeq$14\%) absorption systems in $SA+SB$, which show super solar C-abundances. Comparatively in \AS19, there are eight ($\simeq$15\%) super solar \CIII\ systems in S3. However, when
we restrict ourselves to 
absorbers with \nhi\ $\ge$ $10^{15}$ \cmsq (like the low$-z$ absorbers), we found only one ($\simeq$3\%) super solar system in S3 of \AS19.
This once again is consistent with the fact that the low$-z$ \CIII\ absorption systems are more metal enriched than their high$-z$ counterparts.
\citet{Kim2016} suggested a bimodal distribution of metallicity among high$-z$ \CIII\ systems. This is also barely apparent in panel (b) of Fig.~\ref{fig45}.
The $[C/H]$ distribution of the low$-z$ \CIII\ sample almost lies within the high metallicity branch ($[C/H]$ $>$ $-1.4$) 
of high$-z$ \CIII\ systems.
Interestingly, we do find the paucity of low metallicity ($[C/H]$ $<$ $-$1.6) absorption systems in low$-z$ \CIII\ sample despite of having overlapping \nhi\ ranges. 

\vskip 0.1in
\noindent{\bf Over-density ($\Delta$) and the line-of-sight thickness ($L$):} In panel (c) in Fig.~\ref{fig45}, we have
shown the 
NHDs and CPDs of gas over-density parameter 
($\Delta$ = \nh/$\overline{n_H}$, where $\overline{n_H} = 1.719 \times 10^{-7}$ \cmcb $(1 + z)^3$ is the mean hydrogen density of the universe).
The derived $\Delta$ for the low$-z$ \CIII\ systems are in range, log $\Delta$ \si\ $3 - 4.5$, with a
median over-density, $<$log $\Delta>_{median}$ {\si} $3.74$.
The over-densities for these systems clearly confirms that these absorbers do not resemble the physical properties of typical IGM gas and are
probably originating from high over-dense regions like CGM. 

Hydrodynamical simulations of the low$-z$ \lya\ forest at $z\sim0$
indicate that most of the absorption
with \nhi\ $\sim$ $10^{15}$ \cmsq\ and $T < 10^5 K$ will be associated with the condensed gas phase and
originated from regions with
$\Delta \ge 100$ (i.e, log $\Delta\ge 2.0$) \citep[see Fig.~5 and Fig.~8 of][]{Gaikwad2017a}.
The above inferred range in $\Delta$ is at least an order of magnitude higher
(i.e, log~$\Delta$ = $3-4.5$ for \nhi\ (\cmsq) $= 10^{15}-10^{16.2}$) than that predicted
by $\Delta$ vs. \nhi\ relationship for the \lya\ forest absorbers \citep[see equation 10 of][and values given in their Table 2]{Gaikwad2017a}.
This once again suggests that the \CIII\ absorbers may be related to CGM.
The median $\Delta$ measured for the low$-z$ \CIII\ absorbers is 1.6 dex higher than their high$-z$
counterparts. The large difference in $\Delta$ as compared to \nh\ is mainly due to the strong redshift
dependence of mean density $\overline{n_H}$.

Panel (d) in Fig.~\ref{fig45}
shows the NHDs and CPDs of 
line-of-sight thickness, $L$ $\sim N({\rm H})/n_H$. We find that the $L$ obtained for SA $+$ SB is varying over a large range from 1.3 pc to 10 kpc with a median line-of-sight thickness of 60 pc.
The median $L$ of the low$-z$ \CIII\ systems is \si10 times lower than that measured for the high$-z$ counterparts. 
As the size of the regions may be related to physical parameters such as sound speed, cooling time-scale, etc., (see \AS19),
this difference suggests a strong redshift evolution in the physical parameters (like metallicity and density) of the \CIII\ absorbers.

\vskip 0.1 in
\noindent{\bf PI model predictions for other ions:}
As discussed earlier, for constraining \nh, we
consider other metal ions (\SiIII, \SiII, \OIII\ and \OII) present in the
absorbing gas in our PI models.
We have presented the constrained $[Si/H]$ and $[O/H]$ for sample SA2 in Table~\ref{tab:si} and ~\ref{tab:oxy}, respectively.
In order to show the consistency, we compare the observed 
column density ratios involving {\SiIII} and {\OIII} with the PI model predictions.
In panel (a) and panel (b) of Fig.~\ref{fig_osi}, we plot the observed column density ratios of 
\nce/$N(\SiIII)$ and \nce/$N(\OIII)$ against their respective PI model predicted values. 
The filled circles show detections of the observed ratios whereas the open circles with arrows show the lower limits.
The errors in the observed ratios are shown by vertical bars.
The solid line represents the `y $=$ x' equality line where the observed ratio is exactly equal to the model predicted ratio.
The dotted lines indicate uncertainty range of $\pm$ 0.17 dex around the `y $=$ x' line which arises due to the PI model uncertainties (caused by the UVBs).
We found 20 (resp. 17) \SIII\ (resp. \OIII) detections and 4 (resp. 7) lower limit systems out of the total 24 (resp. 24) \SiIII\ $+$ \CIII\
(resp. \OIII\ $+$ \CIII) systems in the low$-z$ \CIII\ sample.
It is seen from
Fig.~\ref{fig_osi} that for all the systems, the observed column density ratios are consistent within the uncertainty range of $\pm$0.17 dex of the PI model
predicted column density ratios. From the above validations,
it is quite evident that the \nh\ obtained using observed column densities of Si and O ions are well constrained and consistent. 
We have also obtained the $[Si/H]$ and $[O/H]$ for all the systems while reproducing the observed column density of \CII, {\CIII}, \SiII, {\SiIII} and \OII, \OIII, respectively.
The mean
$[Si/C]$ and $[O/C]$ of the absorbers are $\sim 0.09\pm0.17$ and $\sim 0.31\pm0.22$, respectively. 
The mean $[Si/C]$ in the {\CIII} absorbers
are close to the solar value whereas mean $[O/C]$ to be super solar.
However, for some of the \SiIII\ + \CIII\ absorbers (marked in red stars in panel (a) of Fig.~\ref{fig_osi}), we find $[Si/C]$ to be super solar and require slight enhancement of $[Si/H]$ in order to produce the observed \nce/$N(\SiIII)$ in our PI models. Further investigations of such absorbers will be explored in a future work.

\begin{figure*}
  \centering
   \includegraphics[height=5.5cm, trim=2cm 0.0cm 2cm 2cm, clip=true, angle=0]{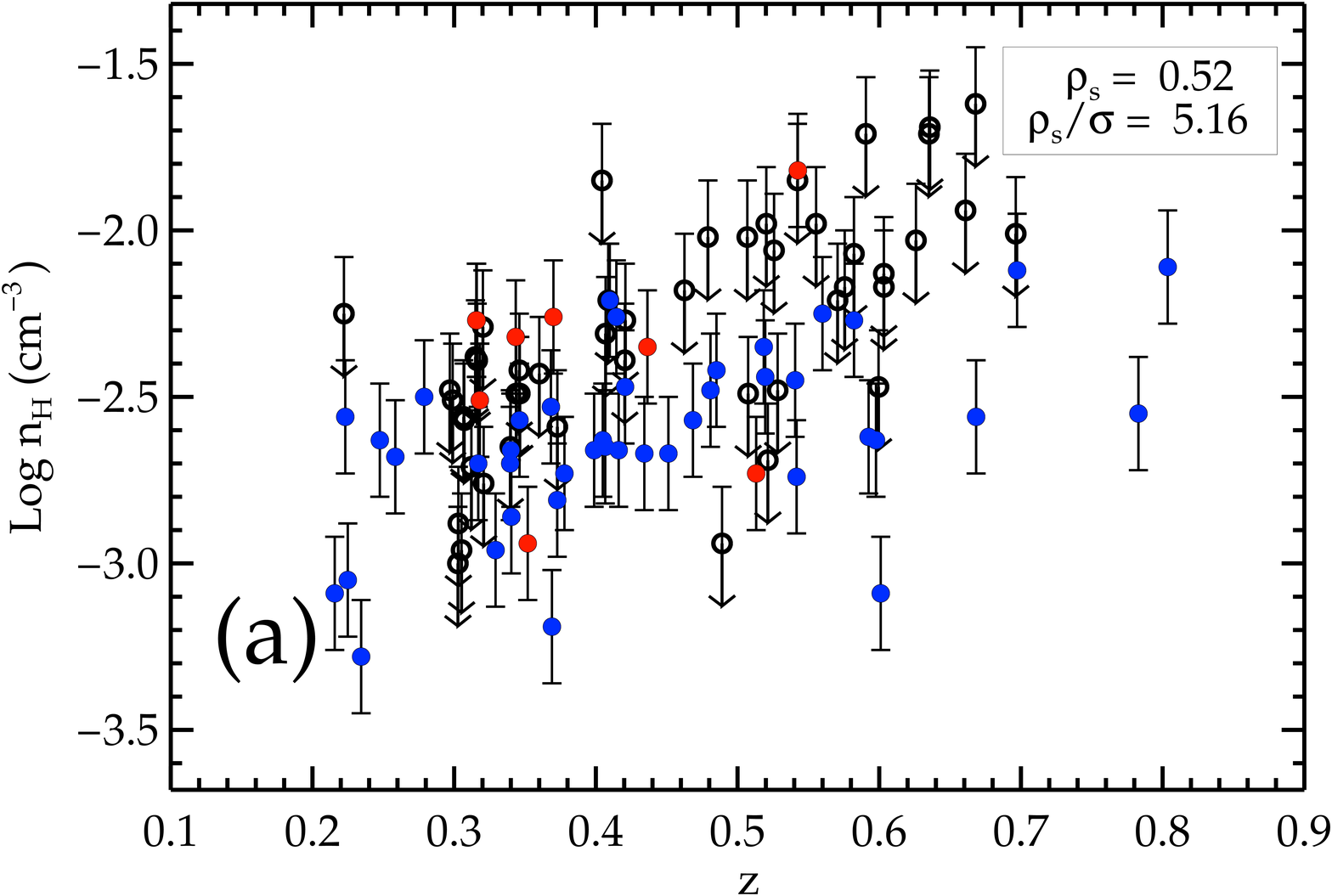}
   \includegraphics[height=5.5cm, trim=2cm 0.0cm 2cm 2cm, clip=true, angle=0]{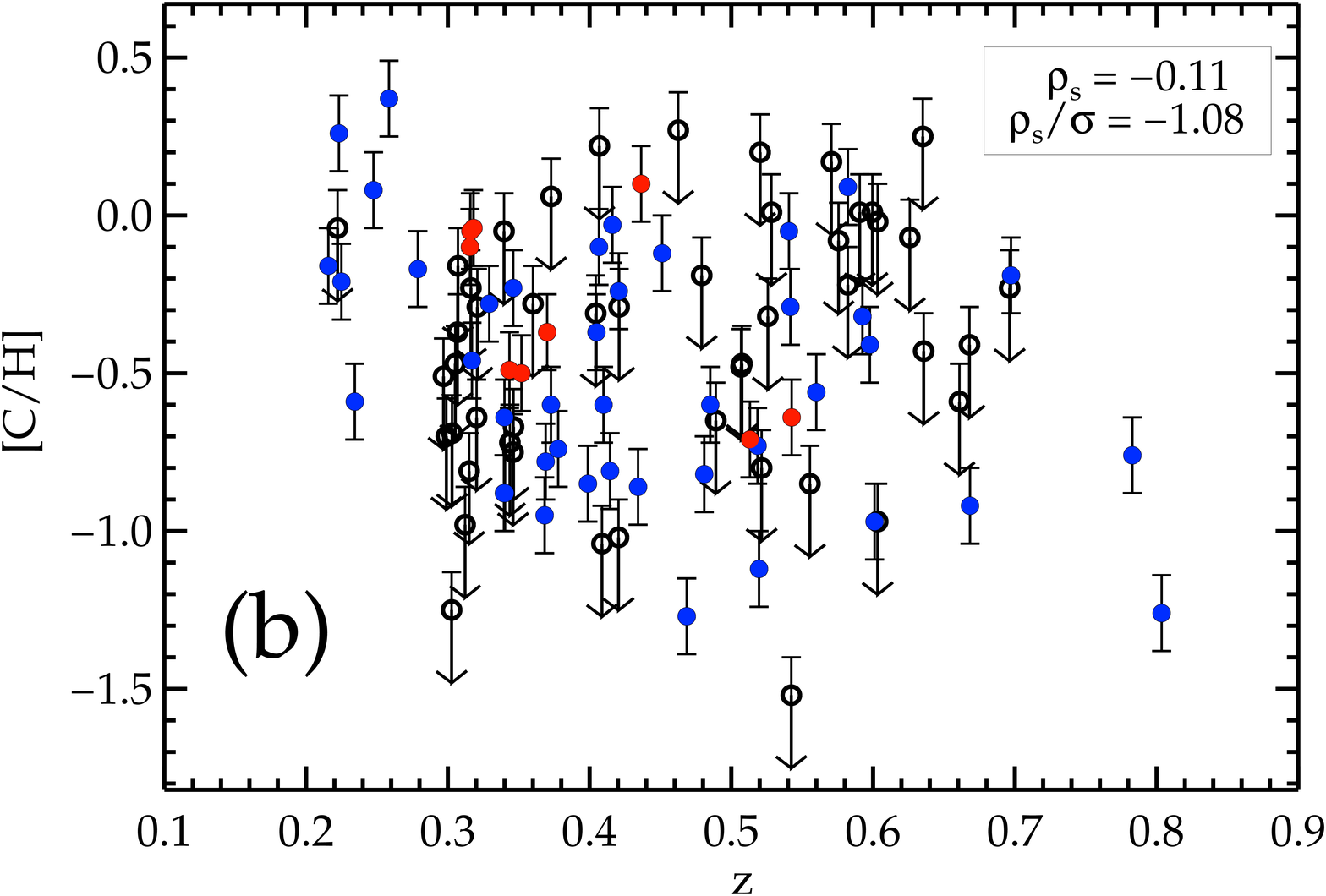}
   \includegraphics[height=5.5cm, trim=2cm 0.0cm 2cm 2cm, clip=true, angle=0]{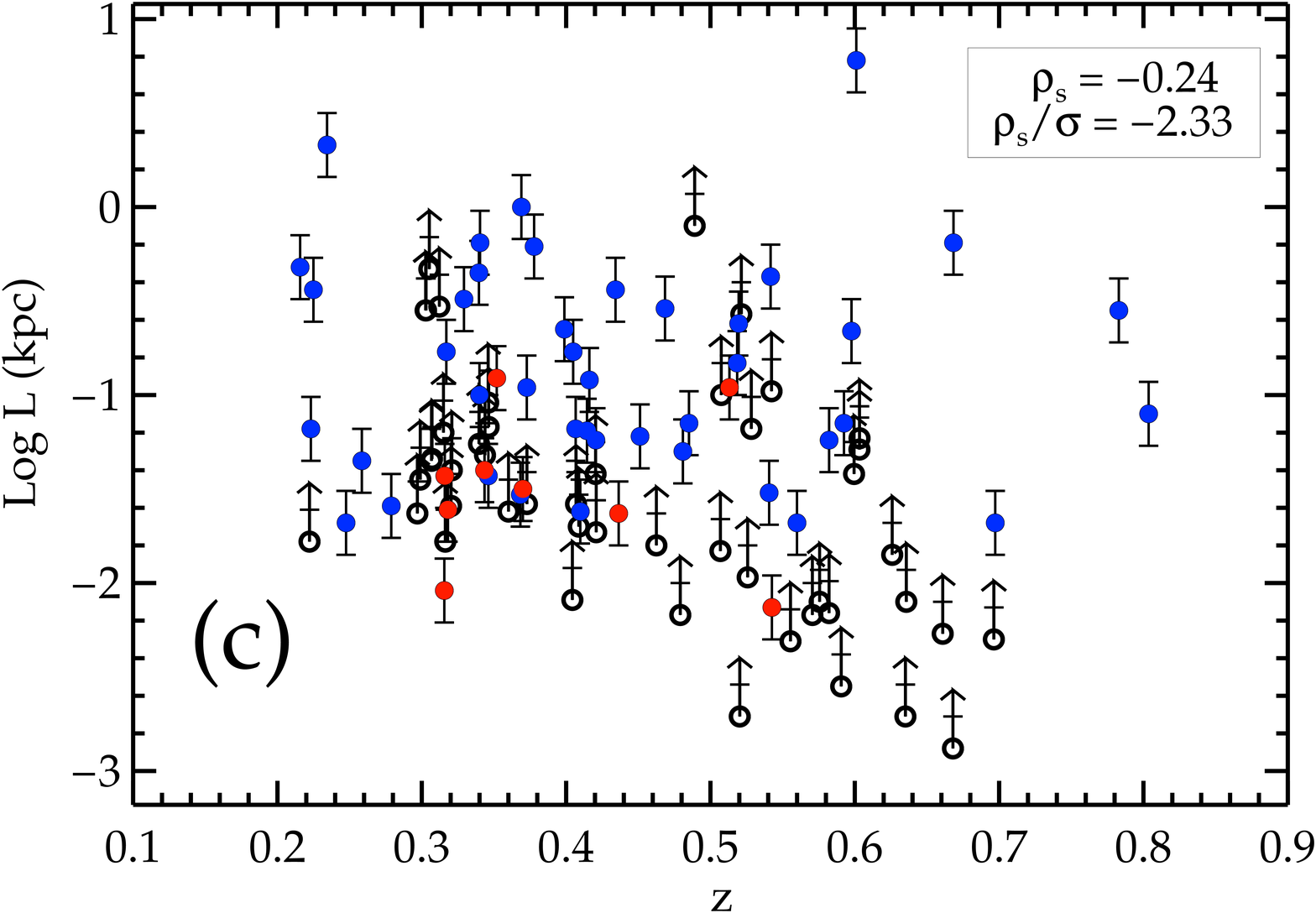}
   \includegraphics[height=5.5cm, trim=2cm 0.0cm 2cm 2cm, clip=true, angle=0]{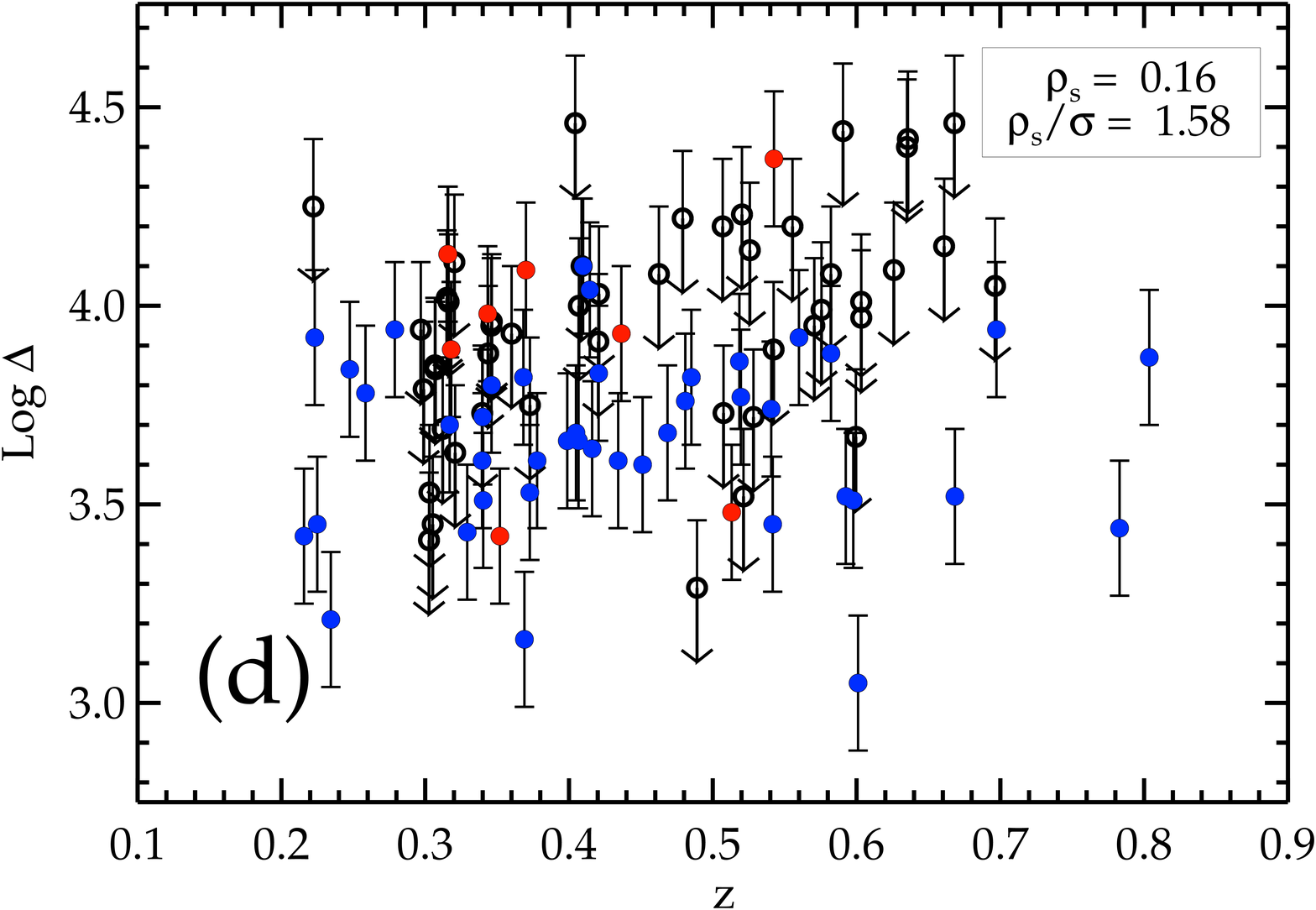}
   \caption{
     The redshift evolution of the model predicted parameters, hydrogen density (\nh), metallicity (C/H),
     line-of-sight thickness ($L$) and over-density ($\Delta$) are shown along with their respective rank correlation coefficients and significance levels. The red and blue solid circles with error bars represent the sub-samples, SA1 and SA2, respectively, whereas, the open circles  with arrows show the limit on the derived parameters for SB.}
\label{fig_zeve}
\end{figure*}
\begin{figure*}
  \centering
  \includegraphics[totalheight=0.172\textheight, trim=0cm 0cm 0cm 0cm, clip=true, angle=0]{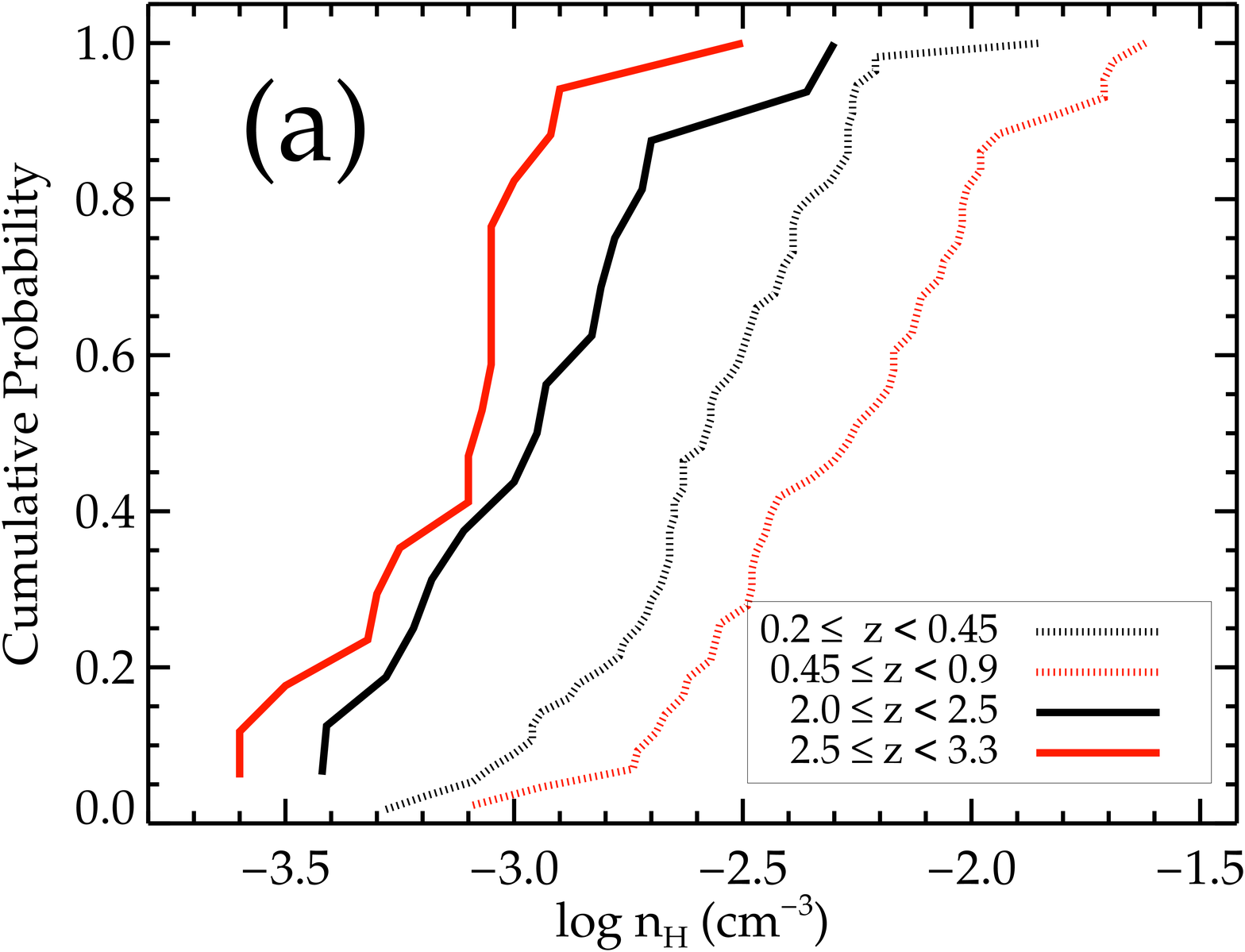}
  \includegraphics[totalheight=0.172\textheight, trim=0cm 0cm 0cm 0cm, clip=true, angle=0]{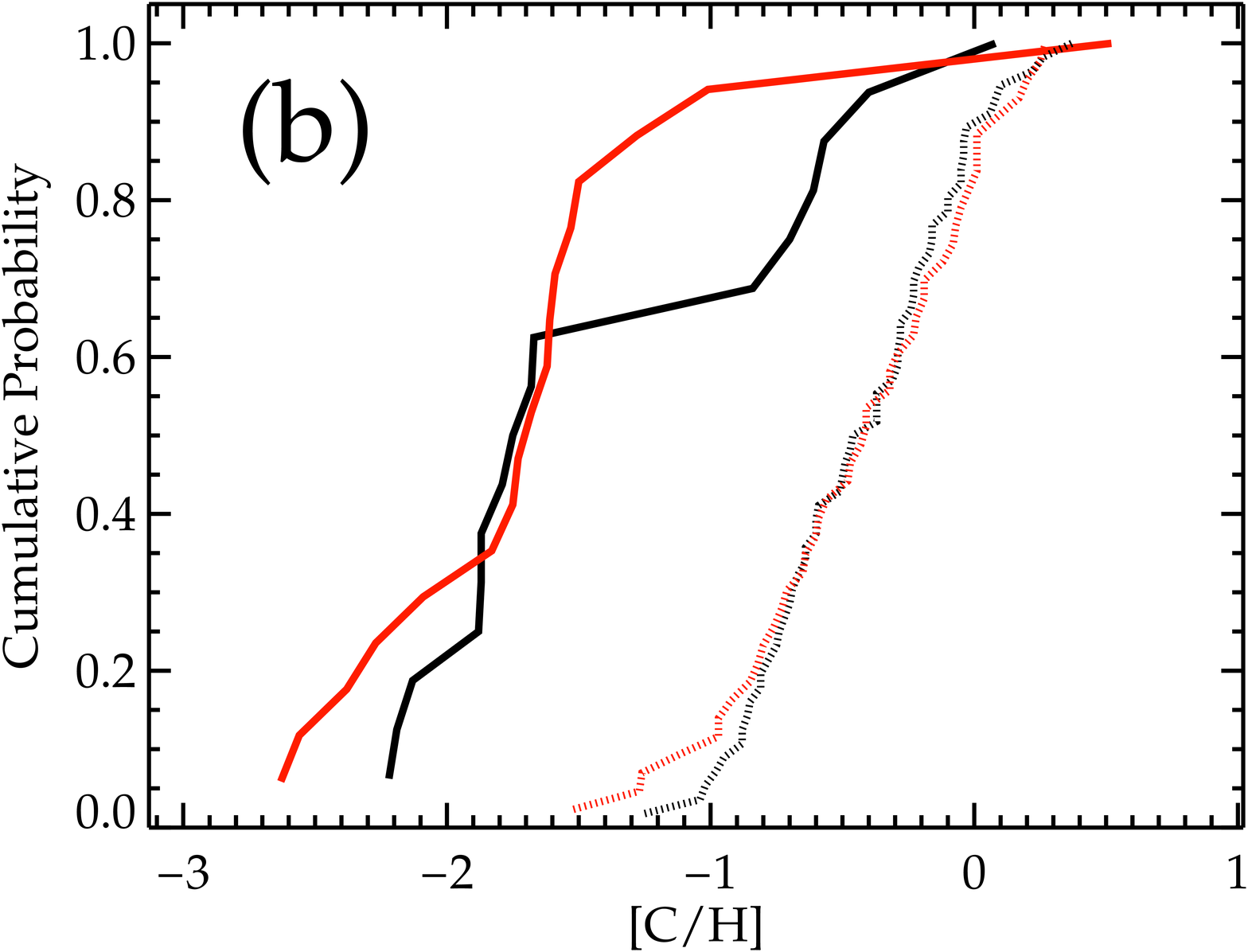}
  \includegraphics[totalheight=0.172\textheight, trim=0cm 0cm 0cm 0cm, clip=true, angle=0]{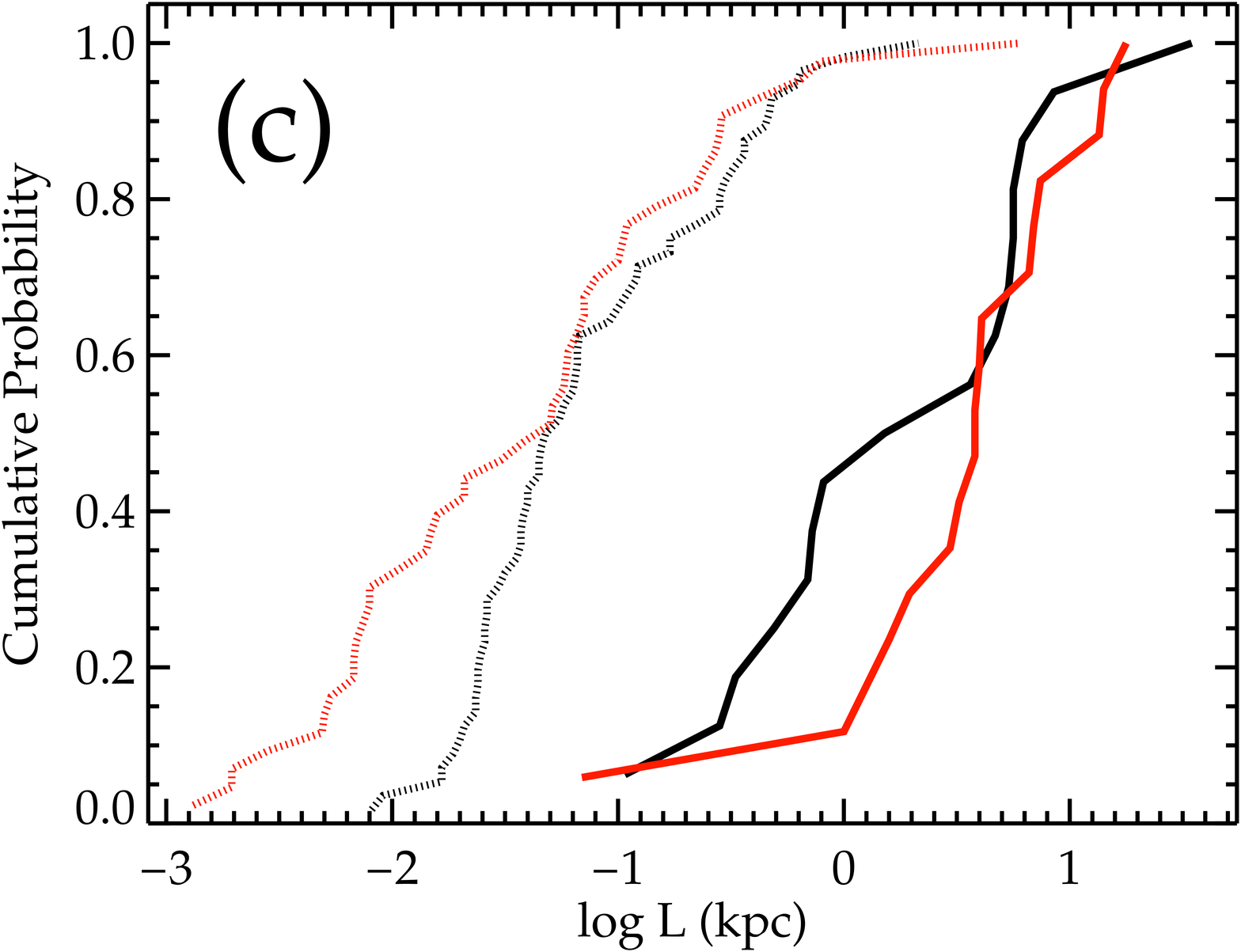}
 \caption{Redshift evolution of derived physical parameters, \nh\  (panel (a)), $[C/H]$ (panel (b)) and  $L$ (panel (c)) for the high$-z$ (solid lines)  and low$-z$ (dotted lines) \CIII\ sample. Black and red lines as shown in the legends in panel (a) represent the different redshift bins used in respective redshift ranges.}
\label{fig_eval}
\end{figure*}

\section{Discussion}
\label{discuss}
\subsection{Evolution of physical parameters with redshift}
\label{red_eval1}
In this section, We discuss redshift evolution of physical parameters, \nh,\, $\Delta$,\, L,\, and $[C/H]$ of the low$-z$ \CIII\ absorption systems obtained using our PI models with the fiducial \KS18\ UVB. 
We use Spearman rank correlation test to study the correlation between different parameters. Since $\approx$ 48\% of the absorption systems in the low$-z$ \CIII\ sample (i.e, for sample SB) have limits on the derived parameters, we perform a survival test with {\sc ASURV package (Astronomy Survival analysis package)} \citep{asruv1} for the correlation analysis. 
{\sc ASURV} package implements the statistical methods for handling censored data with upper limits as presented in \citet{Feigelson1985} and \citet{Isobe1986}.
We have used the bivariate method to determine the correlation between any two derived parameters considering the samples with the limits as the censored sample (see Table~\ref{tab_cor}). 
{\sc ASURV} provides generalizations of Spearman's rank correlation which also allows upper limits in both the variables.

\begin{table}
\caption{Correlation analysis of derived parameters using Spearman rank correlation test and survival test with the {\textit ASURV} package \citep{asruv1}.}
\label{tab_cor}
\begin{tabular}{ccccc}
\hline
\hline
 & \multicolumn{2}{c}{Spearman Rank correlation} & \multicolumn{2}{c}{Survival test} \\ 
Parameters & $\rho_s$ &$\rho_s/\sigma$& $\rho_s$ &$\rho_s/\sigma$\\ \hline
$z$ vs. ${\rm log}$ \nh & +0.52 & +5.16 & +0.33 & +3.27  \\
$z$ vs. $[C/H]$ & $-$0.11 & $-$1.08 & $-$0.16 & $-$1.58  \\ 
$z$ vs. ${\rm log}$ $L$ & $-$0.24 & $-$2.33 & $-$0.19 & $-$1.88  \\ 
$z$ vs. ${\rm log}$ $\Delta$ & +0.16 & +1.58 & +0.10 & +1.01 \\ 
${\rm log}$ $L$ vs. $[C/H]$ & $-$0.47 & $-$4.61 & $-$0.39 & $-$3.86  \\ 
${\rm log}$ $L$ vs. ${\rm log}$ $\nce$ & +0.59 & +5.83 & +0.52 & +5.16  \\ 
${\rm log}$ $L$ vs. ${\rm log}$ $\nhi$ & +0.35 & +3.48 & +0.28 & +2.77  \\ 
$[C/H]$ vs. ${\rm log}$ $\nce$& +0.33 & +3.27  & +0.31 & +3.07 \\ 
$[C/H]$ vs. ${\rm log}$ $\nhi$ & $-$0.44 & $-$4.38  & $-$0.35 & $-$3.47 \\ 
\hline
\end{tabular}
\end{table}
 
In panel (a) of Fig.~\ref{fig_zeve}, we show redshift vs. \nh\
for the low$-z$ \CIII\ absorption systems.
The \nh\ of the {\CIII} systems show an increasing trend with redshift. We find the Spearman rank correlation coefficient, $\rho_s = 0.52$ with a confidence level of 5.16$\sigma$. 
The sub-sample SA shows correlation coefficient $\rho_s = 0.37$ at a confidence level of 2.65$\sigma$ whereas 
SB shows a stronger correlation with $\rho_s = 0.71$ at a confidence level of 4.85$\sigma$. 
Hence, the stronger $\sim 5\sigma$ level correlation observed in the full sample could be due to the presence of lower limit \CIII\ systems in SB.
While the clean \CIII\ absorbers in sub-sample SA (i.e., SA1) do not show any statistically significant correlation between $z$ and \nh, we find similar $\rho_s (= 0.51)$  like the full low$-z$ sample albeit of slightly lower significance level (3.33$\sigma$) for the sub-sample SA2.
The survival analysis  for the full low-$z$ sample shows 3.27$\sigma$ confidence level correlation with $\rho_s = 0.33$ (see Table~\ref{tab_cor}) as seen in the case of SA where we have stringent constraint on the derived \nh.

In panel (a) of Fig.~\ref{fig_eval}, we show the cumulative distribution of \nh\ obtained in two redshift bins
(L1: $0.2\le z < 0.45$ and L2: $0.45\le z < 0.9$) for the low$-z$ \CIII\ absorbers.
We provide the Kolmogorov-Smirnov (KS) statistics with maximum deviation ($D_{KS}$) and the probability of finding
the difference by chance ($p-value$) for different redshift bins in Table~\ref{t2} . 
 A two-sided KS test on \nh\ distributions
of the low$-z$ bins shows a maximum difference between the cumulative distributions, $D_{KS}=0.47$ and p-value, $p=2.23\times10^{-5}$.
\bgroup\small
\begin{table}
\caption{The Kolmogorov-Smirnov (KS) statistics with maximum deviation ($D_{KS}$) and the probability of finding the difference by chance ($p-value$) for different redshift bins.}
\tabcolsep=3pt
\begin{center}
\begin{tabular}{c|cc|cc|cc}
\hline
\hline
& \multicolumn{2}{c|} {L1 vs L2} & \multicolumn{2}{c|} {H1 vs H2} & \multicolumn{2}{c} {low$-z$ vs high$-z$} \\ 
Parameters & $D_{KS}$  & $p-value$ & $D_{KS}$  & $p-value$ & $D_{KS}$  & $p-value$ \\
\hline
\nh\ & 0.47 & 2.23$\times 10^{-5}$ & 0.29 & 0.52 & 0.72 & 4.01$\times 10^{-12}$  \\ 
$[C/H]$ & 0.09 & 0.95 & 0.32 & 0.32 & 0.75 & 3.92$\times 10^{-13}$  \\ 
L & 0.36 & 2.5 $\times 10^{-3}$ & 0.38 & 0.15 & 0.81 & 2.82$\times 10^{-15}$ \\ 
\hline
\end{tabular}
\end{center}
\label{t2}
\end{table}
\egroup
It confirms that $n_H$ shows a mild increasing trend with redshift for $z\le 0.9$. 
As discussed in \S \ref{piresult}, the \nh\ range obtained at high$-z$ is lower than the \nh\ range measured at low$-z$.
A two-sided KS test also indicates statistically significant difference ($D_{KS}=0.72$ and $p=4.01\times10^{-12}$) in \nh\ distribution between low$-z$ and high$-z$ \CIII\ absorbers (see Table~\ref{t2}).
We have also divided the high$-z$ \CIII\ sample of \AS19\ into two redshift bins (H1: $2.0\le z < 2.5$ and H2: $2.5\le z < 3.3$).
From panel (a) of Fig~\ref{fig_eval}, we see that there is no clear \nh\ dependence of $z$
between these two high$-z$ redshift bins.
Thus, our analysis suggests that the redshift evolution of \nh\ for \CIII\ absorbers is not monotonic.
Interestingly this trend we notice is very much similar to the redshift dependence of \pirate\ in the \KS18\ UVBs \citep[see Fig.~4 of][]{Khaire2018}.
We discuss this further in \S \ref{uvbeval}.

We show $[C/H]$ as a function of $z$ in
panel (b) of Fig.~\ref{fig_zeve}.
For the low$-z$ \CIII\ absorption systems, $[C/H]$ lacks a
clear trend with $z$, and we find
$\rho_s = -0.11$ with a confidence level of 1.08$\sigma$. 
The survival test also confirms the same with similar $\rho_s$ = $-0.16$ at a slightly higher significance level of
$1.58\sigma$. The commutative distributions for L1 and L2, as shown in panel (b) of Fig.~\ref{fig_eval} also confirms the same.
As seen from Table~\ref{t2}, the two-sided KS test gives $D_{KS}=0.09$ with a high p-value.
The large spread in $[C/H]$ distribution for the low$-z$ sample indicates
that these absorbers probably trace the CGMs of different types of host galaxies.
Note that, in our analysis of \CIII\ selected absorbers, 
$[C/H]$ solely depends on the 
column densities of C ions and hence, it is slightly different 
from the metallicity reported in \L18.
Similar to \nh,
we find a strong redshift evolution of $[C/H]$ between low$-z$ and high$-z$ \CIII\ absorbers of \AS19 which is also confirmed using a two-sided KS test indicated in Table~\ref{t2} ($D_{KS}=0.75$ with a p-value $3.92\times10^{-13}$). However, we do not see such strong redshift evolution of $[C/H]$ within small redshift intervals for both the cases of the low$-z$ and high$-z$ \CIII\ absorbers.
This suggests that metallicity in the CGM evolves slowly with redshift. In \S \ref{abs_zevo}, we compare the redshift evolution of metallicity seen in \CIII\ systems with other types of quasar absorption systems.

In panel (c) of Fig.~\ref{fig_zeve}, we plot $L$ vs. $z$. We find a mild redshift evolution of $L$ with redshift
($\rho_s = -0.24$ at
a significance level of 2.33$\sigma$). The survival analysis test confirms the same with $\rho_s = -0.19$ at 1.88$\sigma$ level. 
Panel (c) of Fig.~\ref{fig_eval} and K-S test from Table~\ref{t2} also confirm the shallower trend in $L$ as a function of $z$
for the low$-z$ \CIII\ absorption systems. This mild correlation is mainly due to the lower limit absorbers of sub-sample SB.
We see a
large spread in $L$ within
small intervals of redshift which suggests that the low$-z$ \CIII\ absorbers are tracing in-homogeneously mixed population of pc to sub-kilo parsec scale absorbers in the CGM. However, when we compare our $L$ measurements with those of \AS19\ we do see a statistically significant evolution in $L$ with $z$ (see panel (c) of Fig.~\ref{fig_eval}). A two-sided KS test on $L$ for the low$-z$ and high$-z$ samples gives $D_{KS}=0.81$
and $p=2.82\times10^{-15}$, implying a significant difference in $L$ distributions.
It is also clear from this figure that the $L$ distributions are similar for the two high$-z$ bins ($D_{KS}=0.38$ and $p=0.15$)
compared to a slight different distributions of $L$ between the two low$-z$ bins ($D_{KS}=0.36$ and $p=2.5\times10^{-3}$).
The median $L$ of the entire low$-z$ sample is \si51 pc whereas
the median $L$ of the sub-sample SA and SB are $\simeq$75 pc and $\simeq$31 pc, respectively.
We see similar \CIII\ systems
having sub-kilo
parsec scale line-of-sight length predominantly present at $z$ \si\ 2.1 $-$ 2.5 as reported in \AS19. However, the frequency of occurrence of such systems is only 15\% at high$-z$.
Interestingly, when we consider absorbers with \nhi\ $\ge$ $10^{15}$ \cmsq, we find only 6\% of these high$-z$ \CIII\ absorbers with sub-kpc scale line-of-sight thickness.

In panel (d) of Fig.~\ref{fig_zeve}, we plot the redshift evolution of $\Delta$
for the low$-z$ \CIII\ absorption systems.
We do not see any statistically significant
correlation between these two for the low$-z$ \CIII\ sample in contrary
to the strong correlation found for high$-z$ \CIII\ sample of \AS19.
As seen in panel (a) of Fig.~\ref{fig_zeve}, there is a positive correlation between \nh\ and redshift and 
also, there is a strong dependence of $\overline{n_H}$ on 
$z$ as, $\overline{n_H}$ $=$ $1.719 \times 10^{-7}$ \cmcb~$(1+z)^3$. 
Hence,
$\Delta$ becomes almost independent of $z$ 
for low$-z$ \CIII\ sample. 
The survival test also confirms the same with $\rho_s$= 0.10 at a significance level of 1.01 $\sigma$. 
The lack of redshift evolution of \nh\ and $\Delta$ is generally expected if the
\CIII\ absorbers predominantly originate in the CGM of galaxies. In this case, the \nh\ distribution is dominated by the prevailing local physical conditions and not from the cosmological density fluctuation (as in the case of \lya\ absorption from the IGM).

\begin{figure}
  \centering
  \includegraphics[totalheight=0.25\textheight, trim=0.1cm .1cm 0cm 0.559cm, clip=true, angle=0]{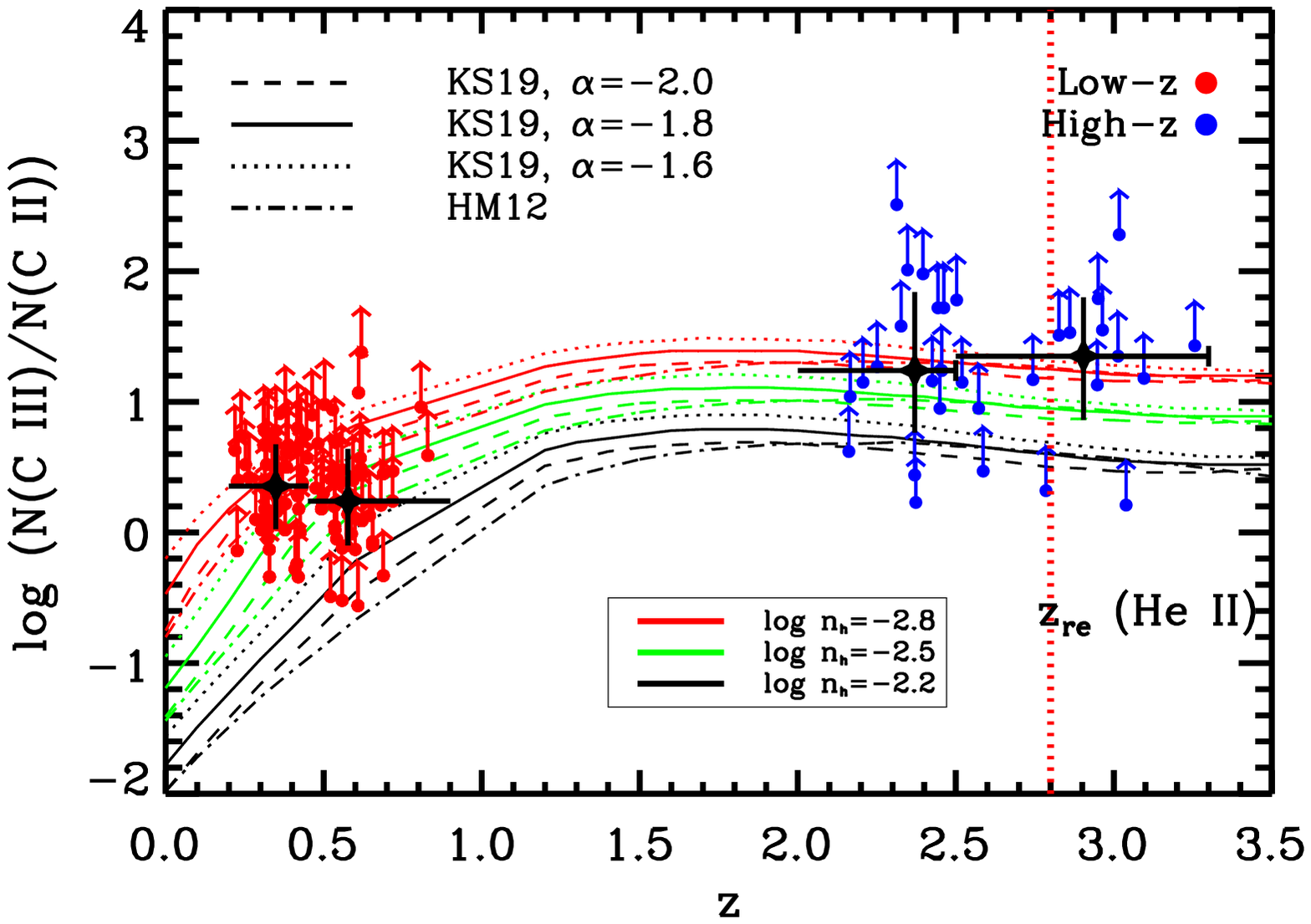}
  \includegraphics[totalheight=0.25\textheight, trim=0.1cm .1cm 0cm 0.559cm, clip=true, angle=0]{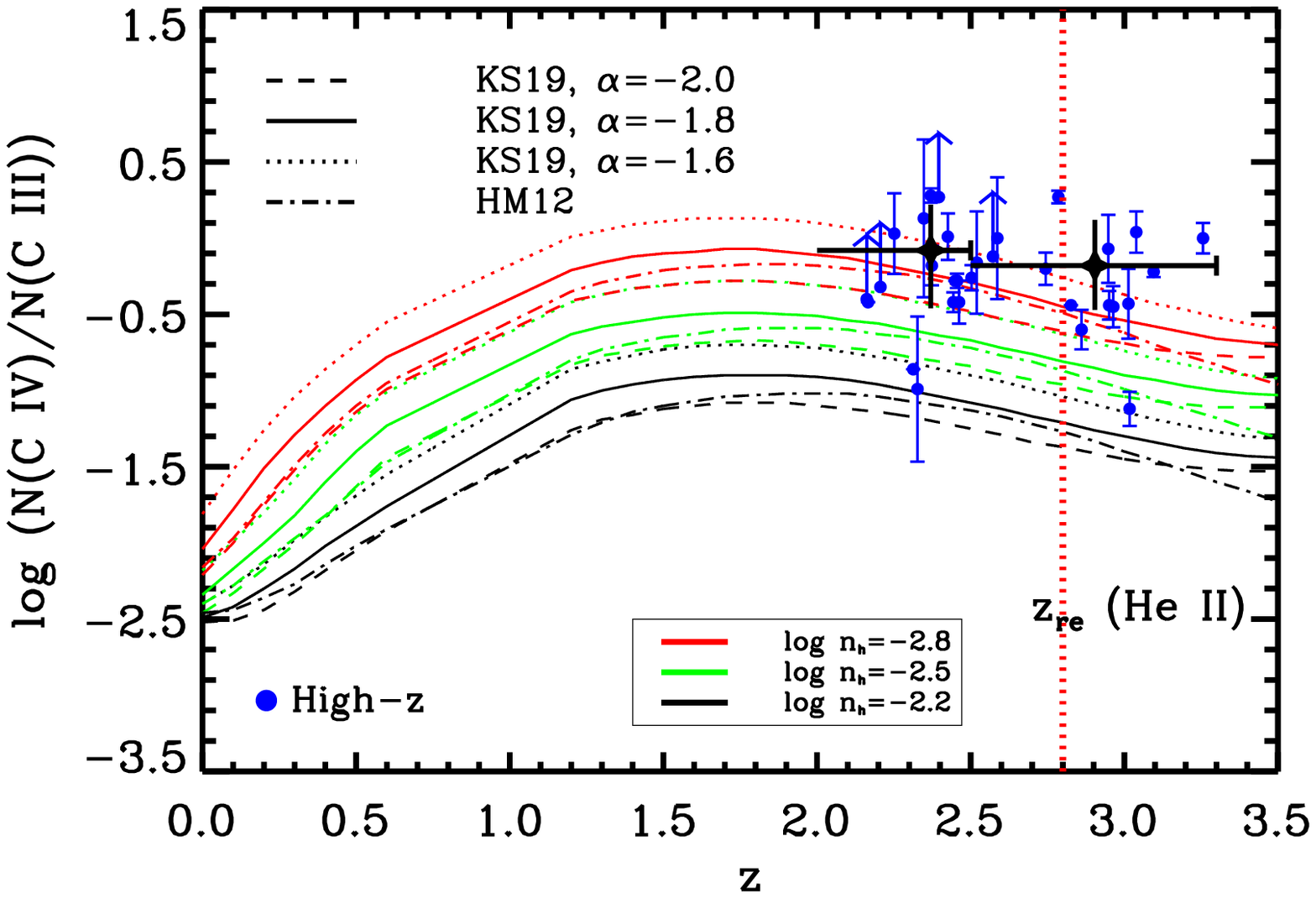}
\caption{{Top panel:
The observed $N$({\CIII})/$N$({\CII})
for low$-z$ and high$-z$ \CIII\ systems versus
redshift. The red and blue circles represent the low$-$ and high$-z$ data, respectively. The thick stars denote the median value of the observed ratio obtained over the extent of each horizontal bar in the redshift range with the vertical bar showing the 1$\sigma$ range in the median values.
The model predicted column density ratios obtained using
different UVBs (\KS18\ UVBs for $\alpha$, $-$1.6, $-$1.8 and $-$2.0 and HM12 UVB) are over-plotted
using different lines as shown in the legends. From bottom to top, \nh\ is decreasing as denoted by the legends. 
The vertical red line shows the reionization redshift ($z_{re}$) of \HeII\ obtained using \KS18\ UVB \citep{Khaire2017}. 
Bottom panel: The observed $N$({\CIV})/$N$({\CIII})
for the low$-$ and high$-z$ \CIII\ systems versus redshift. Symbols and colours are same as the top panel.
Note that the lack of \CIV\ at low$-z$ is just observational bias. Hence, we use the model curves obtained with the median density (log \nh\ $\simeq$ $-$2.5) of the low$-z$ \CIII\ absorption systems along with the 1$\sigma$ ranges (log \nh\
$\simeq$ $-$2.8 and $-$2.2).}}
\label{fig2a}
\end{figure}

\subsubsection{UVB and density evolution with $z$:}
\label{uvbeval}
In \S \ref{red_eval1}, we have discussed about the evolution of \nh\ as a function of $z$.
We notice that the inferred evolution
is not monotonous and to some extent seems to follow the
redshift evolution of the UVB. It is also important to remember while
we use $N$({\CIII})/$N$({\CII}) to measure \nh\ at low$-z$, at high$-z$, we preferred $N$({\CIV})/$N$({\CIII}) instead.
The difference in choice is purely due to observational constraints.
So, in order to study the UVB and \nh\ evolution in a large redshift range, we have therefore combined the low$-z$ ($0.2-0.9$) sample with the previously analysed high$-z$ ($2.1-3.3$)
sample from \AS19 for the absorbers having \nhi\ $\ge$ $10^{15}$ \cmsq. 

In top panel of Fig.~\ref{fig2a}, we show the variation of observed $N$({\CIII})/$N$({\CII}) vs. $z$ using scattered circles.
The red and blue
circles represent the
low and high redshift data, respectively. We have obtained $N$({\CII}) for the high$-z$ \CIII\ systems from \citet{Kim2016}. We have
estimated the upper limits using apparent optical depth method from the spectra for some
of the absorbers where \CII\ was not reported by \citet{Kim2016}. 
We have divided the entire redshift range into four redshift bins,
[0.2, 0.45], [0.45, 0.9], [2, 2.5] and [2.5, 3.3]. 
The ranges of observed \nce/$N (\CII)$ in logarithmic scale corresponding to
the above redshift bins are,
[-0.34, 0.93], [-0.55, 1.38], [0.21, 2.52] and [0.20, 2.31], respectively.
The filled stars indicate the median observed \nce/$N (\CII)$ 
(0.36, 0.24, 1.24 and 1.35) at each median value of the four redshift bins
(0.35, 0.58, 2.37 and 2.91), respectively. 
The median observed \nce/$N (\CII)$ show a mild variation within small
redshift ranges ($0.2-0.9$ and $2.1-3.3$) which
again indicates a slower evolution of \CIII\ absorbers in the CGM as
discussed in \S \ref{red_eval1}.
However, the median $N$({\CIII})/$N$({\CII}) for low$-z$ absorbers is approximately $\simeq$1.2 dex lower than the high$-z$ counterparts which may drive the observed \nh\ evolution at low$-z$. 

To further investigate this, we run a set of constant \nh\ PI models using different UVBs
(\KS18\ UVBs ($\alpha=-1.6,-1.8$ and $-2.0$) and \HM12\ UVB) assuming a constant temperature and metallicity of $2\times10^4$ K and
$0.1$ Z$_\odot$, respectively. 
The stopping criteria for the PI model is set at
($N(H)$) for an assumed \nh\ and taken from \citet{Schaye2001} as, 
$  N(H)_{J} \sim 1.6 \times 10^{21} cm^{-2} {n_H}^{1/2}T_4^{1/2}\bigg({\frac{f_g}{0.16}}\bigg)^{1/2} $,
where, $T_4(K)=T(K)/10^4$ and $f_g$ is the gas mass fraction. 
We have used $f_g=$ 0.16 to evaluate the stopping criteria $N(H)_{J}$ for each \nh\ values.
We have generated grids of column densities of observed ions in the redshift range, 0 to 3.3 (with a step size of 0.1)
for \nh\ in range, $-$1.0 $\le$ \nh\ $\le$ $-$4.0 (with a step size of 0.05).
Note that, this assumption
is valid for deriving the physical parameters of typical IGM \lya\ absorbers
and are simple enough to explain the UVB evolution
and \nh\ variation due to the UVBs (for a brief discussion see, Section~3.2 of \AS19). Here,
the model curves may not predict the true physical \nh\ of the \CIII\ absorbers as derived in our fiducial PI models.

The model generated $N$({\CIII})/$N$({\CII}) for three \nh\ values
(median \nh\ (log \nh\ $= -2.5$; green) of the low$-z$ absorbers along-with the 1$\sigma$ ranges (log \nh\ $= -2.8$; red and $-2.2$; black))
are over plotted in the top panel of
Fig.~\ref{fig2a} with different
line styles as shown in the legends.
It is clear from this figure that for a given \nh\ the
model predicted $N$({\CIII})/$N$({\CII}) is almost one dex lower at $z\approx0.5$ than at $z\approx2.5$
with a steeper slope for $z\leq1.2$. 
As discussed above,
the median observed $N$({\CIII})/$N$({\CII}) in low$-z$ bins are almost constant in the low$-z$
range, $0.2 \le z \le 0.9$. Therefore, this nearly constant $N$({\CIII})/$N$({\CII}) observed for $z\le0.9$ will
require \nh\ to increase mildly with increasing $z$. As seen from this figure, the
range of \nh\ that explains the low$-z$ points under predict 
$N$({\CIII})/$N$({\CII}) at high$-z$. Hence, for a given UVB, one has to decrease \nh\ values compared to that at low$-z$. It is also evident that this difference can not be
accommodated using the change in $\alpha$ (i.e., the quasar SED in the UV range used in the UVB calculations).

Next, to check whether $N$({\CIV}) to \nce\ predict the same \nh\ evolution, we
show the 
the variation of observed $N$({\CIV})/$N$({\CIII}) with redshift in bottom panel of Fig.~\ref{fig2a}.
Note that, 
at low$-z$ ($0.2 \le z \le 0.9$) the \CIV\ is not observable owing to the limitation of $HST/COS$ wavelength coverage.
We have also over plotted the model predicted $N$({\CIV})/$N$({\CIII}) for the same \nh\ ranges as discussed above for
different UVBs indicated by different line-styles as shown in the legends.
It is evident from the figure that the
high$-z$ $N$({\CIV})/$N$({\CIII}) will also require slightly lower \nh\ values as
suggested by $N$({\CIII})/$N$({\CII}).
So for a given UVB, the \nh\ needs to be changed for \CIII\ absorbers at low$-z$ compared to high$-z$ but not in a monotonically way in order to yield the observed column density ratios.

There are few possible biases that could have lead to the \nh\ evolution we note here. 
The low$-z$ sample typically has higher \nhi\ compare to the high$-z$ sample.
In cosmological simulations one finds absorption from a given \nhi\ originating from higher densities at low$-z$.
The \nh\ values discussed here are for a system, i.e., averaged over different individual velocity components.
Also, our measurements assume a single phase for \CII, \CIII, and \CIV. Therefore a better picture will emerge if
we can do the analysis of individual components over the full redshift range including addition data of \CIII\ absorbers at $1.0\le z\le 2.0$. 

\begin{figure*}
  \centering
  \includegraphics[totalheight=0.3\textheight, trim=0cm 0cm 0cm 0cm, clip=true, angle=0]{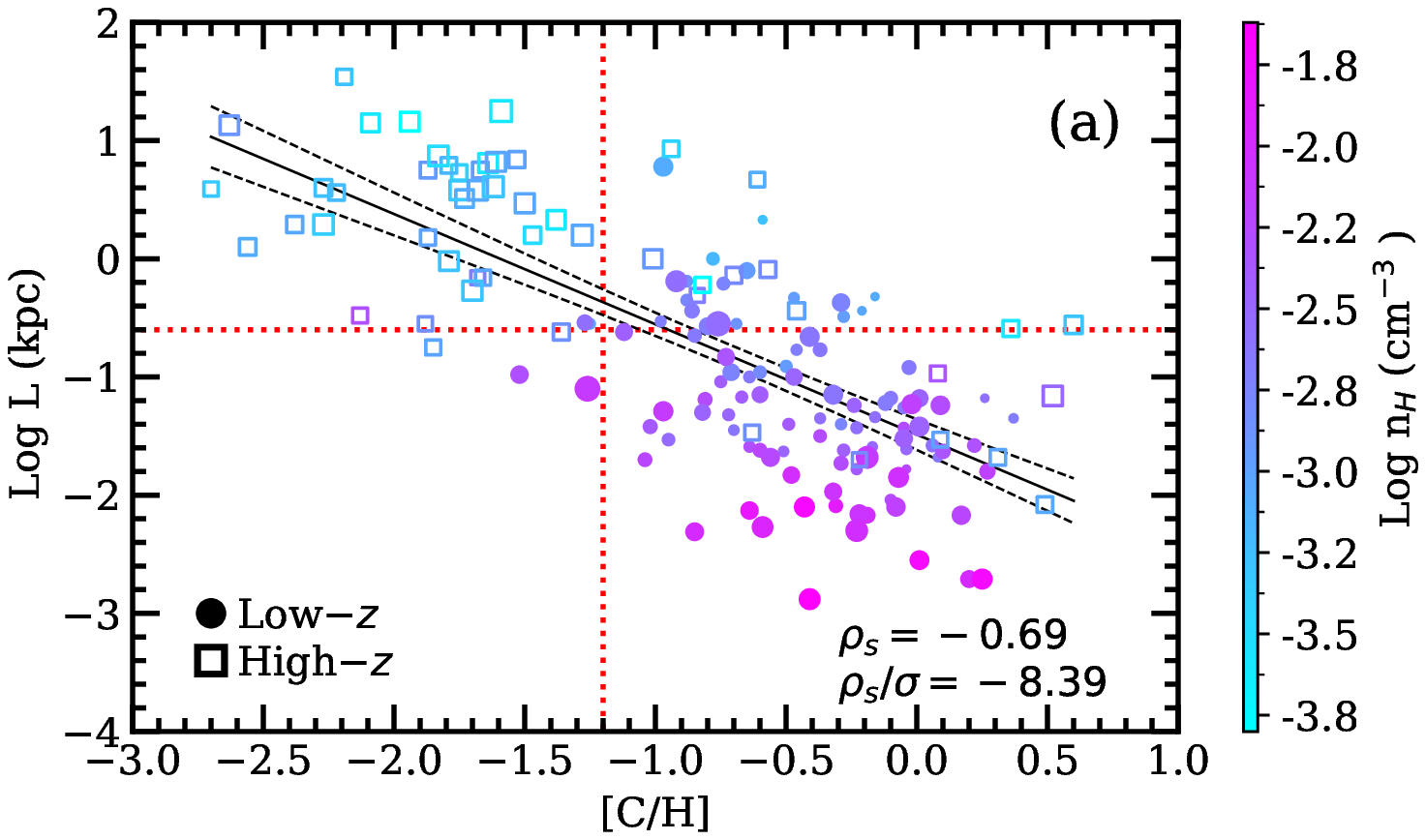}\\
  \includegraphics[totalheight=0.2\textheight, trim=0cm 0cm 0cm 0cm, clip=true, angle=0]{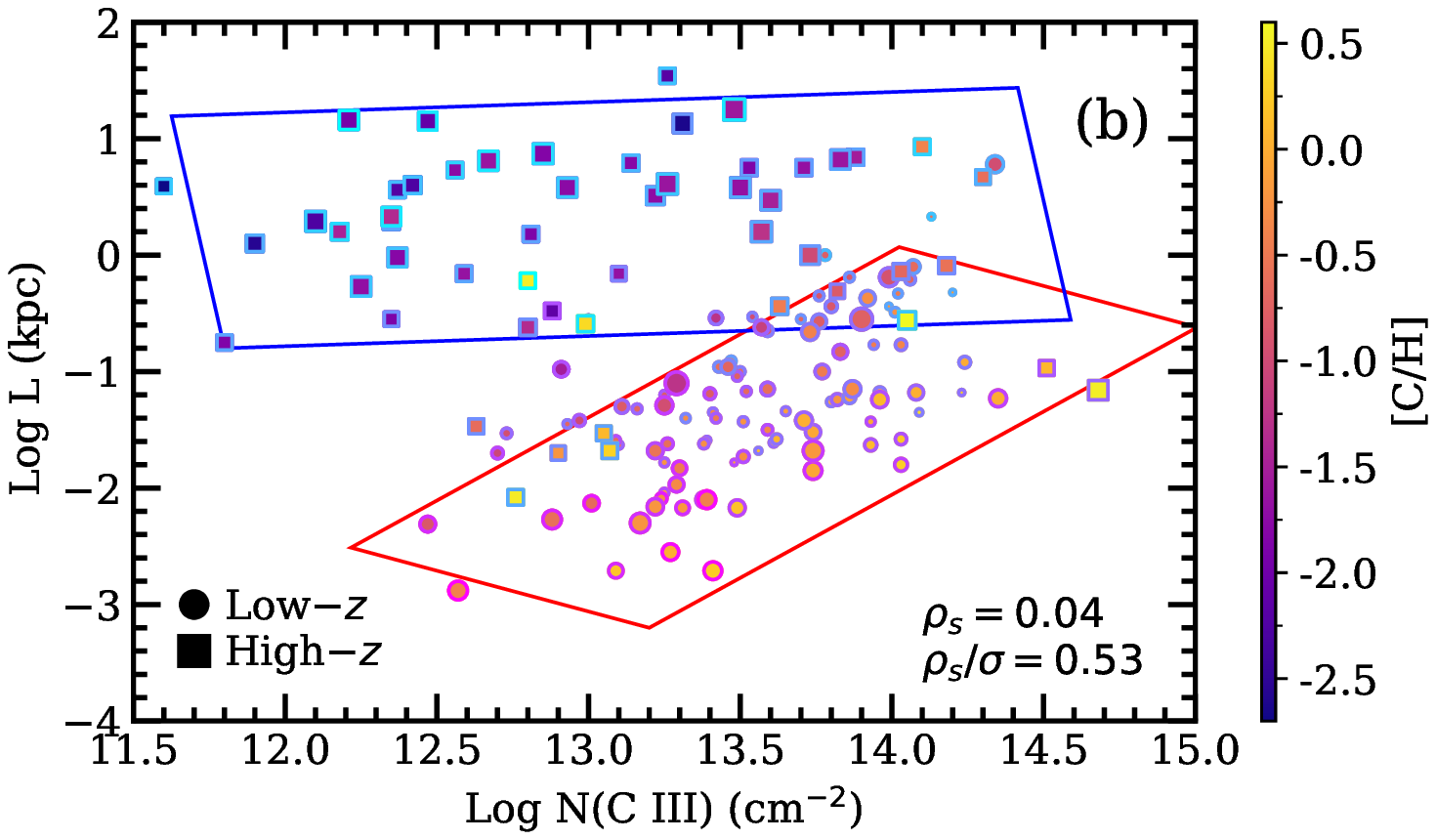}
    \includegraphics[totalheight=0.20\textheight, trim=0cm 0cm 0cm 0cm, clip=true, angle=0]{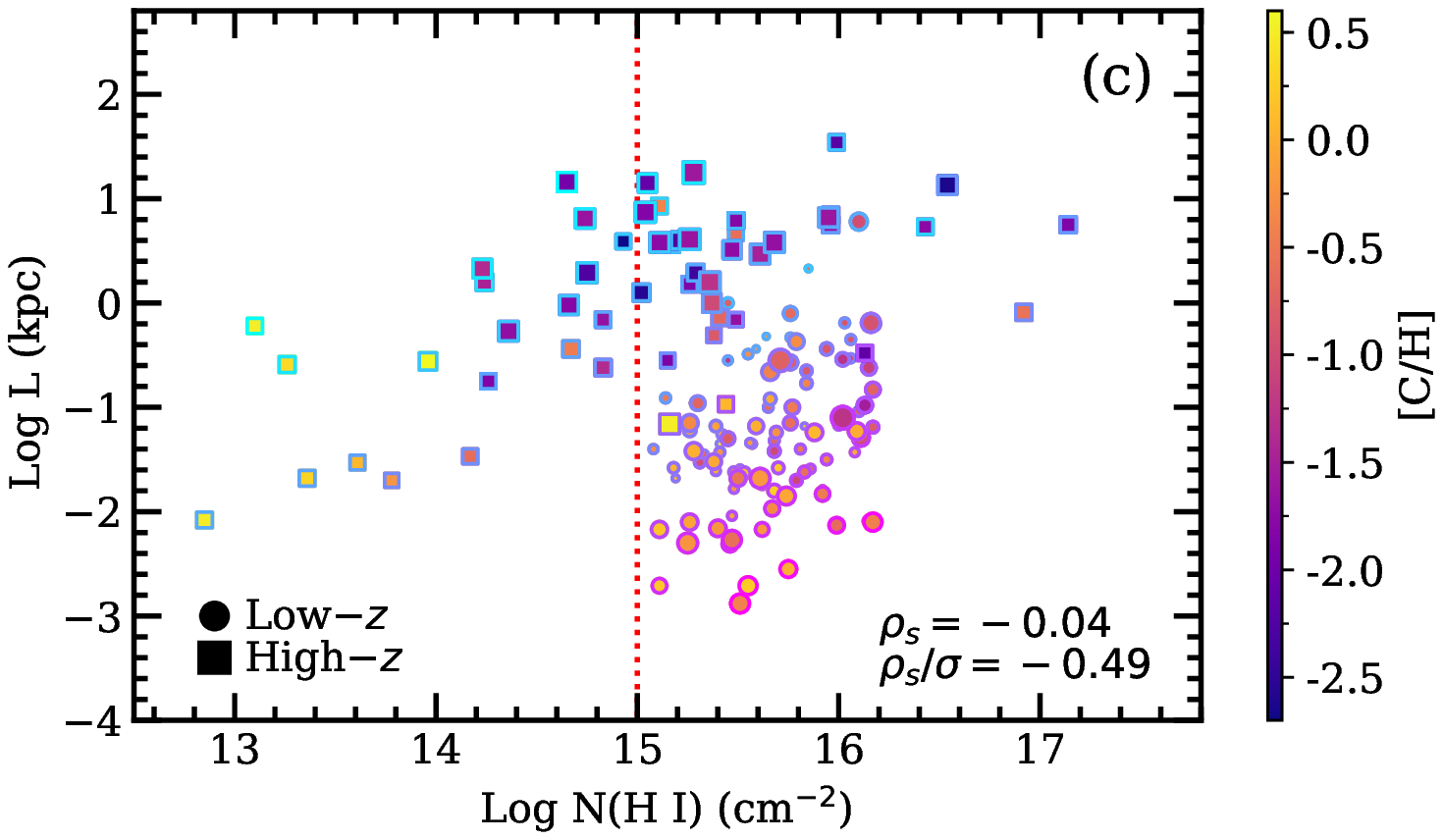}
      \includegraphics[totalheight=0.2\textheight, trim=0cm 0cm 0cm 0cm, clip=true, angle=0]{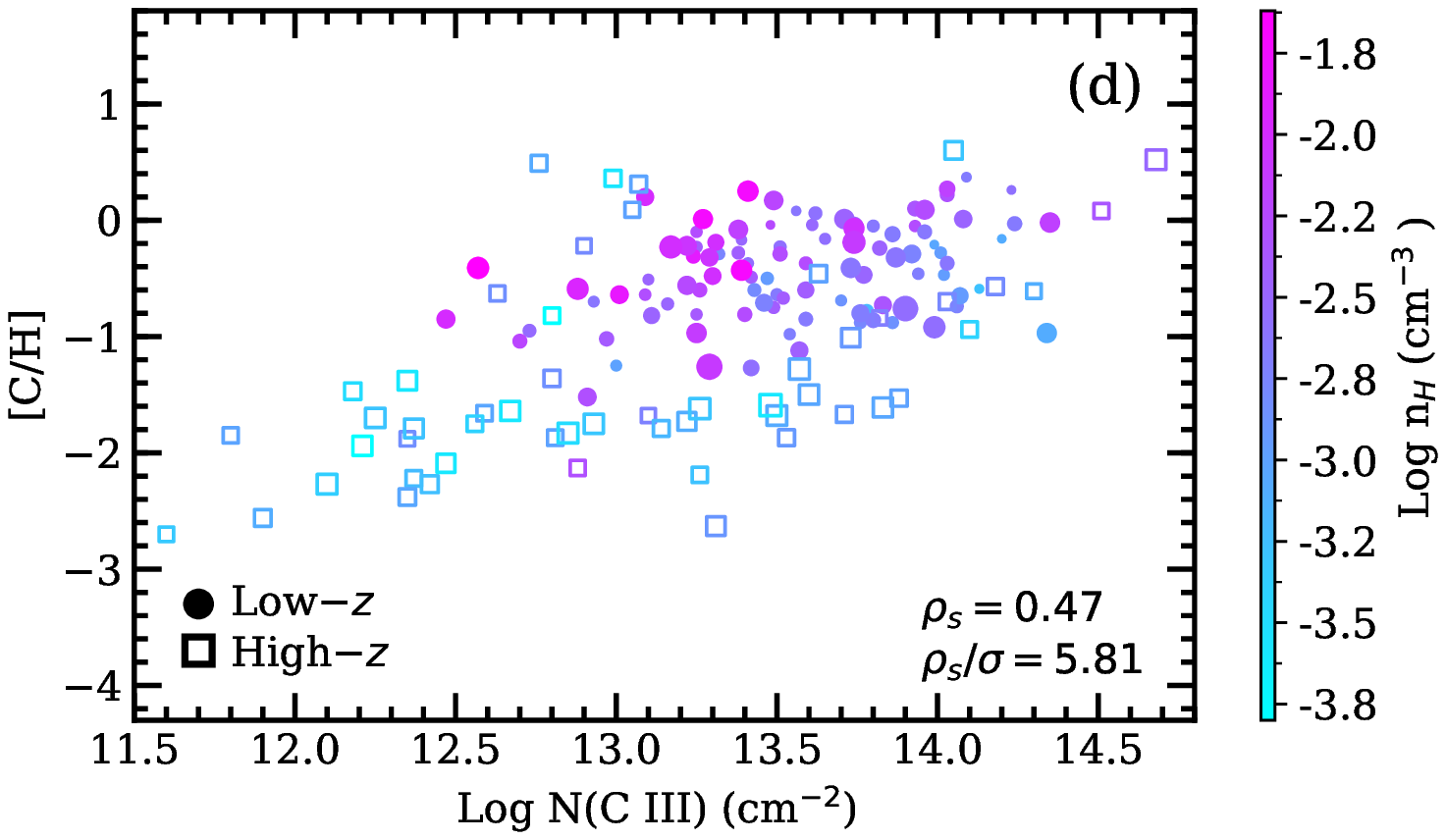}
      \includegraphics[totalheight=0.20\textheight, trim=0cm 0cm 0cm 0cm, clip=true, angle=0]{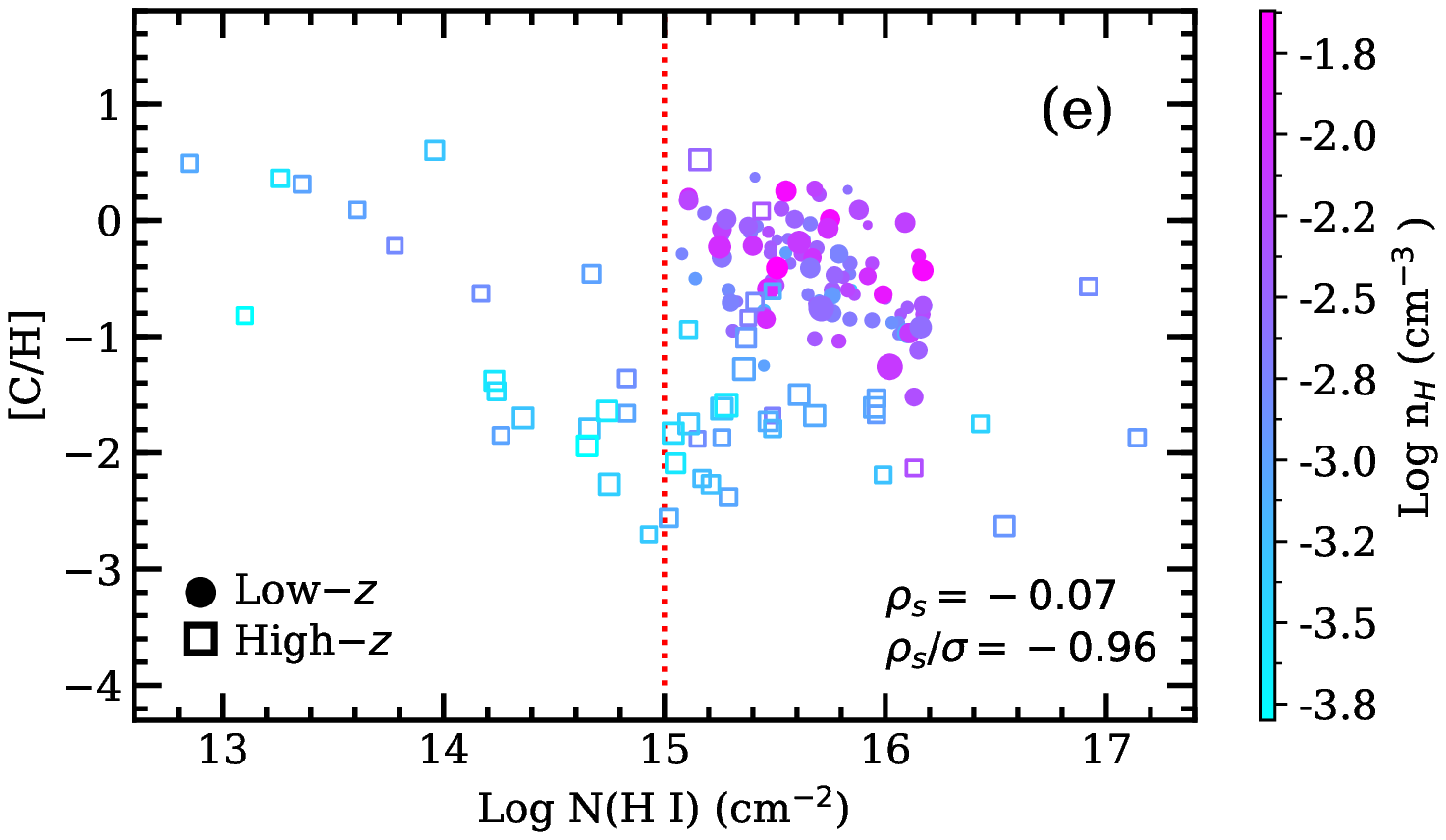}
\caption{{The plots show correlations among
 various model predicted parameters along with
 their rank correlation coefficients and significance levels.
 Panel (a): $L$ vs. $[C/H]$ and the symbols are colour coded with \nh\ values as shown in the side colour bar. The solid line represents the linear regression fit to the data, whereas the dashed lines indicate 1$\sigma$ deviation around the linear regression fit. The dotted horizontal and vertical lines show segregation between high$-[C/H]$ 
 ($[C/H]$ $\ge$ $-$1.2 and log~$L$ $<$ $-$0.6) and low$-[C/H]$ ($[C/H]$ $<$ $-$1.2 and log~$L$ $\ge$ $-$0.6) branch \CIII\ absorbers (see text for details).
Panel (b): $L$ vs. \nce\ with the face of the symbol colour coded with \nh\ values similar to panel (a) whereas the thick outer lines are colour coded with the $[C/H]$ values as shown with the side colour bar.
 The regions marked by red and blue boxes show the high$-[C/H]$ and low$-[C/H]$ branch absorbers, respectively.;
 Panel (c): $L$ vs. \nhi\ and the symbols represent same colour coding as panel (b). The vertical red line show \nhi\ cut off at log \nhi\ (\cmsq) $\ge$ 15.
 The panels (d) and (e) show the $[C/H]$ vs. \nce\ and $[C/H]$ vs. \nhi, respectively, with the side colour bars representing
 log \nh\ values embedded in the symbols.
 The solid circles and squares represent the parameters corresponding to low$-z$ ($0.2 \le z \le 0.9$) and high$-z$ (\AS19; $2.1 \le z \le 3.3$) samples, respectively. The
 symbol sizes increase with the increasing of redshift within the specified redshift range.
 }}
\label{fig_corr}
\end{figure*}

\subsection{Correlations between derived parameters:}
\label{cort}
In \AS19, we have studied the correlations between different derived parameters of \CIII\ absorbers 
for component-wise analysed data (S1 $+$ S2) and system-wide analysed data (S3), separately (see Fig.~10 of \AS19).
We found an interesting anti-correlation between $L$ and $[C/H]$ with significance level, 6.1$\sigma$ and 3.8$\sigma$
for S1 + S2 and S3, respectively.
We explain this anti-correlation using a
simple toy model where $L$ is driven by the cooling length scale ($t_{cool}$) $\times$ $c_s$)
as seen in the cloud fragmentation under thermal instabilities \citep{McCourt2018}. 

In Fig.~\ref{fig_corr}, we show the correlations between various derived parameters of the low$-z$ \CIII\ absorption
systems (indicated by circles).
We also have over-plotted the
same parameters of high$-z$ \CIII\ sample, $S3$ from \AS19 (indicated by squares) in order to make a comparative correlation analysis. 
In each panel, the redshift value increases with the symbol size, for the systems of respective samples.

\subsubsection {{$L$ vs. [{ C/H}]:}}
In panel (a) of Fig.~\ref{fig_corr}, we show the anti-correlation between $L$ and $[C/H]$ for the low$-z$ sample (filled circles) with
colour coded \nh\ values as displayed in the side colour bar. For the low$-z$ \CIII\ absorption systems,
we see an anti-correlation in $L$ vs.$ [{C/H}]$ with Spearman rank correlation coefficient, $\rho_s = -0.47$ at
4.61$\sigma$ level (see, panel (a) of Fig.~\ref{lowzcor} for $L$ vs. $[C/H]$ of the low$-z$ sample only). The survival analysis test
also confirms the anti-correlation between these two parameters with 
$\rho_s$ = $-0.39$ at a significance level of 3.86$\sigma$ (see Table~\ref{tab_cor}). \citet{Muzahid2018} have noted a similar correlation in their $z<0.3$ COS-weak sample.
We also plot the data from high$-z$ \CIII\ sample, $S3$ from \AS19 (open squares) in Fig.~\ref{fig_corr}.
As shown in the figure, the low$-z$ \CIII\ absorption systems are skewed towards the high$-[C/H]$ and low $L$, whereas
the high$-z$ \CIII\ sample has a wider spread in $[C/H]$ and $L$. In the overlapping metallicity range (i.e., $[C/H]\ge-1.2$) the low$-z$ absorbers tend to have smaller $L$ than the high$-z$ counterparts. It is also evident from the figure that the low$-z$ systems with smaller $L$ also have higher \nh.
We observe a strong anti-correlation in $L$ vs. $[{C/H}]$, with Spearman rank correlation value $\rho_s = -0.69$ significant at
8.39$\sigma$ level, for the combined sample of the low$-z$ and high$-z$ \CIII\ absorption systems. 
To show the strong correlation, we have also plotted linear regression fit (LRF) with 1$\sigma$ range
using black solid and dashed lines in Fig.~\ref{fig_corr}.
We obtain a strong dependence of $L$ on $[C/H]$ as:
\begin{equation} \label{eq2}
 log~{L} = (-1.49\pm0.07) + (-0.93\pm0.08) {\rm {[C/H]}}.
\end{equation}
The LRF for the combined sample has a steeper slope compared to the low$-z$ \CIII\ sample
(log~${L} = (-1.60\pm0.10) + (-0.78\pm0.15) {\rm {[C/H]}})$. It is also evident that
the size of the absorption system for a given $[C/H]$ is typically smaller for low$-z$. 
This also implies that the actual sizes of the low$-z$ absorbers are smaller than the predictions of the toy model used by \AS19 for modelling this correlation at high$-z$.
Furthermore, we find that for the combined \CIII\ absorbers, the
correlation remains unchanged ($\rho_s$ = $-$0.68 with 7.78 $\sigma$)
even if we restrict the high$-z$ sample to SLFs only (\nhi\ $>10^{15}$ \cmsq).

\subsubsection{ $L$ vs. \nce\ : }
In panel (b) of Fig.~\ref{fig_corr}, we plot the observed
\nce\ vs. $L$ with the same symbols and colours as mentioned above. 
For the low$-z$ \CIII\ sample, we find a tight correlation between these two
parameters with $\rho_s = 0.59$ at 5.83$\sigma$ (see, panel (b) of Fig.~\ref{lowzcor}) which is
also confirmed using a survival
analysis test as shown in Table~\ref{tab_cor} ($\rho\, =\, 0.52$ at 5.16$\sigma$). However, high$-z$ data points typically have larger $L$
for a given \nce.

While analyzing similar plot of $L$ vs $N$(C~{\sc iv}), \citet{Kim2016} argued
for the presence of two population based on metallicity. Following similar logic
we divide the combined sample into high$-[C/H]$ metallicity ($[C/H]$ $\ge$ $-$1.2 and log~$L$ $<$ $-$0.6) and low
metallicity absorbers ($[C/H]$ $<$ $-$1.2 and log~$L$ $\ge$ $-$0.6). 
From the boxes drawn in panel (b) of Fig.~\ref{fig_corr},
we find a clear segregation of high$-[C/H]$ (red box) and low$-[C/H]$ (blue box) systems in the $L$ vs \nce\ plane.
We see that
the high$-[C/H]$ branch absorbers
show a clear trend of increasing $L$ with
increasing log~$N$(\CIII)
whereas such a trend is missing for the low$-[C/H]$ branch absorbers with log~$L > 0.4$ kpc. 
As a result, two distinct populations in \CIII\ sample on the $L$ vs. \nce\ plane are visible in the combined sample with the low$-z$ \CIII\ sample following the $L$ vs. \nce\ evolution of high$-[C/H]$ branch of the high$-z$ \CIII\ absorbers. While this is 
in line with the discussions presented in \citet{Kim2016}, in our sample the low$-[C/H]$ branch is predominantly dominated by the high$-z$ systems and high$-[C/H]$ by the low$-z$ systems. 
Hence, it will be interesting to increase the number of
\CIII\ selected absorbers (spanning a broader range in $L$ and metallicity, $[C/H]$) to confirm the existence of statistically significant
bimodality in metallicity among the \CIII\ absorption systems.

\subsubsection{$L$ vs. $N(\HI)$ :}
We show the correlation between $L$ and $N(\HI)$ in panel (c) of Fig.~\ref{fig_corr}. 
The low$-z$ and high$-z$ \CIII\ sample have significantly different \nhi\ distributions with \nhi\ in ranges,
$12.8\, \le {\rm log}\, \nhi\, (cm^{-2})\, \le 16.6$ and $15\, \le {\rm log}\, \nhi\,(cm^{-2}) \le 16.2$, respectively.
In the low$-z$ \CIII\ sample, all the absorption systems have \nhi\ $\ge$ $10^{15}$ \cmsq\ compared to only
$\simeq$63\% of the high$-z$ absorption systems.
Similarly, the size distribution of the absorption systems are different with most of the 
low \nhi\ high$-z$ absorbers showing similar $L$ as the low$-z$ absorbers.
Hence, as seen in the figure the low$-z$ and high$-z$ \CIII\ absorption
systems are separable and have
two different distributions on the $L$ vs \nhi\ plane.
We find a statistically significant correlation between $L$ and \nhi\ for the low$-z$ \CIII\ sample
with $\rho_s = 0.35$ (3.48$\sigma$) (see, panel (c) of Fig.~\ref{lowzcor}). 
The low$-z$ \CIII\ absorption systems are skewed towards the lower right end of the 
$L$ vs. \nhi\ plane whereas the high$-z$ \CIII\ absorption systems have a larger spread in \nhi\ and $L$ and 
show a stronger correlation with $\rho_s = 0.52$ at 3.48$\sigma$.
In addition, it is also clear from the figure that analogous of some of the highest metallicity absorbers with smaller $L$ originating from low $N$(H~{\sc i}) systems are not probed in the low$-z$ sample.
For the combined low$-z$ and high$-z$ \CIII\ systems, no correlation
between $L$ and $N(\HI)$ ($\rho_s = -0.04$ at 0.49$\sigma$) was found as anticipated based on well displayed segregation.

\subsubsection {{$[C/H]$ vs. \nce\ : }}
In panel (d) of Fig.~\ref{fig_corr}, we show $[C/H]$ as a function of
\nce\ for all the systems. 
For the low$-z$ \CIII\ sample, we see a moderate correlation with $\rho_s$ = 0.33 (3.27$\sigma$) (see, panel (d) of Fig~\ref{lowzcor})
which is also confirmed from the survival
analysis test with $\rho_s$ = 0.31 (3.07$\sigma$). 
The high$-z$ sample
has a large scatter in \nce\ compared to the low$-z$ sample and shows a strong correlation with $[C/H]$ ($\rho_s = 0.56$ (4$\sigma$)).
It is also evident from the figure that for a given \nce, the low$-z$ \CIII\ absorbers have systematically higher $[C/H]$ compared to high$-z$
absorbers. This is mainly due to the redshift evolution of metallicity among the \CIII\ absorbers as we have discussed in \S \ref{red_eval1}.

\subsubsection {{$[C/H]$ vs. \nhi\ : }}
We show $[C/H]$ as a function of observed \nhi\ for the low$-z$ and high$-z$
absorbers in panel (e)
of Fig.~\ref{fig_corr}. 
We find an anti-correlation between these two quantities with $\rho_s = -0.44$ (at 4.38$\sigma$)
for the low$-z$ \CIII\ absorbers as shown in panel (e) of Fig~\ref{lowzcor}. 
A survival analysis test also confirms the trends
with a $\rho_s$ = $-$0.33 at 3.27$\sigma$ significance.
Note that, this trend is weak compared to the low$-z$ survey
of the COS-Halos sample of \citet{Prochaska2017} ($\rho_s = -0.69$ at 3.80$\sigma$) and the COS-weak sample of
\citet{Muzahid2018} ($\rho_s\, =\, -0.94$ at 5.24$\sigma$).
The high$-z$ \CIII\ absorbers over plotted on panel (e)
of Fig.~\ref{fig_corr} also show a weak anti-correlation ($\rho_s = -0.34$ at 2.49$\sigma$)
with most of the lower-\nhi\ absorbers (\nhi\ $\le\, 10^{14}$ \cmsq) having $[C/H]$ $>\, -0.5$ and higher-\nhi\ absorbers 
(\nhi\ $>\, 10^{14}$ \cmsq) having large scatter in $[C/H]$ in the range, $-2.8 < [C/H] < 0.6$. However, a strong anti-correlation was observed between $[C/H]$ and \nhi\ ($\rho_s = -0.65$ at 5,80$\sigma$)
for the component-wise analysed data, S1 $+$ S2 of \AS19.
Hence, the weak trend of \CIII\ absorption systems could be due to the
possibility of poor small scale metal mixing so that the
low-ionization metal lines need not necessarily be associated with the entirely
observed \nhi\ \citep[for e.g.,][]{Schaye2007}.
{Also, the high$-z$ \CIII\ sample on \nhi\ vs. $[C/H]$ plane gives a strong hint of metallicity evolution for weak \nhi\ (\nhi\ < $10^{15}$ \cmsq\ ) absorbers.
These absorbers are believed to be transients caught in the act and maybe related stripping of metal enriched gas from galaxies or galactic outflows \citep{Schaye2007}. These absorbers seem to be present in group
environments and are less active at a later Universe than they were earlier \citep{Muzahid2018}. A homogeneous sample of such absorbers at low and intermediate redshift could shed light to the detail physical process of such transient absorbers in the CGM.}

\subsubsection{Summary}
We find a strong correlation between $L$ and $[C/H]$ for the low$-z$ \CIII\ absorption
systems as seen in high$-z$ absorbers.
However, the low$-z$ and high$-z$ absorbers show different ranges in $L$ and $[C/H]$ for similar \nhi.
This means, if $L$ is related to the cooling length ($t_{cool}$ $\times$ $c_s$)
as argued by \AS19, the higher density and metallicity could be the reason
for the small $L$ seen in the low$-z$ absorbers having same \nhi.
We also notice that the trend seen in
the $L$ vs \nce\ plane is consistent with the two population idea suggested by \citet{Kim2016}. We find when appropriately divided in $L$ and $[C/H]$ ranges,
the high metallicity branch exhibits a correlation between $L$ and \nce, while the
low metallicity branch does not show such correlation. Unfortunately, in the combined sample, low metallicity
range (resp. high metallicity range) is dominated by the high$-z$ (resp. low$-z$) absorbers.
Confirming this with an increased number of \CIII\ absorbers spanning
the $L$ and \nce\ phase uniformly will have interesting implications in
understanding the metallicity evolution of CGM gas . 

\subsection{Size metallicity relationship}
\begin{figure}
  \centering
  \includegraphics[totalheight=0.22\textheight, trim=0.0cm 0.0cm 0cm 0.0cm, clip=true, angle=0]{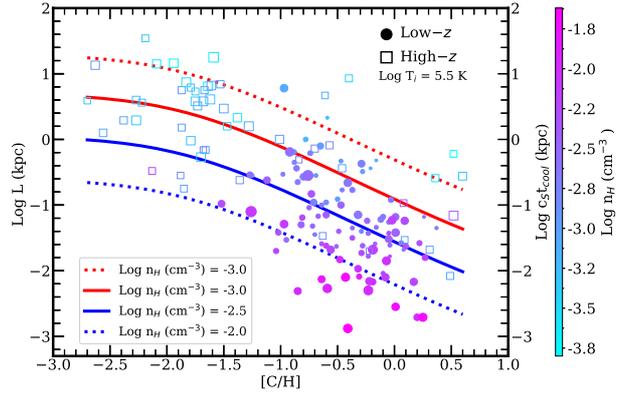}
 \caption{Comparison of
$L$ vs. $[C/H]$ of the combined \CIII\ systems for
a simple isochoric cooling case. The symbols and colours are same as panel (a) of Fig.~\ref{fig_corr}. The cooling
length scale ($c_s \times t_{cool}$) obtained from our simple toy model with different constant \nh\ (i.e., for the median \nh\ of low$-z$ (log \nh\ $\simeq$ -3) and high$-z$ (log $\simeq$ -2.5)
with 1$\sigma$ upper and lower range, respectively) are indicated with continuous lines as shown in the legends. We use an initial
temperature, log $T_i$ (K) = 5.5 in our model.}
\label{fig_cool}
\end{figure}

The strongest correlation we notice in our study
is the one between $L$ vs. $[C/H]$. While this correlation is seen for both high$-z$ and low$-z$ samples, the slopes of the distributions are different. In particular for a
given metallicity, $L$ is smaller for low$-z$ systems as
these systems tend to have higher \nh\ compare to the high$-z$ systems. In
\AS19, this correlation is explained through a simple toy model where the cloud size is decided by the cooling length.

To explore this further, we obtained $t_{cool}$ $\times$ $c_s$ for the `isochoric' case with our simple toy model as discussed in \AS19.
In Fig.~\ref{fig_cool}, we show the PI model predicted $[C/H]$ vs. $L$ in scattered points with
over plotted `isochoric' model predictions of $t_{cool}$ $\times$ $c_s$ for
different \nh\ (i.e., for the median \nh\ of low$-z$ (log \nh\ $\simeq$ -3) and high$-z$ (log $\simeq$ -2.5)
with 1$\sigma$ upper and lower range, respectively). We use an initial
temperature, log $T_i$ (K) = 5.5 in our model. As shown in \AS19, the model curves roughly follow the observed anti-correlation
between $[C/H]$ vs. $L$ for the high$-z$. 

The slight enhancement in \nh\ at low$-z$ explains that
the size of the low$-z$ absorbers is less than that of the high$-z$ absorbers
for a given metallicity. However, for the low$-z$, the observed slope is steeper than what
is predicted by the model. This is mainly due to the presence of small size clouds near solar metallicity.
Also for a given metallicity the predicted range in $L$ is narrower than what we observe.
In our toy model, to increase the slope, we have to decrease the cooling time for high metallicity gas.
This can naturally happen if we introduce a correlation between the initial density and metallicity
in our models. In our data, we do see a moderate (i.e., 2$\sigma$) correlation between \nh\ and $[C/H]$.
Another possibility is the
fragmentation of a larger high metallicity parent cloud to reach pressure equilibrium with its surroundings at a
final cooling time scale \citep{McCourt2018}.
However, it is important to remember that in this work, we use
total column densities (and not the column densities measured for individual components) to derive \nh, $[C/H]$ and $L$.
Thus, the inferred length can also be related to the number of individual components along the line-of-sight.

\subsection{Redshift evolution of $[C/H]$}
\label{abs_zevo}
\begin{figure}
  \centering
  \includegraphics[totalheight=0.25\textheight, trim=0.2cm 0.4cm 0cm 0.15cm, clip=true, angle=0]{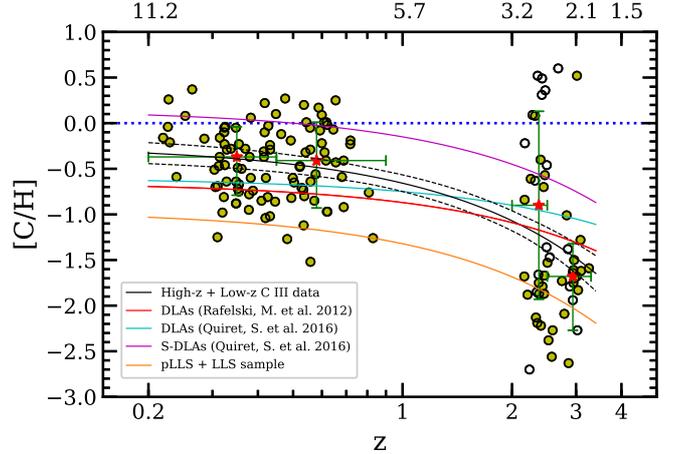}
 \caption{{$[C/H]$ versus redshift (time since big bang). The scattered points show the combined
 \CIII\ systems whereas the yellow filled circles are \CIII\ systems with \nhi\ $\ge$ $10^{15}$ \cmsq. The thick stars denote the median value of the [C/H]
 obtained over the extent of the horizontal bar in the redshift range and the vertical bar denote 1$\sigma$ range in shown median values.
 The black solid line denotes the linear regression fit (LRF) to $[C/H]$ vs. $z$ and the black dotted lines indicate 1$\sigma$ deviation
 around the LRF. The red, cyan, magenta and orange lines show the LRF fits from the measurements of cosmic metallicity
 of DLAs from \citet{Rafelski2012}, \citet{Quiret2016}, sub$-$DLAs from \citet{Quiret2016} and pLLSs + LLSs from \citet{Lehner2013, Lehner2016, wotta19}, respectively.}}
\label{fig_met}
\end{figure}

Understanding the redshift evolution of
metallicity of a particular class of absorption systems (like damped \lya\ systems (DLAs) and metal line systems)
is very important as it is intimately related to the redshift
evolution of star formation and associated feedback processes in galaxies.
Generally, DLAs, sub$-$DLAs (s$-$DLAs), Lyman limit systems (LLSs) and partial-LLSs (pLLSs) are used to study
the evolution of metallicity of gas around the galaxies
which show an increasing trend with age of the Universe \citep{Rafelski2012, Lehner2013, Lehner2016,Quiret2016, wotta19}.
In this section, we compare the redshift evolution of $[C/H]$ of \CIII\ absorbers with the redshift evolution of metallicity of DLAs, s$-$DLAs, and pLLS $+$ LLS systems. As we have discussed before,
the \CIII\ absorbers studied here are optically thin to Lyman continuum radiation and
typically have \nhi\ $<$ 10$^{17}$ cm$^{-2}$.

In Fig.~\ref{fig_met}, we show the $[C/H]$ evolution
of \CIII\ absorbers as a function of redshift (and time since big bang).
At high$-z$, the scatter in metallicity is larger (over $\simeq$3 dex spread in [C/H]) compared to
the low$-z$ ($\simeq$2 dex spread in [C/H]). 
As seen from panel (c) of Fig.~\ref{fig_corr}, most of the systems with super-solar metallicity at
high$-z$ have \nhi\ $<$ 10$^{14}$ cm$^{-2}$. Such systems are not represented in the low$-z$ sample
due to the \nhi\ cutoff used for the sample selection. When restricting to
the same \nhi\ range for low$-z$ and high$-z$, we find the scatter at high$-z$ is reduced considerably.

Fig.~\ref{fig_met} suggests a
strong evolution in metallicity between $z \simeq 2.5$ 
and $z\simeq 1.0$ (the age of the universe between $\simeq$ 2.5 to 5.8 Gyr). The evolution 
becomes relatively slower in later times as seen from the median
[C/H] values in Fig.~\ref{fig_met} (thick stars). We find a strong anti-correlation between
$z$ and $[C/H]$ with $\rho_s=-0.44$ at a significance level of
5.40$\sigma$. The LRF to $[C/H]$ vs. $z$ gives a slope, $-0.41\pm0.05$ with intercept, $-0.25\pm0.08$. 

We compare the redshift
evolution of $[C/H]$ of the \CIII\ absorbers with the cosmic metallicity ($Z$) evolution of DLAs \citep{Rafelski2012,Quiret2016},
sub$-$DLAs \citep{Quiret2016} and pLLSs + LLSs \citep{Lehner2013, Lehner2016, wotta19}. The redshift
evolution of cosmic metallicity of DLAs measured by \citet{Rafelski2012} shows a shallower slope 
[$<Z>\, = (-0.22\pm0.03)z - (0.65\pm0.09)$] upto $z$ $\simeq$ 5 (denoted by red line in Fig.~\ref{fig_met}).
Similarly, a LRF to $<Z>$ vs. $z$ for DLAs measured by \citet{Quiret2016}
in the redshift, $0<z<4$ shows a slope of $-0.15\pm0.03$ with a higher intercept, $-(0.6\pm0.13)$ (denoted by cyan line)
whereas the LLSs in their data show a more elevated slope
with redshift, $<Z> = (-0.3\pm0.07)z - (0.15\pm0.31)$ (denoted by magenta line). 
The low$-z$ (z < 1) and high$-z$ $(2.3<z<3.4)$ pLLSs + LLSs are selected from \citet{Lehner2013, wotta19}
and \citet{Lehner2016}, respectively. The LRF to the combined pLLSs + LLSs sample across redshift range, $0.2<z<3.4$
gives a slope, $-0.36\pm0.10$ with an intercept, $-0.95\pm0.12$ (denoted by orange line in Fig.~\ref{fig_met}).
As seen from Fig.~\ref{fig_met},
the slope we measure for the combined low$-z$ and high$-z$ \CIII\ samples are higher than that measured for DLAs but consistent within the error range measured for sub$-$DLAs for
high$-[C/H]$ branch and pLLSs + LLSs for low$-[C/H]$ branch of the \CIII\ absorbers.
\noindent \section {Summary}
\label{result}
We have constructed a sample of \CIII\ absorption systems in
optically thin H~{\sc i} clouds using the COS CGM compendium (CCC) sample of
\L18\ at low$-z$ i.e., in the range, $0.2 \le z \le 0.9$.
We find a total of 99 \CIII\ absorption systems
from the CCC sample with \nhi\ (\cmsq) in the range, $15.0 \le {\rm log}\, \nhi\ \le 16.2$.
The observed N(\HI), N(\CII), N(\CIII) and column densities of other
available metal ions along with their associated errors in the measured values have been adapted from the \L18\ measurements.
We performed photoionization models for each of these individual \CIII\ systems using \CLOUDY\ with the updated
\KS18\ UVB ($\alpha = -1.8$) as the incident radiation field.
We have discussed the
implications of the observed and derived properties of the \CIII\ absorbers systematically.
We have also combined the low$-z$ systems with the high$-z$ \CIII\ absorption systems from \AS19\ to study the various correlation between the physical and chemical properties of the optically thin \CIII\ absorption systems.
We have also shown the redshift evolution of the inferred properties of the \CIII\ absorbers for a wide
redshift range from $0-3.3$ in compliment with our PI models. Our main results are as follows:

\begin{itemize}
\item{} 
Under the single-phase PI models, the derived hydrogen density (\nh) of the low$-z$ optically thin
\CIII\ absorption systems
ranges from $10^{-3.4}$ - $10^{-1.6}$ \cmcb\ with a median value of 
$10^{-2.5}$ \cmcb. The median \nh\ of the low$-z$ \CIII\ systems is 0.6 dex higher than the high$-z$
systems analysed in \AS19. When we use four sub-samples divided in redshift bins, the redshift evolution of \nh\ is not monotonous. 

\item{} 
Metallicity, denoted as {\it [C/H]}, for the low$-z$ absorbers show a unimodal distribution ranged from
-1.6 to 0.4, with a median value of -0.46. 
The distribution of $[C/H]$ for the low and high redshift \CIII\ systems show a large spread which indicates
that the absorbers are originating from the CGM of host galaxies having significantly different
physical and chemical properties.
 We found a strong redshift evolution of $[C/H]$ between
low$-z$ and high$-z$ absorbers.
For the \CIII\ absorbers with \nhi\ (\cmsq) > $10^{15}$, 
we found that the low$-z$ systems have $\approx$14\% absorbers with super solar metallicity and are more metal enriched
than the high$-z$ counterparts which have only $\approx$3\% super solar systems.
\item{} 
The over-density of the low$-z$ \CIII\ systems show a range from $3-4.5$ with a median value of 3.74 which suggests that
the absorbers are indeed originating from the high over-dense regions like CGM. The large difference in the median $\Delta$ between
low$-z$ and high$-z$ \CIII\ absorbers
($\approx$1.6 dex) is mainly due to the redshift evolution of the mean density, $\overline{n_H}$.

\item{} 
The derived line-of-sight thickness ($L$) of the low$-z$ \CIII\ systems are ranging from 1.3 pc to 10 kpc with a median value of 0.6 kpc which is $\sim$10 times lower than the high$-z$ counterparts. The difference in physical parameters (like \nh, [C/H], $\Delta$ and $L$) of the absorbers suggest a strong redshift evolution of these quantities between
low$-z$ and high$-z$ \CIII\ systems. The distribution of $L$ is also significantly different between high$-z$ and low$-z$ \CIII\ systems. While there are significant numbers of sub-kpc size absorbers (78\%) present at low$-z$, only 6\% of the absorbers at high$-z$ show such sub-kpc scale line-of-sight thickness.
This clearly indicates that the low$-z$ \CIII\ absorbers are metal enriched, dense and compact clouds compared to the high$-z$ counterparts.

\item{}
We find a statistically significant anti-correlation between $L$ vs. $[C/H]$ for the combined \CIII\ sample, which suggests that the highest metallicity systems are the tiniest in size.
However, unlike high$-z$ \CIII\ absorbers, the low$-z$ absorbers are typically smaller in size
for a given $[C/H]$ as predicted by the toy model used in \AS19. 
This may be due to the slight enhancement in \nh\ at low$-z$ and existence of moderate level (i.e., 2$\sigma$)
correlation between \nh\ and $[C/H]$ in which case the clouds will have a
shorter cooling time scale (or cooling length).
Another possibility of the existence of such small scale absorbers could be the
fragmentation of a larger high metallicity cloud where the parent cloud is fragmented to tiny cloudlets to
reach pressure equilibrium with its surroundings efficiently at a
final cooling time scale \citep{McCourt2018}.

\item{}
We see a clear segregation of the high $[C/H]$ and low $[C/H]$ branch absorbers occupied in
distinct regions of $L$ vs. \nce\ plane consistent with the two populations idea suggested by \citet{Kim2016}.
We find that the high$-[C/H]$ branch absorbers ($[C/H]>-1.2$) with $L<0.25$ kpc show a strong
correlation with \nce, whereas, the low$-[C/H]$ branch absorbers do not exhibit such correlation with \nce.
However, the high$-[C/H]$ branch absorbers are mostly contributed by low$-z$ systems and low$-[C/H]$ by high$-z$ systems.
Hence, an increased number of \CIII\ absorbers spanning in the $L$ vs. \nce\ plane uniformly is required to confirm the existence of a statistically significant bimodality in metallicity.
The results will have implications in understanding the metal enrichment of the CGM gas and its evolution. 

\item{}
Finally, we compare the redshift evolution of
[C/H] with the different class of absorption systems (DLAs, s$-$DLAs, pLLLs and LLSs).
As seen in Fig.~\ref{fig_met}, the metallicity of the \CIII\ absorbers show a strong evolution between $z\approx2.5$ and $z\approx1$ while the age of the Universe was between $\approx2.5$ Gyr to 5.8 Gyr. The evolution is relatively slower at a later stage of the Universe. The linear regression fit to the $z$ vs. $[C/H]$ gives a slope $-0.41\pm0.05$ with intercept,
$-0.25\pm0.08$. This is steeper than that measured for DLAs \citep{Rafelski2012, Quiret2016}. Interestingly, the elevated slope 
of the \CIII\ absorbers are within the error range of the measured slope of s$-$DLAs from \citet{Quiret2016} and
pLLSs + LLSs sample across redshift from \citet{Lehner2013, Lehner2016, wotta19}. 
However, we reiterate the fact that our results
may be illusive owing to the lack of the low$-$\nh\ absorbers at low$-z$ as well as lack of \CIII\ absorbers in the
intermediate redshift ($0.9\le z \le 2.1$).

To make further progress in this field, it will be important to search for such \CIII\ absorbers in the intermediate redshift
ranges (i.e., $1.0\le z\le 2.0$) and at higher redshift (i.e., $z>3$) with uniform coverage in \nhi. This
will allow to create a homogeneous \CIII\ sample across a large redshift ranges. 
\end{itemize}

\section*{Acknowledgments}
The authors wish to thank the anonymous referee for providing
valuable comments and suggestions for improving the manuscript.
AM acknowledges the Department of Science and Technology (DST), Government of India for financial support through DST-INSPIRE fellowship (IF150845). AM and ACP wish to thank
the Inter-University Centre for Astronomy and Astrophysics (IUCAA), India
for providing hospitality and travel grants. 

\section*{DATA AVAILABILITY}
The data underlying this article are available in
VizieR On-line Data Catalog: J/ApJ/866/33, at \url{https://cdsarc.unistra.fr/viz-bin/cat/J/ApJ/866/33}.
\def\aj{AJ}%
\def\actaa{Acta Astron.}%
\def\araa{ARA\&A}%
\def\apj{ApJ}%
\def\apjl{ApJ}%
\def\apjs{ApJS}%
\def\ao{Appl.~Opt.}%
\def\apss{Ap\&SS}%
\def\aap{A\&A}%
\def\aapr{A\&A~Rev.}%
\def\aaps{A\&AS}%
\def\azh{AZh}%
\def\baas{BAAS}%
\def\bac{Bull. astr. Inst. Czechosl.}%
\def\caa{Chinese Astron. Astrophys.}%
\def\cjaa{Chinese J. Astron. Astrophys.}%
\def\icarus{Icarus}%
\def\jcap{J. Cosmology Astropart. Phys.}%
\def\jrasc{JRASC}%
\def\mnras{MNRAS}%
\def\memras{MmRAS}%
\def\na{New A}%
\def\nar{New A Rev.}%
\def\pasa{PASA}%
\def\pra{Phys.~Rev.~A}%
\def\prb{Phys.~Rev.~B}%
\def\prc{Phys.~Rev.~C}%
\def\prd{Phys.~Rev.~D}%
\def\pre{Phys.~Rev.~E}%
\def\prl{Phys.~Rev.~Lett.}%
\def\pasp{PASP}%
\def\pasj{PASJ}%
\def\qjras{QJRAS}%
\def\rmxaa{Rev. Mexicana Astron. Astrofis.}%
\def\skytel{S\&T}%
\def\solphys{Sol.~Phys.}%
\def\sovast{Soviet~Ast.}%
\def\ssr{Space~Sci.~Rev.}%
\def\zap{ZAp}%
\def\nat{Nature}%
\def\iaucirc{IAU~Circ.}%
\def\aplett{Astrophys.~Lett.}%
\def\apspr{Astrophys.~Space~Phys.~Res.}%
\def\bain{Bull.~Astron.~Inst.~Netherlands}%
\def\fcp{Fund.~Cosmic~Phys.}%
\def\gca{Geochim.~Cosmochim.~Acta}%
\def\grl{Geophys.~Res.~Lett.}%
\def\jcp{J.~Chem.~Phys.}%
\def\jgr{J.~Geophys.~Res.}%
\def\jqsrt{J.~Quant.~Spec.~Radiat.~Transf.}%
\def\memsai{Mem.~Soc.~Astron.~Italiana}%
\def\nphysa{Nucl.~Phys.~A}%
\def\physrep{Phys.~Rep.}%
\def\physscr{Phys.~Scr}%
\def\planss{Planet.~Space~Sci.}%
\def\procspie{Proc.~SPIE}%
\let\astap=\aap
\let\apjlett=\apjl
\let\apjsupp=\apjs
\let\applopt=\ao

\bibliographystyle{mnras}
\bibliography{paper}
 \appendix 
\section{Correlations among various model predicted parameters of the low$-z$ \CIII\ sample}\label{coraa}
\begin{figure*}
  \centering
    \includegraphics[totalheight=0.3\textheight, trim=0cm 0cm 0cm 0cm, clip=true, angle=0]{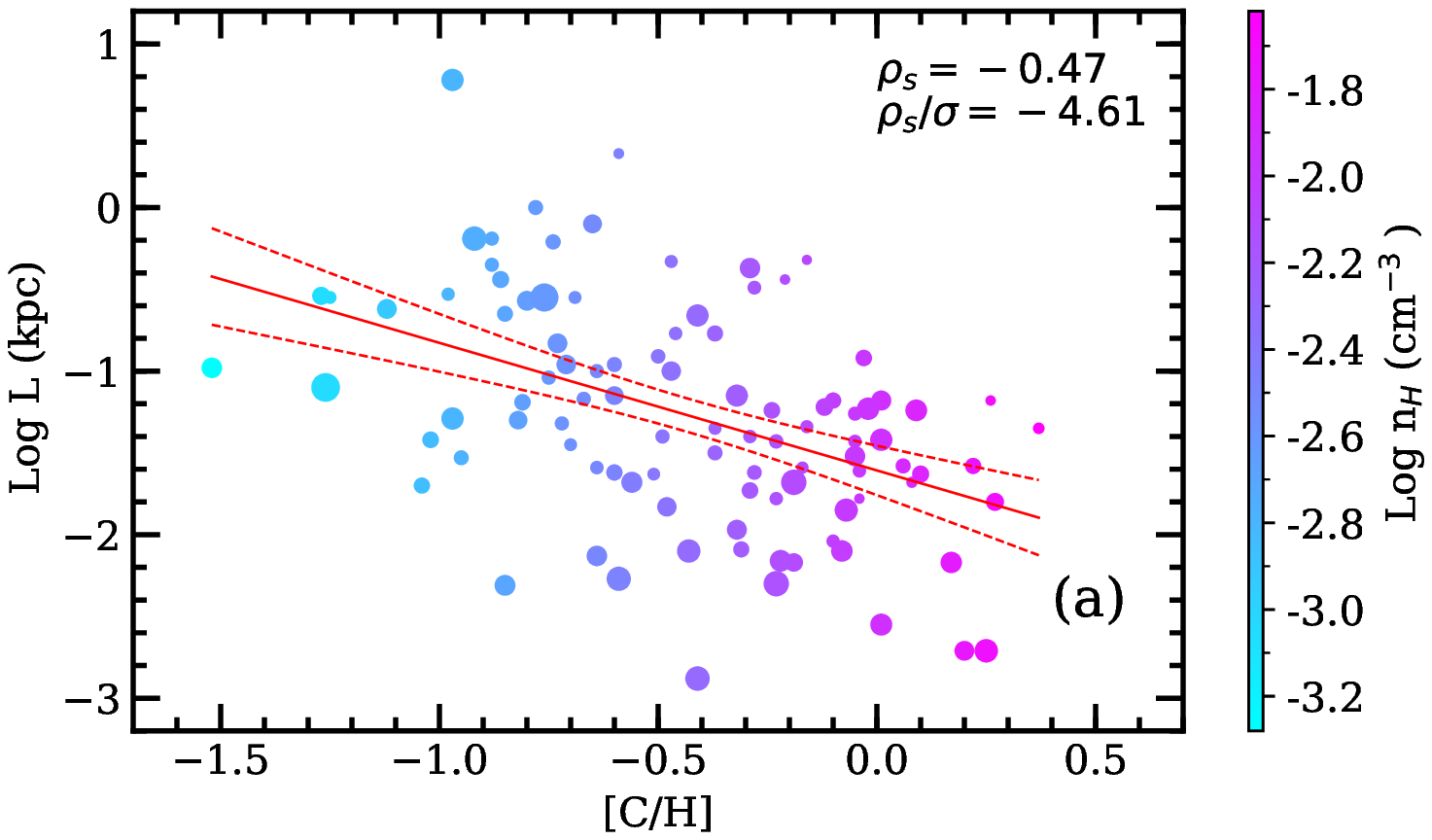}\\
  \includegraphics[totalheight=0.2\textheight, trim=0cm 0cm 0cm 0cm, clip=true, angle=0]{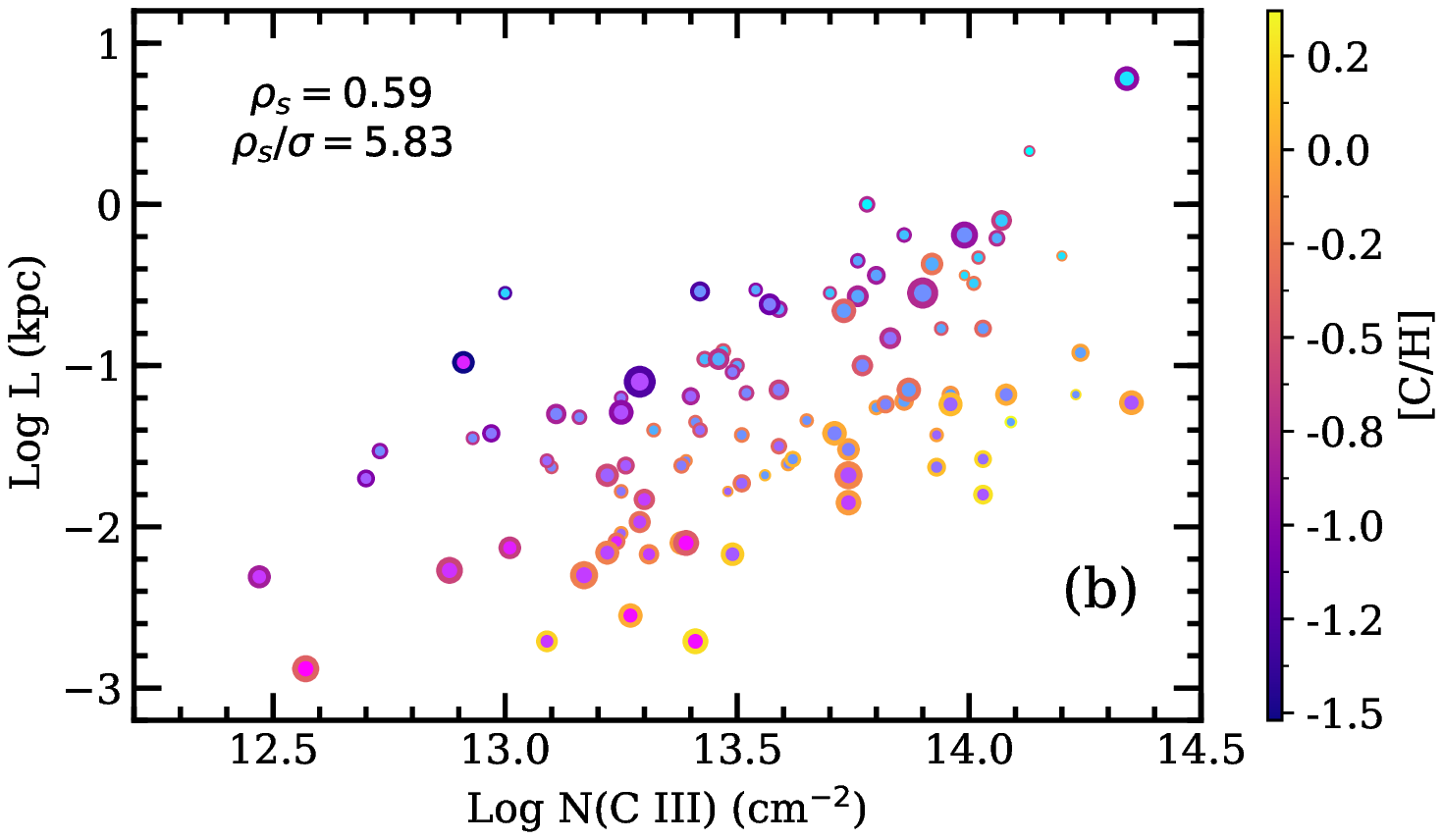}
    \includegraphics[totalheight=0.20\textheight, trim=0cm 0cm 0cm 0cm, clip=true, angle=0]{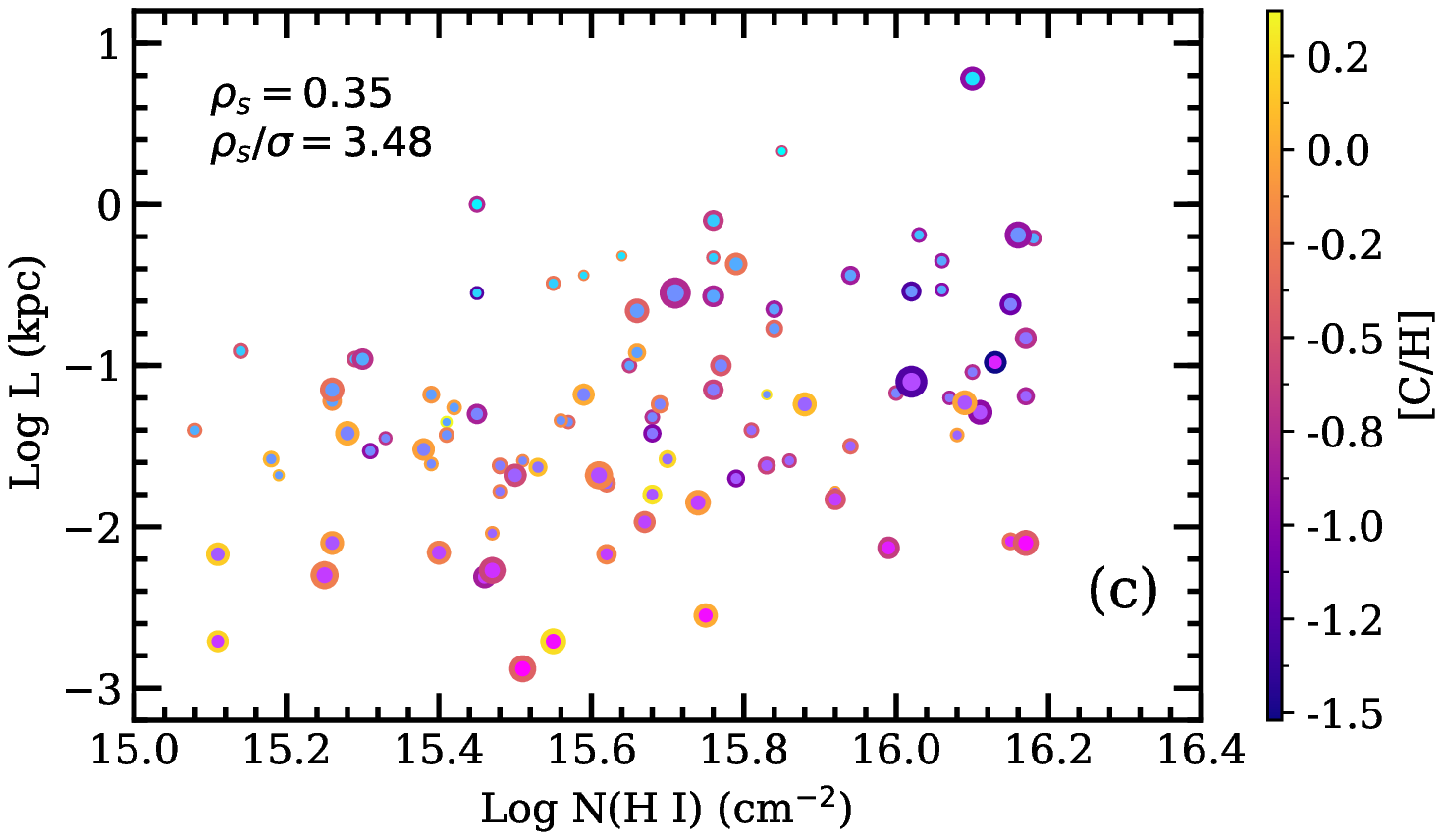}
      \includegraphics[totalheight=0.2\textheight, trim=0cm 0cm 0cm 0cm, clip=true, angle=0]{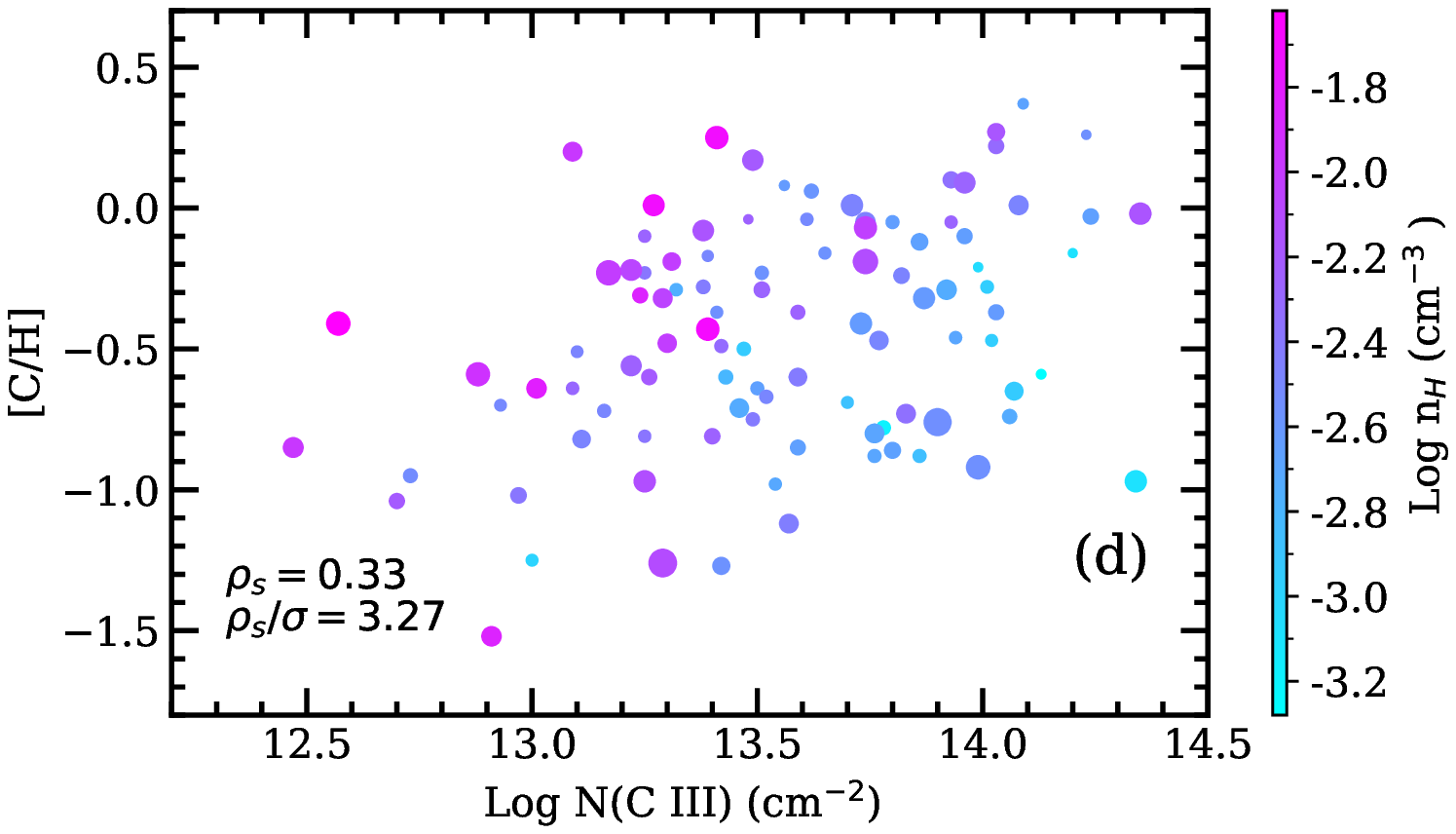}
      \includegraphics[totalheight=0.20\textheight, trim=0cm 0cm 0cm 0cm, clip=true, angle=0]{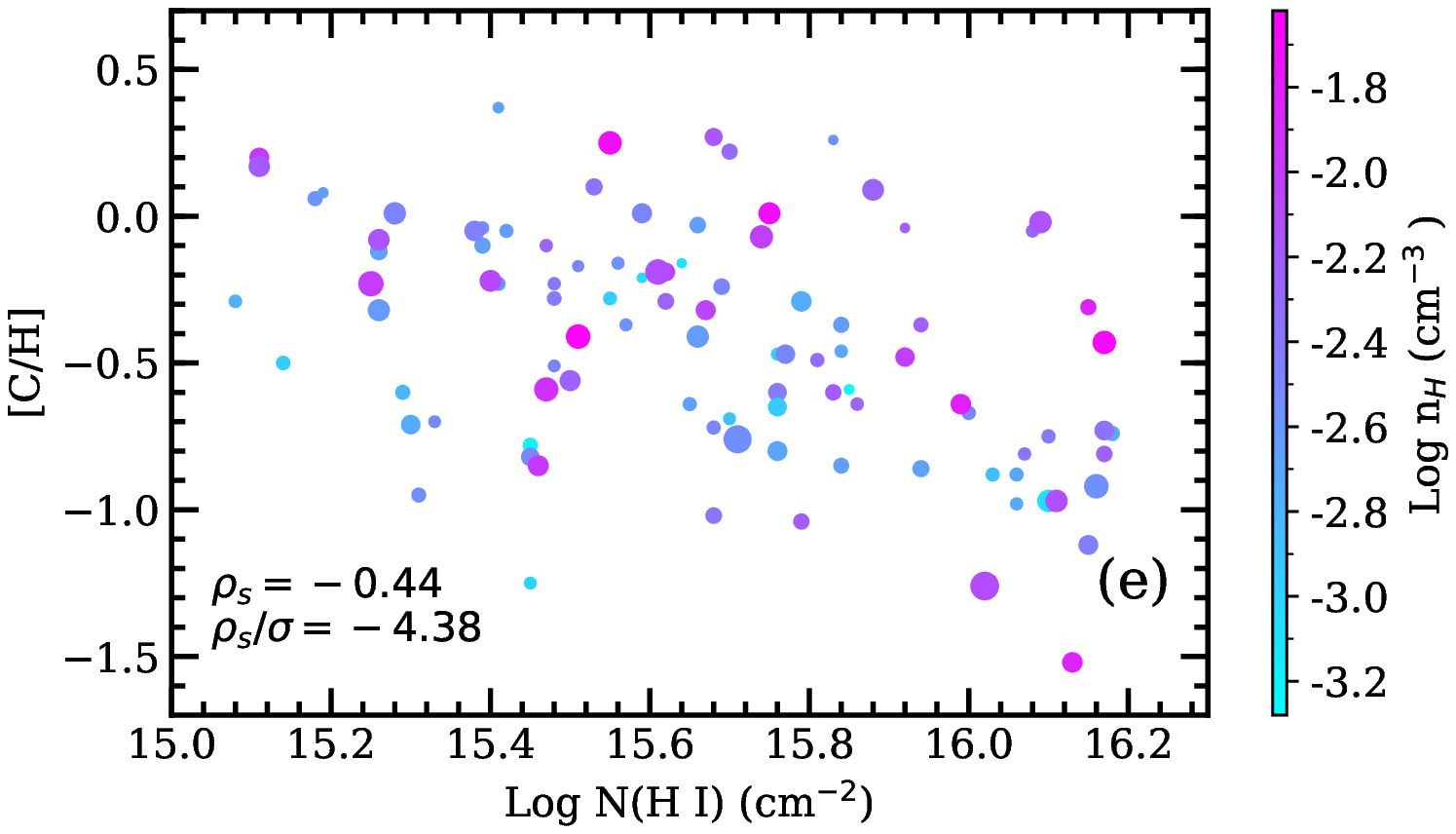}
  \caption{The plots show correlations among
 various model predicted parameters of the low$-z$ \CIII\ sample. We show the Spearman
 rank correlation coefficients ($\rho_s$) and significance levels ($\rho_s$/$\sigma$) in the legends.
 Panel (a): $L$ vs. $[C/H]$ and the symbols are colour coded with log \nh\ values as shown in the side colour bar.
 The solid line shows the LRF to the data and the dashed solid lines indicate the 1$\sigma$ deviation
 around the LRF.
 Middle panels: $L$ vs. \nce\ (panel (b)) and $L$ vs. \nhi\ (panel (c)) with additional colour codings of $[C/H]$ values (thick rings) as shown 
 in the side colour bars. 
 Bottom panels: $[C/H]$ vs. \nce\ (panel (d))
 and $[C/H]$ vs. \nhi\ (panel (e)) with the same colour coding of \nh\ as used in the top panel.
 The symbol sizes increase with the increasing of redshifts for all the figures.}
\label{lowzcor}
\end{figure*}

In this appendix, we show the correlation between derived parameters of the low$-z$ absorbers
that are discussed in \S \ref{cort} (indicated by circles in Fig.~\ref{fig_corr}). In panel (a) of Fig.~\ref{lowzcor}, we present the
anti-correlation between $L$ and $[C/H]$. The symbols are colour coded with log \nh\ values as shown in the side colour bar.
We show the correlation between $L$ vs. \nce\ and $L$ vs. \nhi\ in panel (b) and panel (c) of Fig.~\ref{lowzcor}, respectively.
In addition to the colour coding of \nh\ as used in panel (a), we colour code $[C/H]$ values (thick rings) as
shown in the side colour bars of these panels.
In panel (d) and panel (e) of Fig.~\ref{lowzcor}, we
plot the correlation between $[C/H]$ vs. \nce\ and $[C/H]$ vs. \nhi, respectively.
The symbol sizes increase with the increasing of
redshifts for the figures in all panels.
The correlation coefficient with significance level as indicated in the legends of each panels of Fig.~\ref{lowzcor}
are discussed in \S \ref{cort}.
\section{Details of the low$-z$ sample and Photoionization (PI) model outputs}\label{piout}
In this appendix, we provide the details of the low$-z$ absorbers which are analysed in this work along with the results of PI models.
In Table~\ref{pitable}, we give the details of the observed column densities of \CIII\ and \CII\ ions along with
the flags in the measurements. We also provide the results of PI models for each on these absorbers. 
In Table~\ref{tab:si} and ~\ref{tab:oxy}, we provide the details of the absorbers for which there are associations of other metal ion combinations
(i.e., sample SA2 of the low$-z$ absorbers).
\newpage
\renewcommand{\arraystretch}{1.02}
\scriptsize
\onecolumn
\begin{center}
\begin{longtable}{lccccccccccccc}
\caption{Summary of low$-z$ \CIII\ sample and photoionization (PI) model predicted parameters.} \label{pitable} \\
\hline
\hline
&  & \multicolumn{6}{c}{Observed } & \multicolumn{6}{c}{PI model} \\ 
\cline{3-8} \noalign{\smallskip} \cline{9-14}  \\
IDs & $z_{abs}$ & log \nhi\ &log $N$(\CIII) & Flag & log $N$(\CII) & Flag & Flag on &log \nh\ & $[C/H]$ & {log \it{L}} & log $\Delta$ &log $N$(\CIII) &log $N$(\CII) \\ 
&  & (\cmsq) & (\cmsq) &  & (\cmsq) & & $\frac{N(\CIII)}{N(\CII)}$ & (\cmcb) &  & (kpc) &  & (\cmsq) & (\cmsq) \\ 
(1) & (2) & (3) & (4) & (5) & (6) & (7) & (8) & (9) & (10) & (11) & (12) & (13) & (14) \\ \hline
\endfirsthead

\multicolumn{14}{c}%
{{\bfseries \tablename\ \thetable{}} ({\it {continued}})} \\

\hline
\hline
&  & \multicolumn{6}{c}{Observed } & \multicolumn{6}{c}{PI model} \\ 
\cline{3-8} \noalign{\smallskip} \cline{9-14}  \\
IDs & $z_{abs}$ & log \nhi\ &log $N$(\CIII) & Flag & log $N$(\CII) & Flag & Flag on &log \nh\ & $[C/H]$ & {log \it{L}} & log $\Delta$ &log $N$(\CIII) &log $N$(\CII) \\ 
&  & (\cmsq) & (\cmsq) &  & (\cmsq) & & $\frac{N(\CIII)}{N(\CII)}$ & (\cmcb) &  & (kpc) &  & (\cmsq) & (\cmsq) \\ 
(1) & (2) & (3) & (4) & (5) & (6) & (7) & (8) & (9) & (10) & (11) & (12) & (13) & (14) \\ \hline
\endhead

\hline \multicolumn{7}{r}{{{{\bfseries \tablename\ \thetable{}} ({\it {continued}})}}} \\
\endfoot

\hline
\endlastfoot
1 & 0.21945 & $15.64\pm0.04$ & $>14.20$ & $-2$ & $<13.57$ & $-1$ & $-2*$ & $-3.09$ & $-0.16$ & $-0.32$ & $3.42$ & $14.25$ & $13.42$ \\ 
2 & 0.22620 & $15.92\pm0.09$ & $13.48\pm0.09$ & $0$ & $<13.62$ & $-1$ & $-2$ & $-2.25$ & $-0.04$ & $-1.78$ & $4.25$ & $13.53$ & $13.58$ \\ 
3 & 0.22719 & $15.83\pm0.01$ & $>14.23$ & $-2$ & $13.84\pm0.08$ & $0$ & $-2*$ & $-2.56$ & $+0.26$ & $-1.18$ & $3.92$ & $14.24$ & $13.93$ \\ 
4 & 0.22902 & $15.59\pm0.04$ & $>13.99$ & $-2$ & $<13.59$ & $-1$ & $-2*$ & $-3.05$ & $-0.21$ & $-0.44$ & $3.45$ & $14.13$ & $13.32$ \\ 
5 & 0.23881 & $15.85\pm0.03$ & $>14.13$ & $-2$ & $<13.39$ & $-1$ & $-2*$ & $-3.28$ & $-0.59$ & $+0.33$ & $3.21$ & $14.26$ & $13.18$ \\ 
6 & 0.25247 & $15.19\pm0.01$ & $13.56\pm0.01$ & $0$ & $<13.04$ & $-1$ & $-2*$ & $-2.63$ & $+0.08$ & $-1.68$ & $3.84$ & $13.54$ & $13.14$ \\ 
7 & 0.26382 & $15.41\pm0.02$ & $>14.09$ & $-2$ & $<13.67$ & $-1$ & $-2*$ & $-2.68$ & $+0.37$ & $-1.35$ & $3.78$ & $14.12$ & $13.66$ \\ 
8 & 0.28495 & $15.51\pm0.06$ & $13.39\pm0.06$ & $0$ & $<13.29$ & $-1$ & $-2*$ & $-2.50$ & $-0.17$ & $-1.59$ & $3.94$ & $13.49$ & $13.18$ \\ 
9 & 0.30371 & $15.48\pm0.06$ & $13.1\pm0.06$ & $0$ & $<13.08$ & $-1$ & $-2$ & $-2.48$ & $-0.51$ & $-1.63$ & $3.94$ & $13.13$ & $12.81$ \\ 
10 & 0.30579 & $15.33\pm0.06$ & $12.93\pm0.06$ & $0$ & $<12.83$ & $-1$ & $-2$ & $-2.51$ & $-0.70$ & $-1.45$ & $3.79$ & $13.01$ & $12.51$ \\ 
11 & 0.30963 & $15.45\pm0.12$ & $13.00\pm0.12$ & $0$ & $<12.97$ & $-1$ & $-2$ & $-3.00$ & $-1.25$ & $-0.55$ & $3.41$ & $13.02$ & $12.13$ \\ 
12 & 0.30995 & $15.70\pm0.05$ & $>13.70$ & $-2$ & $<13.03$ & $-1$ & $-2$ & $-2.88$ & $-0.69$ & $-0.55$ & $3.53$ & $13.70$ & $12.93$ \\ 
13 & 0.31230 & $15.76\pm0.05$ & $14.02\pm0.05$ & $0$ & $<13.24$ & $-1$ & $-2$ & $-2.96$ & $-0.47$ & $-0.33$ & $3.45$ & $14.07$ & $13.22$ \\ 
14 & 0.31387 & $15.57\pm0.06$ & $13.41\pm0.06$ & $0$ & $<13.23$ & $-1$ & $-2$ & $-2.56$ & $-0.37$ & $-1.35$ & $3.85$ & $13.49$ & $13.07$ \\ 
15 & 0.31426 & $15.56\pm0.04$ & $13.65\pm0.04$ & $0$ & $<13.32$ & $-1$ & $-2$ & $-2.57$ & $-0.16$ & $-1.34$ & $3.84$ & $13.71$ & $13.27$ \\ 
16 & 0.31950 & $16.06\pm0.03$ & $13.54\pm0.03$ & $0$ & $<13.02$ & $-1$ & $-2$ & $-2.71$ & $-0.98$ & $-0.53$ & $3.69$ & $13.59$ & $12.99$ \\ 
17 & 0.32246 & $16.07\pm0.04$ & $13.25\pm0.04$ & $0$ & $<13.30$ & $-1$ & $-2$ & $-2.38$ & $-0.81$ & $-1.20$ & $4.02$ & $13.33$ & $13.08$ \\ 
18 & 0.32307 & $15.47\pm0.02$ & $13.25\pm0.02$ & $0$ & $13.17\pm0.09$ & $0$ & $0$ & $-2.27$ & $-0.10$ & $-2.04$ & $4.13$ & $13.27$ & $13.14$ \\ 
19 & 0.32340 & $16.08\pm0.01$ & $13.93\pm0.01$ & $0$ & $13.81\pm0.02$ & $0$ & $0$ & $-2.27$ & $-0.05$ & $-1.43$ & $4.13$ & $13.94$ & $13.80$ \\ 
20 & 0.32377 & $15.48\pm0.10$ & $13.25\pm0.1$ & $0$ & $<13.22$ & $-1$ & $-2$ & $-2.39$ & $-0.23$ & $-1.78$ & $4.01$ & $13.33$ & $13.07$ \\ 
21 & 0.32451 & $15.84\pm0.01$ & $>13.94$ & $-2$ & $<13.22$ & $-1$ & $-2*$ & $-2.70$ & $-0.46$ & $-0.77$ & $3.70$ & $13.91$ & $13.28$ \\ 
22 & 0.32565 & $15.39\pm0.01$ & $13.61\pm0.01$ & $0$ & $13.21\pm0.04$ & $0$ & $0*$ & $-2.51$ & $-0.04$ & $-1.61$ & $3.89$ & $13.61$ & $13.21$ \\ 
23 & 0.32793 & $15.86\pm0.08$ & $13.09\pm0.08$ & $0$ & $<13.22$ & $-1$ & $-2$ & $-2.29$ & $-0.64$ & $-1.59$ & $4.11$ & $13.18$ & $13.01$ \\ 
24 & 0.32848 & $15.08\pm0.04$ & $13.32\pm0.04$ & $0$ & $<13.66$ & $-1$ & $-2$ & $-2.76$ & $-0.29$ & $-1.40$ & $3.63$ & $13.37$ & $12.70$ \\ 
25 & 0.33719 & $15.55\pm0.04$ & $>14.01$ & $-2$ & $<13.59$ & $-1$ & $-2*$ & $-2.96$ & $-0.28$ & $-0.49$ & $3.43$ & $14.09$ & $13.19$ \\ 
26 & 0.34793 & $16.06\pm0.01$ & $13.76\pm0.01$ & $0$ & $<13.40$ & $-1$ & $-2*$ & $-2.70$ & $-0.88$ & $-0.35$ & $3.61$ & $13.82$ & $13.10$ \\ 
27 & 0.34800 & $15.42\pm0.04$ & $13.8\pm0.04$ & $0$ & $<13.42$ & $-1$ & $-2$ & $-2.65$ & $-0.05$ & $-1.26$ & $3.73$ & $13.85$ & $13.27$ \\ 
28 & 0.34833 & $15.65\pm0.02$ & $13.5\pm0.02$ & $0$ & $<13.03$ & $-1$ & $-2*$ & $-2.66$ & $-0.64$ & $-1.00$ & $3.72$ & $13.51$ & $12.91$ \\ 
29 & 0.34873 & $16.03\pm0.01$ & $>13.86$ & $-2$ & $<13.08$ & $-1$ & $-2*$ & $-2.86$ & $-0.88$ & $-0.19$ & $3.51$ & $13.89$ & $13.08$ \\ 
30 & 0.35197 & $15.81\pm0.06$ & $13.42\pm0.06$ & $0$ & $13.26\pm0.19$ & $0$ & $0$ & $-2.32$ & $-0.49$ & $-1.40$ & $3.98$ & $13.47$ & $13.15$ \\ 
31 & 0.35231 & $15.68\pm0.09$ & $13.16\pm0.09$ & $0$ & $<13.10$ & $-1$ & $-2$ & $-2.49$ & $-0.72$ & $-1.32$ & $3.88$ & $13.24$ & $12.82$ \\ 
32 & 0.35460 & $16.10\pm0.12$ & $>13.49$ & $-2$ & $<13.24$ & $-1$ & $-2$ & $-2.42$ & $-0.75$ & $-1.04$ & $3.95$ & $13.54$ & $13.20$ \\ 
33 & 0.35471 & $15.41\pm0.14$ & $13.51\pm0.14$ & $0$ & $<13.31$ & $-1$ & $-2*$ & $-2.57$ & $-0.23$ & $-1.43$ & $3.80$ & $13.56$ & $13.06$ \\ 
34 & 0.35493 & $16.00\pm0.09$ & $>13.52$ & $-2$ & $<13.18$ & $-1$ & $-2$ & $-2.49$ & $-0.67$ & $-1.17$ & $3.96$ & $13.51$ & $13.17$ \\ 
35 & 0.36074 & $15.14\pm0.01$ & $13.47\pm0.01$ & $0$ & $12.56\pm0.06$ & $0$ & $0*$ & $-2.94$ & $-0.50$ & $-0.91$ & $3.42$ & $13.47$ & $12.56$ \\ 
36 & 0.36921 & $15.48\pm0.06$ & $13.38\pm0.06$ & $0$ & $<13.04$ & $-1$ & $-2$ & $-2.43$ & $-0.28$ & $-1.62$ & $3.93$ & $13.43$ & $13.05$ \\ 
37 & 0.37771 & $15.31\pm0.09$ & $12.73\pm0.09$ & $0$ & $<12.71$ & $-1$ & $-2*$ & $-2.53$ & $-0.95$ & $-1.53$ & $3.82$ & $12.78$ & $12.25$ \\ 
38 & 0.37845 & $15.45\pm0.02$ & $13.78\pm0.02$ & $0$ & $<12.85$ & $-1$ & $-2*$ & $-3.19$ & $-0.78$ & $+0.00$ & $3.16$ & $13.80$ & $12.56$ \\ 
39 & 0.37954 & $15.94\pm0.05$ & $13.59\pm0.05$ & $0$ & $13.37\pm0.10$ & $0$ & $0$ & $-2.26$ & $-0.37$ & $-1.50$ & $4.09$ & $13.59$ & $13.37$ \\ 
40 & 0.38233 & $15.29\pm0.04$ & $13.43\pm0.04$ & $0$ & $<12.82$ & $-1$ & $-2*$ & $-2.81$ & $-0.60$ & $-0.96$ & $3.53$ & $13.45$ & $12.62$ \\ 
41 & 0.38243 & $15.18\pm0.08$ & $13.62\pm0.08$ & $0$ & $<13.12$ & $-1$ & $-2$ & $-2.59$ & $+0.06$ & $-1.58$ & $3.75$ & $13.69$ & $13.13$ \\ 
42 & 0.38759 & $16.18\pm0.02$ & $>14.06$ & $-2$ & $<13.39$ & $-1$ & $-2*$ & $-2.73$ & $-0.74$ & $-0.21$ & $3.61$ & $14.15$ & $13.37$ \\ 
43 & 0.40936 & $15.84\pm0.04$ & $13.59\pm0.04$ & $0$ & $<13.87$ & $-1$ & $-2*$ & $-2.66$ & $-0.85$ & $-0.65$ & $3.66$ & $13.67$ & $12.92$ \\ 
44 & 0.41510 & $16.15\pm0.13$ & $13.24\pm0.13$ & $0$ & $<13.48$ & $-1$ & $-2$ & $-1.85$ & $-0.31$ & $-2.09$ & $4.46$ & $13.29$ & $13.46$ \\ 
45 & 0.41565 & $15.84\pm0.01$ & $>14.03$ & $-2$ & $<13.38$ & $-1$ & $-2*$ & $-2.63$ & $-0.37$ & $-0.77$ & $3.68$ & $14.05$ & $13.38$ \\ 
46 & 0.41747 & $15.39\pm0.01$ & $13.96\pm0.01$ & $0$ & $<13.16$ & $-1$ & $-2*$ & $-2.65$ & $-0.10$ & $-1.18$ & $3.66$ & $13.93$ & $13.19$ \\ 
47 & 0.41774 & $15.70\pm0.01$ & $>14.03$ & $-2$ & $13.75\pm0.05$ & $0$ & $-2$ & $-2.31$ & $+0.22$ & $-1.58$ & $4.00$ & $14.08$ & $13.75$ \\ 
48 & 0.41969 & $15.79\pm0.12$ & $12.7\pm0.12$ & $0$ & $<13.04$ & $-1$ & $-2$ & $-2.21$ & $-1.04$ & $-1.70$ & $4.10$ & $12.77$ & $12.55$ \\ 
49 & 0.42079 & $15.83\pm0.03$ & $13.26\pm0.03$ & $0$ & $<13.08$ & $-1$ & $-2*$ & $-2.21$ & $-0.60$ & $-1.62$ & $4.10$ & $13.30$ & $13.05$ \\ 
50 & 0.42559 & $16.17\pm0.11$ & $13.40\pm0.11$ & $0$ & $<13.40$ & $-1$ & $-2*$ & $-2.26$ & $-0.81$ & $-1.19$ & $4.04$ & $13.49$ & $13.19$ \\ 
51 & 0.42731 & $15.66\pm0.03$ & $>14.24$ & $-2$ & $<13.55$ & $-1$ & $-2*$ & $-2.66$ & $-0.03$ & $-0.92$ & $3.64$ & $14.21$ & $13.52$ \\ 
52 & 0.43177 & $15.68\pm0.04$ & $12.97\pm0.04$ & $0$ & $<12.62$ & $-1$ & $-2$ & $-2.39$ & $-1.02$ & $-1.42$ & $3.91$ & $12.95$ & $12.52$ \\ 
53 & 0.43196 & $15.69\pm0.03$ & $>13.82$ & $-2$ & $<13.35$ & $-1$ & $-2*$ & $-2.47$ & $-0.24$ & $-1.24$ & $3.83$ & $13.84$ & $13.33$ \\ 
54 & 0.43221 & $15.62\pm0.09$ & $13.51\pm0.09$ & $0$ & $<13.13$ & $-1$ & $-2$ & $-2.27$ & $-0.29$ & $-1.73$ & $4.03$ & $13.46$ & $13.15$ \\ 
55 & 0.44609 & $15.94\pm0.01$ & $>13.80$ & $-2$ & $<13.05$ & $-1$ & $-2*$ & $-2.67$ & $-0.86$ & $-0.44$ & $3.61$ & $13.85$ & $13.03$ \\ 
56 & 0.44837 & $15.53\pm0.05$ & $13.93\pm0.05$ & $0$ & $13.36\pm0.15$ & $0$ & $0$ & $-2.35$ & $+0.10$ & $-1.63$ & $3.93$ & $13.90$ & $13.48$ \\ 
57 & 0.46370 & $15.26\pm0.02$ & $13.86\pm0.02$ & $0$ & $<12.96$ & $-1$ & $-2*$ & $-2.67$ & $-0.12$ & $-1.22$ & $3.60$ & $13.82$ & $13.05$ \\ 
58 & 0.47547 & $15.68\pm0.08$ & $>14.03$ & $-2$ & $13.69\pm0.16$ & $0$ & $-2$ & $-2.18$ & $+0.27$ & $-1.80$ & $4.08$ & $14.03$ & $13.76$ \\ 
59 & 0.48168 & $16.02\pm0.10$ & $13.42\pm0.1$ & $0$ & $<12.74$ & $-1$ & $-2*$ & $-2.57$ & $-1.27$ & $-0.54$ & $3.68$ & $13.45$ & $12.69$ \\ 
60 & 0.49263 & $15.62\pm0.06$ & $13.31\pm0.06$ & $0$ & $<13.13$ & $-1$ & $-2$ & $-2.02$ & $-0.19$ & $-2.17$ & $4.22$ & $13.30$ & $13.18$ \\ 
61 & 0.49450 & $15.45\pm0.02$ & $13.11\pm0.02$ & $0$ & $<12.87$ & $-1$ & $-2*$ & $-2.48$ & $-0.82$ & $-1.30$ & $3.76$ & $13.22$ & $12.55$ \\ 
62 & 0.49888 & $15.76\pm0.06$ & $13.59\pm0.06$ & $0$ & $<13.29$ & $-1$ & $-2*$ & $-2.42$ & $-0.60$ & $-1.15$ & $3.82$ & $13.64$ & $13.05$ \\ 
63 & 0.50297 & $15.76\pm0.08$ & $>14.07$ & $-2$ & $<13.09$ & $-1$ & $-2$ & $-2.94$ & $-0.65$ & $-0.10$ & $3.29$ & $14.12$ & $13.01$ \\ 
64 & 0.52147 & $15.92\pm0.05$ & $13.30\pm0.05$ & $0$ & $<13.79$ & $-1$ & $-2$ & $-2.02$ & $-0.48$ & $-1.83$ & $4.20$ & $13.37$ & $13.21$ \\ 
65 & 0.52208 & $15.77\pm0.06$ & $13.77\pm0.06$ & $0$ & $<13.36$ & $-1$ & $-2$ & $-2.49$ & $-0.47$ & $-1.00$ & $3.73$ & $13.85$ & $13.19$ \\ 
66 & 0.52794 & $15.30\pm0.01$ & $13.46\pm0.01$ & $0$ & $12.52\pm0.19$ & $0$ & $0$ & $-2.73$ & $-0.71$ & $-0.96$ & $3.48$ & $13.42$ & $12.50$ \\ 
67 & 0.53351 & $16.17\pm0.10$ & $>13.83$ & $-2$ & $<13.63$ & $-1$ & $-2*$ & $-2.35$ & $-0.73$ & $-0.83$ & $3.86$ & $13.90$ & $13.34$ \\ 
68 & 0.53459 & $16.15\pm0.11$ & $13.57\pm0.11$ & $0$ & $<13.52\pm0.16$ & $-1$ & $-2*$ & $-2.44$ & $-1.12$ & $-0.62$ & $3.77$ & $13.64$ & $12.96$ \\ 
69 & 0.53538 & $15.11\pm0.06$ & $13.09\pm0.06$ & $0$ & $<13.07$ & $-1$ & $-2$ & $-1.98$ & $+0.20$ & $-2.71$ & $4.23$ & $13.20$ & $13.06$ \\ 
70 & 0.53649 & $15.76\pm0.02$ & $13.76\pm0.02$ & $0$ & $<13.07$ & $-1$ & $-2$ & $-2.69$ & $-0.80$ & $-0.57$ & $3.52$ & $13.77$ & $12.87$ \\ 
71 & 0.54105 & $15.67\pm0.09$ & $13.29\pm0.09$ & $0$ & $<13.34$ & $-1$ & $-2$ & $-2.06$ & $-0.32$ & $-1.97$ & $4.14$ & $13.37$ & $13.14$ \\ 
72 & 0.54368 & $15.59\pm0.02$ & $>14.08$ & $-2$ & $<13.67$ & $-1$ & $-2$ & $-2.48$ & $+0.01$ & $-1.18$ & $3.72$ & $14.17$ & $13.49$ \\ 
73 & 0.55647 & $15.38\pm0.06$ & $>13.74$ & $-2$ & $<13.35$ & $-1$ & $-2*$ & $-2.45$ & $-0.05$ & $-1.52$ & $3.74$ & $13.79$ & $13.19$ \\ 
74 & 0.55753 & $15.79\pm0.04$ & $>13.92$ & $-2$ & $13.44\pm0.10$ & $0$ & $-2*$ & $-2.74$ & $-0.29$ & $-0.37$ & $3.45$ & $14.34$ & $13.42$ \\ 
75 & 0.55820 & $16.13\pm0.09$ & $12.91\pm0.09$ & $0$ & $<13.43$ & $-1$ & $-2$ & $-1.85$ & $-1.52$ & $-0.98$ & $3.89$ & $12.99$ & $12.48$ \\ 
76 & 0.55845 & $15.99\pm0.07$ & $13.01\pm0.07$ & $0$ & $13.13\pm0.13$ & $0$ & $0$ & $-1.82$ & $-0.64$ & $-2.13$ & $4.37$ & $13.04$ & $13.04$ \\ 
77 & 0.57183 & $15.46\pm0.16$ & $12.47\pm0.16$ & $0$ & $<12.57$ & $-1$ & $-2$ & $-1.98$ & $-0.85$ & $-2.31$ & $4.20$ & $12.56$ & $12.38$ \\ 
78 & 0.57649 & $15.50\pm0.06$ & $13.22\pm0.06$ & $0$ & $<13.08$ & $-1$ & $-2*$ & $-2.25$ & $-0.56$ & $-1.68$ & $3.92$ & $13.30$ & $12.81$ \\ 
79 & 0.58756 & $15.11\pm0.08$ & $13.49\pm0.08$ & $0$ & $<13.5$ & $-1$ & $-2$ & $-2.21$ & $+0.17$ & $-2.17$ & $3.95$ & $13.57$ & $13.13$ \\ 
80 & 0.59276 & $15.26\pm0.08$ & $13.38\pm0.08$ & $0$ & $<12.98$ & $-1$ & $-2$ & $-2.17$ & $-0.08$ & $-2.10$ & $3.99$ & $13.42$ & $13.02$ \\ 
81 & 0.59954 & $15.40\pm0.11$ & $13.22\pm0.11$ & $0$ & $<13.35$ & $-1$ & $-2$ & $-2.07$ & $-0.22$ & $-2.16$ & $4.08$ & $13.30$ & $12.99$ \\ 
82 & 0.59956 & $15.88\pm0.01$ & $>13.96$ & $-2$ & $13.85\pm0.02$ & $0$ & $-2*$ & $-2.27$ & $+0.09$ & $-1.24$ & $3.88$ & $14.29$ & $13.84$ \\ 
83 & 0.60827 & $15.75\pm0.14$ & $13.27\pm0.14$ & $0$ & $<13.83$ & $-1$ & $-2$ & $-1.71$ & $+0.01$ & $-2.55$ & $4.44$ & $13.35$ & $13.41$ \\ 
84 & 0.61025 & $15.26\pm0.04$ & $>13.87$ & $-2$ & $<12.80$ & $-1$ & $-2*$ & $-2.62$ & $-0.32$ & $-1.15$ & $3.52$ & $13.81$ & $12.84$ \\ 
85 & 0.61566 & $15.66\pm0.04$ & $>13.73$ & $-2$ & $13.16\pm0.09$ & $0$ & $-2*$ & $-2.63$ & $-0.41$ & $-0.66$ & $3.51$ & $14.12$ & $13.16$ \\ 
86 & 0.61751 & $15.28\pm0.04$ & $>13.71$ & $-2$ & $13.22\pm0.06$ & $0$ & $-2$ & $-2.47$ & $+0.01$ & $-1.42$ & $3.67$ & $13.84$ & $13.19$ \\ 
87 & 0.61913 & $16.10\pm0.04$ & $>14.34$ & $-2$ & $<12.96$ & $-1$ & $-2*$ & $-3.09$ & $-0.97$ & $+0.78$ & $3.05$ & $14.42$ & $12.95$ \\ 
88 & 0.62138 & $16.11\pm0.14$ & $13.25\pm0.14$ & $0$ & $<13.16$ & $-1$ & $-2$ & $-2.13$ & $-0.97$ & $-1.29$ & $4.01$ & $13.38$ & $12.98$ \\ 
89 & 0.62143 & $16.09\pm0.01$ & $>14.35$ & $-2$ & $13.90\pm0.02$ & $0$ & $-2$ & $-2.17$ & $-0.02$ & $-1.23$ & $3.97$ & $14.36$ & $13.92$ \\ 
90 & 0.64492 & $15.74\pm0.16$ & $13.74\pm0.16$ & $0$ & $<13.61$ & $-1$ & $-2$ & $-2.03$ & $-0.07$ & $-1.85$ & $4.09$ & $13.80$ & $13.48$ \\ 
91 & 0.65440 & $15.55\pm0.06$ & $13.41\pm0.06$ & $0$ & $<13.51$ & $-1$ & $-2$ & $-1.71$ & $+0.25$ & $-2.71$ & $4.40$ & $13.45$ & $13.47$ \\ 
92 & 0.65506 & $16.17\pm0.06$ & $13.39\pm0.06$ & $0$ & $<13.46$ & $-1$ & $-2$ & $-1.69$ & $-0.43$ & $-2.10$ & $4.42$ & $13.40$ & $13.42$ \\ 
93 & 0.68121 & $15.47\pm0.08$ & $12.88\pm0.08$ & $0$ & $<12.67$ & $-1$ & $-2$ & $-1.94$ & $-0.59$ & $-2.27$ & $4.15$ & $12.94$ & $12.67$ \\ 
94 & 0.68857 & $15.51\pm0.13$ & $12.57\pm0.13$ & $0$ & $<12.90$ & $-1$ & $-2$ & $-1.62$ & $-0.41$ & $-2.88$ & $4.46$ & $12.68$ & $12.75$ \\ 
95 & 0.68895 & $16.16\pm0.02$ & $>13.99$ & $-2$ & $<13.54$ & $-1$ & $-2*$ & $-2.56$ & $-0.92$ & $-0.19$ & $3.52$ & $14.11$ & $13.17$ \\ 
96 & 0.71796 & $15.25\pm0.10$ & $13.17\pm0.10$ & $0$ & $<12.93$ & $-1$ & $-2$ & $-2.01$ & $-0.23$ & $-2.30$ & $4.05$ & $13.23$ & $12.84$ \\ 
97 & 0.71896 & $15.61\pm0.11$ & $>13.74$ & $-2$ & $13.27\pm0.10$ & $0$ & $-2*$ & $-2.12$ & $-0.19$ & $-1.68$ & $3.94$ & $13.81$ & $13.29$ \\ 
98 & 0.80775 & $15.71\pm0.04$ & $>13.90$ & $-2$ & $<12.94$ & $-1$ & $-2*$ & $-2.55$ & $-0.76$ & $-0.55$ & $3.44$ & $13.96$ & $12.86$ \\ 
99 & 0.82917 & $16.02\pm0.05$ & $13.29\pm0.05$ & $0$ & $<12.70$ & $-1$ & $-2*$ & $-2.11$ & $-1.26$ & $-1.10$ & $3.87$ & $13.35$ & $12.68$ \\ 
\hline  
\end{longtable}
\end{center} 
\begin{tablenotes}[para, centering]
Note- Col ID (5): Flag on the column density measurement of \CIII. Col ID (7): Flag on the column density measurement of \CII. Flag {\lq{0}\rq} represents detections whereas {\lq{$-2$}\rq} and {\lq{$-1$}\rq} indicate lower limits and upper limits, respectively. Col ID (8): Flag on observed \nce$/N(\CII)$. Flag {\lq{0}\rq} indicates well constrained observed \nce$/N(\CII)$ (sample SA1), {\lq{$-2*$}\rq} indicates lower limits on observed \nce$/N(\CII)$ with other metal ions associations (sample SA2)  and {\lq{$-2$}\rq} indicates lower limits on observed \nce$/N(\CII)$ without presence of other metal ion combinations (sample SB).
\end{tablenotes}
 
\scriptsize
\newpage
\scriptsize
\begin{table} 
\scriptsize
\caption{Summary of \CIII\ + \SiIII\ selected absorbers in sub-sample SA2 and PI model predicted parameters.} 
\tabcolsep=3pt
\begin{tabular}{lcccccccc}
\hline
\hline
 & \multicolumn{ 5}{c}{Observed } & \multicolumn{ 3}{c}{PI model} \\
\cline{3-6} \noalign{\smallskip} \cline{7-9}  
IDs &$z_{abs}$ & log $N$(\SiIII) & Flag &log  $N$(\SiII) & Flag & $[Si/H]$ &log  $N$(\SiIII) &log  $N$(\SiII) \\
& & (\cmsq) &  & (\cmsq) & \multicolumn{1}{l}{} &  & (\cmsq) & (\cmsq)  \\
(1) & (2) & (3) & (4) & (5) & (6) & (7) & (8) & (9) \\ \hline
1& 0.21945 & $12.99\pm0.06$ & 0 & $<12.55$ & $-1$ &$-$0.30 & 12.99 & 12.4 \\
3 & 0.22719 & $12.84\pm0.05$ & 0 & $13.05\pm0.04$ & 0 & $+0.10$ & 12.91 & 13.08 \\
4 &0.22902 & $12.84\pm0.14$ & 0 & $<12.60$ & $-1$ &$-$0.29 & 12.93 & 12.37 \\
5 & 0.23881 & $12.85\pm0.04$ & 0 & $<12.10$ & $-1$ &$-$0.75 & 12.87 & 12.00 \\
6 & 0.25247 & $12.16\pm0.09$ & 0 & $<12.03$ & $-1$ &$-$0.24 & 12.10 & 12.09 \\
7 & 0.26382 & $12.99\pm0.09$ & 0 & $<12.84$ & $-1$ & $+$0.27 & 12.9 & 12.85 \\
8 & 0.28495 & $12.73\pm0.11$ & 0 & $<12.79$ & $-1$ & $+$0.14 & 12.63 & 12.79 \\
21 & 0.32451 & $12.99\pm0.11$ & 0 & $<12.91$ & $-1$ &$-$0.30 & 12.95 & 12.69 \\
$22^{@}$ & 0.32565 & $12.58\pm0.06$ & 0 & $12.57\pm0.17$ & 0 & $+$0.03 & 12.55 & 12.57 \\
25 & 0.33719 & $13.47\pm0.17$ &$-$2 & $<12.76$ & $-1$ & $+$0.16 & 13.39 & 12.75 \\
26 & 0.34793 & $12.91\pm0.05$ & 0 & $12.48\pm0.12$ & $-1$ &$-$0.75 & 12.85 & 12.43 \\
28 & 0.34833 & $12.55\pm0.05$ & 0 & $<12.14$ & $-1$ &$-$0.62 & 12.54 & 12.17 \\
29 & 0.34873 & $12.89\pm0.02$ & 0 & $<12.18$ & $-1$ &$-$0.95 & 12.81 & 12.17 \\
33 & 0.35471 & $12.95\pm0.17$ & 0 & $<12.60$ & $-1$ & $+$0.09 & 12.89 & 12.66 \\
$35^{@}$ & 0.36074 & $12.65\pm0.02$ & 0 & $<11.84$ & $-1$ &$-$0.31 & 12.58 & 11.85 \\
$38^*$ & 0.37845 & $13.22\pm0.02$ & 0 & $<12.02$ & $-1$ &$-$0.17 & 13.16 & 12.02 \\
45  & 0.41565 & $13.18\pm0.09$ & 0 & $<12.70$ & $-1$ &$-$0.26 & 13.09 & 12.72 \\
46 & 0.41747 & $12.68\pm0.06$ & 0 & $<12.51$ & $-1$ &$-$0.25 & 12.64 & 12.27 \\
$49^*$ &0.42079 & $12.31\pm0.09$ & 0 & $<12.59$ & $-1$ &$-$0.47 & 12.34 & 12.42 \\
50 & 0.42559 & $13.15\pm0.20$ &$-$2 & $<13.19$ & $-1$ &$-$0.22 & 12.91 & 13.08 \\
$51^*$ & 0.42731 & $13.29\pm0.22$ &$-$2 & $<13.19$ & $-1$ & $+$0.19 & 13.28 & 13.07 \\
53 & 0.43196 & $13.00\pm0.14$ &$-$2 & $<12.85$ & $-1$ &$-$0.07 & 12.91 & 12.78 \\
$55^*$ &0.44609 & $12.74\pm0.05$ & 0 & $<12.41$ & $-1$ &$-$0.87 & 12.8 & 11.99 \\
57 &0.46370 & $12.67\pm0.11$ & 0 & $<12.74$ & $-1$ &$-$0.12 & 12.71 & 12.25 \\
\hline 
\end{tabular} 
\label{tab:si} 
\begin{tablenotes}[para, centering]
Note:- 1. Col ID: (4) Flag in the column density measurement of \SiIII\ (6)  Flag in the column density measurement of \SiII. Flag {\lq{0}\rq} indicate detections whereas {\lq{$-2$}\rq} and {\lq{$-1$}\rq} indicate lower limits and upper limits, respectively. \\
2. The absorbers with IDs. 22 and 35 marked by {\lq{@}\rq} are included in SA1 as they have clean detection of \CIII\ $+$ \CII.\\
3. The absorbers marked with {\lq{*}\rq} have both \SiIII\ and \OIII\ detections. The same IDs are also marked in Table~\ref{tab:oxy}.
\end{tablenotes}       
\end{table}   

\begin{table} 
\scriptsize
\caption{Summary of \CIII\ + \OIII\ selected absorbers in sub-sample SA2 and PI model predicted parameters.}  
\tabcolsep=3pt
\begin{tabular}{lcccccccc}
\hline
\hline
 & \multicolumn{ 5}{c}{Observed } & \multicolumn{ 3}{c}{PI model} \\
\cline{3-6} \noalign{\smallskip} \cline{7-9} 
IDs & $z_{abs}$ &log  $N$(\OIII) & Flag &log  $N$(\OII) & Flag & $[O/H]$ &log  $N$(\OIII) &log  $N$(\OII) \\
& & (\cmsq) &  & (\cmsq) & \multicolumn{1}{l}{} &  & (\cmsq) & (\cmsq)  \\
(1) & (2) & (3) & (4) & (5) & (6) & (7) & (8) & (9)\\ \hline
37 & 0.37771 & $13.79\pm0.12$ & 0 & $<13.38$ &$-1$ &$-$0.24 & 13.68 & 13.37 \\ 
$38^*$ & 0.37845 & $14.49\pm0.08$ & 0 & $<13.78$ &$-1$ &$-$0.20 & 14.51 & 13.16 \\ 
40 & 0.38233 & $14.02\pm0.09$ & 0 & $<13.42$ &$-1$ &$-$0.24 & 14.03 & 13.25 \\ 
42 &0.38759 & $14.83\pm0.13$ &$-$2 & $<14.11$ &$-1$ &$-$0.30 & 14.81 & 14.10 \\ 
43 &0.40936 & $14.28\pm0.09$ & 0 & $<13.59$ &$-1$ &$-$0.52 & 14.21 & 13.58 \\ 
$49^*$ &0.42079 & $13.77\pm0.09$ & 0 & $<13.74$ &$-1$ &$-$0.31 & 13.67 & 13.80 \\ 
$51^*$ &0.42731 & $14.71\pm0.08$ & 0 & $<13.91$ &$-1$ & $+$0.20 & 14.66 & 14.05 \\ 
$55^*$ &0.44609 & $14.37\pm0.03$ & 0 & $<13.45$ &$-1$ &$-$0.70 & 14.23 & 13.47 \\ 
59 &0.48168 & $14.17\pm0.04$ & 0 & $13.54\pm0.10$ & 0 &$-$0.78 & 14.16 & 13.50 \\ 
61 &0.49450 & $13.52\pm0.08$ & 0 & $<12.98$ &$-1$ &$-$0.78 & 13.47 & 12.97 \\ 
62 &0.49888 & $14.25\pm0.09$ & 0 & $<13.85$ &$-1$ &$-$0.24 & 14.21 & 13.80 \\ 
67 &0.53351 & $14.50\pm0.21$ &$-$2 & $14.12\pm0.27$ & 0 &$-$0.33 & 14.50 & 14.12 \\ 
68 &0.53459 & $14.25\pm0.15$ & 0 & $<13.69$ &$-1$ &$-$0.75 & 14.23 & 13.68 \\ 
73 &0.55647 & $14.53\pm0.17$ &$-$2 & $<13.87$ &$-1$ & $+$0.38 & 14.44 & 13.94 \\ 
74 &0.55753 & $14.75\pm0.04$ &$-$2 & $<13.68$ &$-1$ &$-$0.18 & 14.73 & 13.68 \\ 
78 &0.57649 & $13.89\pm0.16$ & 0 & $<13.64$ &$-1$ &$-$0.15 & 13.90 & 13.63 \\ 
82 &0.59956 & $14.38\pm0.01$ & 0 & $14.05\pm0.02$ & 0 &$-$0.08 & 14.41 & 14.06 \\ 
84 &0.61025 & $14.31\pm0.05$ & 0 & $<13.57$ &$-1$ & $+$0.04 & 14.31 & 13.41 \\ 
85 &0.61566 & $14.59\pm0.03$ &$-$2 & $13.54\pm0.08$ & 0 &$-$0.20 & 14.54 & 13.56 \\ 
87 &0.61913 & $15.18\pm0.03$ &$-$2 & $<13.47$ &$-1$ &$-$0.36 & 15.13 & 13.44 \\ 
95 &0.68895 & $14.90\pm0.04$ &$-$2 & $<13.86$ &$-1$ &$-$0.42 & 14.86 & 13.83 \\ 
97 &0.71896 & $14.04\pm0.10$ & 0 & $<13.66$ &$-1$ &$-$0.23 & 13.96 & 13.65 \\ 
98 &0.80775 & $14.41\pm0.02$ & 0 & $<13.29$ &$-1$ &$-$0.49 & 14.41 & 13.24 \\ 
99 &0.82917 & $13.42\pm0.09$ & 0 & $<13.15$ &$-1$ &$-$1.37 & 13.45 & 12.93 \\ 
\hline 
\end{tabular} 
\label{tab:oxy} 
\begin{tablenotes}[para, centering]
Note- Col ID: (4) Flag in the column density measurement of \OIII\ (6)  Flag in the column density measurement of \OII. Flag {\lq{0}\rq} indicate detections whereas {\lq{$-2$}\rq} and {\lq{$-1$}\rq} indicate lower limits and upper limits, respectively. 
\end{tablenotes}        
\end{table}   

\bsp
\label{lastpage1}
\end{document}